\documentclass[12pt]{article}
\usepackage{epsfig,amssymb}

\newcommand{\bq}{\begin{equation}}
\newcommand{\eq}{\end{equation}}
\newcommand{\bqa}{\begin{eqnarray}}
\newcommand{\eqa}{\end{eqnarray}}
\newcommand{\ben}{\begin{enumerate}}
\newcommand{\een}{\end{enumerate}}
\newcommand{\bc}{\begin{center}}
\newcommand{\ec}{\end{center}}
\newcommand{\bqb}{\begin{eqnarray*}}
\newcommand{\eqb}{\end{eqnarray*}}

\def\pr#1#2#3{ Phys. Rev. ${\bf{#1}}$, #2 (#3)}
\def\prl#1#2#3{ Phys. Rev. Lett. ${\bf{#1}}$, #2 (#3)}
\def\pl#1#2#3{ Phys. Lett. ${\bf{#1}}$, #2 (#3)}
\def\prep#1#2#3{ Phys. Rep. ${\bf{#1}}$, #2 (#3)}

\def\np#1#2#3{ Nucl. Phys. ${\bf{#1}}$, #2 (#3)}
\def\zp#1#2#3{ Z. f. Phys. ${\bf{#1}}$, #2 (#3)}

\def\arnps#1#2#3{Ann. Rev. Nucl. Part. Sci. ${\bf{#1}}$, #2 (#3) }

\def\ie{{\it i.e.\/}}
\def\eg{{\it e.g.\/}}

\def\etal{{\it et.al.\/}}

\global\nulldelimiterspace = 0pt

\def\wtil#1{\widetilde{#1}}

\def\L{ {\cal L }}

\def\A{ {\cal A }}

\def\V{ {\cal V }}
\def\O{ {\cal O }}
\def\G{ {\cal G }}
\def\sw{s_W}
\def\cw{c_W}

\def\mzd{m_Z^2}

\begin{document}
\pagenumbering{arabic}
\thispagestyle{empty}
\def\thefootnote{\fnsymbol{footnote}}
\setcounter{footnote}{1}

\begin{flushright}
PM/00-16 \\
THES-TP 2000/05 \\
hep-ph/0005269 \\
May 2000\\

 \end{flushright}
\vspace{2cm}
\begin{center}
{\Large\bf Off-shell structure of the anomalous $Z$ and
$\gamma$ self-couplings}\footnote{Partially supported by the European
Community grant ERBFMRX-CT96-0090.}
 \vspace{1.5cm}  \\
{\large G.J. Gounaris$^a$, J. Layssac$^b$  and F.M. Renard$^b$}\\
\vspace{0.7cm}
$^a$Department of Theoretical Physics, Aristotle
University of Thessaloniki,\\
Gr-54006, Thessaloniki, Greece.\\
\vspace{0.2cm}
$^b$Physique
Math\'{e}matique et Th\'{e}orique,
UMR 5825\\
Universit\'{e} Montpellier II,
 F-34095 Montpellier Cedex 5.\\
\vspace{0.2cm}

\vspace*{1cm}

{\bf Abstract}
\end{center}

We establish the general off-shell structure of the three
neutral gauge boson self-couplings $V_1^*V_2^*V_3^*$,
with applications to the
$Z^*Z^*Z^*,~Z^*Z^*\gamma^*,\gamma^*\gamma^*Z^*$ cases. New coupling
forms appear which do not exist when two gauge bosons are on-shell.
We give the contribution arising from a fermionic triangle loop.
It covers both the standard model (SM) and possible new physics (NP)
contributions like those arising in the MSSM. For what concerns
NP contributions with a high scale, we discuss the validity of
an effective Lagrangian involving a limited set of parameters.
Finally we write the general expression of the
$V_1^*V_2^*V_3^*$-vertex
contribution to the $e^+e^- \to (f\bar f)+(f'\bar f')$ amplitude.\par

\def\thefootnote{\arabic{footnote}}
\setcounter{footnote}{0}
\clearpage

\section{Introduction}

The phenomenological description of Neutral Anomalous Gauge Couplings
(NAGC) among the photon and $Z$ was established in
\cite{Hagn, GG} and used for the discussion of their observability
at various types of present and future colliders, see
\cite{Baur, Wudka, work, Biebel, Alcaraz}. It has recently
been reexamined  and examples of new physics (NP) contributions
have been discussed \cite{neut, modzg}.
After the first events obtained at the TEVATRON \cite{TEVZg},
experimental data are now being collected at LEP2 \cite{expLEP2}
through the processes $e^+e^-\to ZZ$ and $Z\gamma$. New
possibilities will be offered by linear $e^+e^-$ colliders
LC \cite{LC} and CLIC \cite{CLIC}.\par

The description used  in \cite{Hagn, neut, modzg} applies to the
case where one neutral gauge boson $V^*$ is
off-shell\footnote{This off-shell
state is below indicated by an asterisk.}
 and  coupled to $e^+e^-$, ($V^*=\gamma^*$ or $Z^*$);
while the other two neutral  gauge bosons $ZZ$ or $Z\gamma$
are on-shell.
However, a large set of events collected at LEP2
\cite{LEP2off}, consists of 4-fermion states (like $l\bar lq\bar
q$),
in which the invariant mass of the $l\bar l$ or $q\bar q$ pair varies
from about $10~GeV$ up to the $Z$ mass. For   analyzing  these
events through the processes
$e^+e^-\to Z^*Z^*, ~ Z^*\gamma^*, ~\gamma^*\gamma^*$,
taking into account\footnote{Note that electromagnetic gauge
invariance prohibits any $\gamma^*\gamma^*\gamma^*$ vertex.}
contributions from
$V^*\to Z^*Z^*,Z^*\gamma^*,\gamma^*\gamma^*$;
one needs a description of the off-shell $V^*_iV^*_jV^*_k$
vertex. The usual two-particle-on-shell
vertices  for $Z^*ZZ$, $Z^*Z\gamma$,
$\gamma^*ZZ$, $\gamma^* Z\gamma$, which are forced by Bose
statistics to vanish whenever $V^*$ goes on-shell, are not adequate
to describe  $V^*V^*V^*$,  since additional $q^2_i$ dependences and
new coupling forms may be generated, which cannot be ignored.
Some attempts to treat these off-shell effects exist in the literature
for the $V^*\to Z^*Z^*$ case \cite{AFS}, but a complete treatment
is still lacking.\par

It is the purpose of this paper to present and discuss the general
description of the $V^*_iV^*_jV^*_k$ off-shell couplings.
We proceed in several steps.\par

In Section 2 and the  Appendices A and B, we establish the most
general form for a $V^*V^*V^*$ vertex involving   three
off shell neutral gauge bosons (NGB). For completeness,
we also include   the "scalar"
$q.V$ terms, contributing in the case that one off-shell
$Z$ decays to a  heavy fermion  pair, through its
axial coupling.
The only assumptions used are Lorentz invariance, Bose statistics
and $U(1)_{em}$ invariance; separately for the CP-conserving and  the
CP-violating cases. We make explicit applications to the $Z^*Z^*Z^*$,
$Z^*Z^*\gamma^*$ and $\gamma^*\gamma^*Z^*$ couplings, and we point out
the new coupling forms which do not exist when two particles are
on-shell, thus making contact with the previous
description \cite{neut, modzg}. These general vertices apply
to any SM or NP contribution.\par

In Section 3 we consider an  effective Lagrangian parametrization
which could apply to the case that the NP scale $\Lambda$
is very high; \ie~ $\Lambda \gg m_Z$. We
show that the effective Lagrangian previously considered in
\cite{neut} when two NGB are
on-shell, already contains some of the  off-shell forms; but
new operators must be added in order to describe all  possible
ones. These operators involve higher dimensions, so a hierarchy
may appear  among the various possible off-shell
effects, which is quite natural in this $\Lambda \gg m_Z$ case.\par

In Section 4 we look for a possible dynamical origin of these
couplings. Virtual SM or NP contributions may  indeed
generate various off-shell  NAGC. We describe them  by
generalizing  the procedure of  \cite{modzg} based on triangle
fermionic loops, already considered in \cite{FMR, BBCD}.
In Appendix C we give the  complete
expression of the off-shell $V^*_iV^*_jV^*_k$ vertices  generated by such
fermionic loops. This is useful for the computation of the SM and
the MSSM or NP contributions, and it also allows to illustrate how the
type of
off-shell effects changes, as the NP scale increases from  the
100 GeV level  to the multi-TeV one. Typical figures are presented,
illustrating the dependence of the various neutral gauge couplings
on  the off-shell masses, the relative size of these  couplings as
compared to their on-shell values, and the range of the
NP scales  for which an
effective Lagrangian description in terms of low dimension
operators,  is adequate.\par

In Section 5, we write, for completeness,
the general structure of the $V^*V^*V^*$ contribution
to 4 fermion amplitude, $e^+e^- \to (f\bar f)+(f'\bar f')$,
including all off-shell contributions.\par

The results are summarized in Section 6, where the
conclusions are also given.

\section{Description of off-shell neutral self-boson couplings}

The general procedure for determining the off-shell $V^*_1V^*_2V^*_3$
couplings is described in Appendix A
for the CP-conserving couplings and in Appendix B for
the CP-violating ones. We use the notations of Fig.1 for
the general off-shell
$V^{\alpha}_1(q_1)V^{\beta}_2(q_2)V^{\mu}_3(q3)$ vertex
(all $q_i$ being outgoing momenta\footnote{In
previous works \cite{neut, modzg}
$P\equiv-q_3$ was used for the initial off-shell boson.}).
The results can be summarized as follows:

\vspace{0.5cm}
\subsection{  $Z^*Z^*Z^*$ couplings:}

\underline{There are six CP-conserving independent forms}
listed in Appendix A, which are
multiplied by six  coupling functions denoted
as
\[
f^{Z^*Z^*Z^*}_i(s_1,s_2,s_3), ~~ (i=1-3),~ \mbox{and}
~~ g^{Z^*Z^*Z^*}_i(s_1,s_2,s_3),~~ (i=1-3)~.
\]
As in (\ref{ZZZ-CPc-ap})
we write the vertex interaction as
\bqa
\Gamma^{Z^*Z^*Z^*}_{\alpha\beta\mu}(q_1, q_2, q_3) &= &
i \sum_{i=1}^3 I^{Z^*Z^*Z^*,i}_{\alpha\beta\mu}
f^{Z^*Z^*Z^*}_i(s_1,s_2,s_3) \nonumber \\
&+ & i \sum_{i=1}^3 J^{Z^*Z^*Z^*,i}_{\alpha\beta\mu}
g^{Z^*Z^*Z^*}_i(s_1,s_2,s_3) ~ , \label{ZZZ-CPc}
\eqa
where the kinematics are defined in Fig.\ref{vvv-fig}.\par

The three $I^i$ and the three $J^i$ forms are given in
(\ref{ZZZ-CPc-forms}). We note that the
 $J^i$ forms, associated to  $g_i$,
involve at least one scalar $q.V$-factor, and they are thus
are called "scalar". In contrast  to them, the three
$I^i$ terms  associated to
$f_i$, do not involve $q.V$-factors and they are called "transverse".
These $f_i,~ g_i$ are functions of $s_1,s_2,s_3$
and satisfy the  Bose symmetry
relations presented in (\ref{ZZZ-CPc-Bose}) of  Appendix A.
We note in particular from them, that $f_3(s_1, s_2, s_3)$ is fully
antisymmetric. \par

In  case two of the $Z$'s are on-shell, say \eg~
 $s_1=s_2=m^2_Z$, Bose statistics forces
$f^{Z^*Z^*Z^*}_2$, $f^{Z^*Z^*Z^*}_3$ to vanish, leaving
only one
non-vanishing transverse coupling, corresponding  to
$f^Z_5$ defined in \cite{Hagn, neut, modzg} and satisfying
\bq
f^{Z^*Z^*Z^*}_1(m^2_Z,m^2_Z,s_3)\equiv {s_3-m^2_Z\over
m^2_Z}f^Z_5(s_3) ~ , \label{ZZZ-CPc-f-on}
\eq
where we have emphasized the fact that generally $f^Z_5$ is
not necessarily constant, but rather a  form-factor depending
on\footnote{A similar emphasis of their form-factor nature
is made in this Section for all NAGC defined in
\cite{Hagn, neut, modzg}.} $s_3$.
In this on-shell case there remains also one "scalar" term
\[
g^{Z^*Z^*Z^*}_3(m^2_Z,m^2_Z,s_3) ~~~ ,
\]
which  contributes only when the off-shell $Z^*$ couples to a heavy
fermion pair (like \eg~ $t \bar t$) at a "mass"-squared $s_3$.
Such terms had been previously neglected.\par

Thus, comparing the on- and off-shell situations, we remark
that in the off-shell case we have  in addition two more
"transverse" couplings
and another two   "scalar" ones.\\

\noindent
\underline{In the CP-violating case}
 there exist  14 independent forms, listed in Appendix B.
Defining the kinematics as before through Fig.\ref{vvv-fig},
we write (compare (\ref{ZZZ-CPv-ap}))
\bqa
\Gamma^{Z^*Z^*Z^*}_{\alpha\beta\mu}(q_1, q_2, q_3)&= &i\sum_{i=1}^4
\tilde{I}^{Z^*Z^*Z^*,i}_{\alpha\beta\mu}
\tilde{f}^{Z^*Z^*Z^*}_i(s_1,s_2,s_3) \nonumber \\
&+ & i \sum_{i=1}^{10} \tilde{J}^{Z^*Z^*Z^*,i}_{\alpha\beta\mu}
\tilde{g}^{Z^*Z^*Z^*}_i(s_1,s_2,s_3) ~ , \label{ZZZ-CPv}
\eqa
where the four $\tilde{I}^i$ are transverse,
while the 10 $\tilde{J}^i$ are scalar.
They are listed in (\ref{ZZZ-CPv-forms-t},
\ref{ZZZ-CPv-forms-s}) and imply
the Bose  constraints
(\ref{ZZZ-CPv-Bose-t}, \ref{ZZZ-CPv-Bose-s})
for the corresponding coupling
functions $(\tilde f_i, ~\tilde g_j)$.\par

In case two of the $Z$'s are on-shell ($s_1=s_2=\mzd $), then
 $\tilde{f}_{1},~\tilde{f}_{4}$ vanish,  while the
other two  transverse functions  are opposite to each  other,
because of Bose symmetry. So only one transverse combination remains,
related  to the coupling constant
$f^Z_4$ defined in \cite{ Hagn, neut, modzg}, through
\bq
 \tilde{f}^{Z^*Z^*Z^*}_2(m^2_Z,m^2_Z,s_3)=
-\tilde{f}^{Z^*Z^*Z^*}_3(m^2_Z,m^2_Z,s_3)
= {m^2_Z-s_3\over2m^2_Z}f^Z_4(s_3) ~ , \label{ZZZ-CPv-f-on}
\eq
\noindent
and  the two scalar ones
\[
\tilde{g}^{Z^*Z^*Z^*}_1(m^2_Z,m^2_Z,s_3)~~,~~~
\tilde{g}^{Z^*Z^*Z^*}_{6}(m^2_Z,m^2_Z,s_3) ~
.
\]

Comparing with the results of Appendix B and with those
of the on-shell treatment of \cite{ Hagn, neut, modzg},
we conclude that in the general CP-violating off-shell case,
there are in addition
 two transverse and eight scalar terms.

\vspace{0.5cm}
\subsection{  $Z^*Z^*\gamma^*$ couplings:}

Now, there are \underline{five CP-conserving independent forms}
defined in Appendix A through, (compare
(\ref{ZZgamma-CPc-ap}, \ref{ZZgamma-CPc-forms}))
\bqa
\Gamma^{Z^*Z^*\gamma^*}_{\alpha\beta\mu}(q_1, q_2, q_3)&= &
i \sum_{i=1}^3 I^{Z^*Z^*\gamma^*,i}_{\alpha\beta\mu}
f^{Z^*Z^*\gamma^*}_i(s_1,s_2,s_3) \nonumber \\
& + & i \sum_{i=1,2}J^{Z^*Z^*\gamma^*,i}_{\alpha\beta\mu}
g^{Z^*Z^*\gamma^*}_i(s_1,s_2,s_3) ~~ .
\eqa
Three of them,   $f^{Z^*Z^*\gamma^*}_i(s_1,s_2,s_3)$, $(i=1,2,3)$
are transverse; while  the CVC constraint
$q^{\mu}_3~\Gamma^{Z^*Z^*\gamma^*}_{\alpha\beta\mu}(s_1,s_2,s_3)=0$
reduces the  number of the "scalar" terms to
the two ones $g^{Z^*Z^*\gamma^*}_i(s_1,s_2,s_3)$, $(i=1,2)$.
These functions are submitted to the ($Z^*Z^*$) Bose symmetry
relations appearing  in  (\ref{ZZgamma-CPc-Bose}).\par

In case the two $Z$'s are on-shell ($s_1=s_2=m^2_Z$), Bose symmetry
forces two of the transverse functions to vanish, while
the two "scalar" ones
become  inefficient, as they are proportional to
 $q^{\alpha}_1$ or $q^{\beta}_2$. Thus, we end up with
 only one (transverse) coupling, corresponding
to $f^{\gamma}_5$ defined in \cite{ Hagn, neut, modzg}:
\bq
f^{Z^*Z^*\gamma^*}_1(m^2_Z,m^2_Z,s_3)\equiv
{s_3\over m^2_Z}f^{\gamma}_5(s_3) ~ ~. \label{ZZgamma-CPc-f1-on}
\eq \par

If  only one  $Z$ and the photon are on-shell
(\ie~ $s_1=m^2_Z$, $s_3=0$),  we remain instead  with
two transverse combinations corresponding to the
couplings $h^{Z}_{3,4}$
defined in \cite{ Hagn, neut, modzg}:
\bqa
f^{Z^*Z^*\gamma^*}_2(m^2_Z,s_2,0) & = &
{m^2_Z-s_2\over m^2_Z}~[h^{Z}_3(s_2)+{m^2_Z-s_2\over
4m^2_Z}h^{Z}_4(s_2)]~~, \nonumber \\
f^{Z^*Z^*\gamma^*}_3(m^2_Z,s_2,0) & =&
{m^2_Z-s_2\over 2m^4_Z}~ h^{Z}_4(s_2) ~~ ,
\label{ZZgamma-CPc-f23-on}
\eqa
\noindent
and  one "scalar" term
\[
g^{Z^*Z^*\gamma^*}_2(m^2_Z,s_2,0) ~~,
\]
since the other scalar term  contains a factor
$q^{\alpha}_1$ making it   inefficient on-shell.\par

Thus, in the general off-shell case, the three
transverse functions can be considered as
a  generalization (due to the $(s_1, s_2, s_3)$-dependence),
of the three on-shell couplings $f^{\gamma}_5,
h^{Z}_3, h^{Z}_4$. There are also
two scalar functions, previously
neglected.\\

\noindent
\underline{In the CP-violating case},
there are nine coupling forms,  of which the four
$\tilde{I}^{Z^*Z^*\gamma^*,i}_{\alpha\beta\mu}$ (i=1-4)
are  transverse, while the  five
$\tilde{J}^{Z^*Z^*\gamma^*,i}_{\alpha\beta\mu}$ (i=1-5)
are scalar. They are listed (\ref{ZZgamma-CPv-forms}).
In terms of them, the corresponding neutral gauge self interactions
is defined through (compare (\ref{ZZgamma-CPv-ap}))
\bqa
\Gamma^{Z^*Z^*\gamma^*}_{\alpha\beta\mu}(q_1, q_2, q_3)
&= & i \sum_{i=1}^4
\tilde{I}^{Z^*Z^*\gamma^*,i}_{\alpha\beta\mu}
\tilde{f}^{Z^*Z^*\gamma^*}_i(s_1,s_2,s_3) \nonumber \\
 &+ & i \sum_{i=1}^5 \tilde{J}^{Z^*Z^*\gamma^*,i}_{\alpha\beta\mu}
\tilde{g}^{Z^*Z^*\gamma^*}_i(s_1,s_2,s_3) ~ .
\label{ZZgamma-CPv}
\eqa\par

For $\gamma^*\to ZZ$ with the two $Z$'s being on-shell
($s_1=s_2=m^2_Z$), $\tilde{f}_{1,3,4}$ vanish because of
Bose symmetry; compare (\ref{ZZgamma-CPv-Bose}). In such a case the
only remaining coupling is a transverse one  related to $f^{\gamma}_4$
defined in \cite{ Hagn, neut, modzg} through
\bq
\tilde{f}^{Z^*Z^*\gamma^*}_2(m^2_Z,m^2_Z,s_3) = -~{s_3\over2m^2_Z}
 f^{\gamma}_4(s_3) ~ . \label{ZZgamma-CPv-f-on}
\eq
\noindent
No scalar term remains because $q^{\alpha}_1$ , $q^{\beta}_2$ give
no on-shell contribution.\par

For $Z^*\to Z\gamma$ with one real $Z$ ($s_1=m^2_Z$)
and one real $\gamma$ ($s_3=0$), $\tilde{f}_1$ vanishes and
$\tilde{f}_2$ is related to $\tilde{f}_4$ because of the
CVC constraint. We  thus end up with the two transverse functions
related to the $h^{Z}_{1,2}$ couplings
defined in \cite{ Hagn, neut, modzg} by
\bqa
\tilde{f}^{Z^*Z^*\gamma^*}_2(m^2_Z,s_2,0)
& = & -(s_2-m^2_Z)\tilde{f}^{Z^*Z^*\gamma^*}_4(m^2_Z,s_2,0)
={(s_2-m^2_Z)^2\over8m^4_Z} h^{Z}_2(s_2) ~ , \nonumber \\
\tilde{f}^{Z^*Z^*\gamma^*}_3(m^2_Z,s_2,0) & = &
{s_2-m^2_Z\over2m^2_Z}\left [-h^{Z}_1(s_2)
+{s_2-m^2_Z\over4m^2_Z}h^{Z}_2(s_2) \right ] ~ ,
\label{ZZgamma-CPv-f23-on}
\eqa
\noindent
and the two scalar combinations
\[
(\tilde{g}^{Z^*Z^*\gamma^*}_1(m^2_Z,s_2,0)
-\tilde{g}^{Z^*Z^*\gamma^*}_2(m^2_Z,s_2,0)) ~~,~ ~
(\tilde{g}^{Z^*Z^*\gamma^*}_3(m^2_Z,s_2,0)
-\tilde{g}^{Z^*Z^*\gamma^*}_4(m^2_Z,s_2,0)) ~,
\]
previously neglected.\par

So the general off-shell case involves two more transverse couplings
and three more scalar ones.\par

\vspace{0.5cm}
\subsection{ $\gamma^*\gamma^*Z^*$ couplings}

 There are four invariant forms \underline{in the CP-conserving case},
listed in Appendix A (compare (\ref{gammagammaZ-CPc-ap},
\ref{gammagammaZ-CPc-forms}))
\bqa
\Gamma^{\gamma^*\gamma^*Z^*}_{\alpha\beta\mu}(q_1, q_2, q_3) &= &
 i\sum_{i=1}^3 I^{\gamma^*\gamma^*Z^*,i}_{\alpha\beta\mu}
f^{\gamma^*\gamma^*Z^*}_i(s_1,s_2,s_3) \nonumber \\
&+ &i J^{\gamma^*\gamma^*Z^*,1}_{\alpha\beta\mu}
g^{\gamma^*\gamma^*Z^*}_1(s_1,s_2,s_3) ~ , \label{gammagammaZ-CPc}
\eqa
\noindent
including again the
three transverse functions $f^{\gamma^*\gamma^*Z^*}_i(s_1,s_2,s_3)$ (i=1-3),
but  only  one  scalar  $g^{\gamma^*\gamma^*Z^*}_1(s_1,s_2,s_3)$.
We note that this reduction of the number of scalar forms is
due to the two CVC constraints
\[
q^{\alpha}_1~
\Gamma^{\gamma^*\gamma^*Z^*}_{\alpha\beta\mu}(s_1,s_2,s_3)=
q^{\beta}_2~
\Gamma^{\gamma^*\gamma^*Z^*}_{\alpha\beta\mu}(s_1,s_2,s_3)=0 ~ ,
\]
\noindent
and the  Bose symmetry between  the two photons.\par

When one photon and one $Z$ are on-shell ($s_2=0$, $s_3=m^2_Z$),
these forms reduce to two independent transverse ones
corresponding to the couplings $h^{\gamma}_{3,4}$
defined in \cite{ Hagn, neut, modzg}:
\bqa
f^{\gamma^*\gamma^*Z^*}_1(s_1,0,m^2_Z)& = & {s_1\over2m^2_Z}
~[ h^{\gamma}_{3}(s_1)-~{s_1\over2m^2_Z}h^{\gamma}_{4}(s_1)]
~ , \nonumber \\
f^{\gamma^*\gamma^*Z^*}_2(s_1,0,m^2_Z) & = & {s_1\over2m^2_Z}
~[ h^{\gamma}_{3}(s_1)+{m^2_Z-2s_1\over2m^2_Z}h^{\gamma}_{4}(s_1)]
 ~ , \nonumber \\
f^{\gamma^*\gamma^*Z^*}_3(s_1,0,m^2_Z)
&= & -~{s_1\over2m^4_Z}~ h^{\gamma}_{4}(s_1) ~ ,
\label{gammagammaZ-CPc-f-on}
\eqa
and  one previously neglected scalar term
\[
g^{\gamma^*\gamma^*Z^*}_1(s_1,0,m^2_Z) ~ .
\]
So one sees that the general off-shell situation has one more
(transverse) form
than in the previously studied on-shell case.\\

\noindent
\underline{In the CP-violating case} there are only six forms
\bqa
\Gamma^{\gamma^*\gamma^*Z^*}_{\alpha\beta\mu}(q_1, q_2, q_3)
& = & i \sum_{i=1} ^4
\tilde{I}^{\gamma^*\gamma^*Z^*,i}_{\alpha\beta\mu}
\tilde{f}^{\gamma^*\gamma^*Z^*}_i(s_1,s_2,s_3) \nonumber \\
& + & i\sum_{i=1,2}\tilde{J}^{\gamma^*\gamma^*Z^*,i}_{\alpha\beta\mu}
\tilde{g}^{\gamma^*\gamma^*Z^*}_i(s_1,s_2,s_3) ~ ,
\eqa
\noindent
four of which are  transverse
$\tilde{f}^{\gamma^*\gamma^*Z^*}_i(s_1,s_2,s_3)$ (i=1-4)
and two scalar ones
$\tilde{g}^{\gamma^*\gamma^*Z^*}_i(s_1,s_2,s_3)$ (i=1,2);
see (\ref{gammagammaZ-CPv-forms}, \ref{gammagammaZ-CPv-Bose}))
in  Appendix B.\par

When one photon and one $Z$ are on-shell, ($s_2=0$, $s_3=m^2_Z$,
$q^{\beta}_2\equiv q^{\mu}_3\equiv0$), one remains with only
two independent transverse forms related
to the couplings $h^{\gamma}_{1,2}$
defined in \cite{ Hagn, neut, modzg}:
\bqa
\tilde{f}^{\gamma^*\gamma^*Z^*}_1(s_1,0,m^2_Z) &= &
{s_1\over2m^2_Z} ~[h^{\gamma}_1(s_1)
-~{s_1-m^2_Z\over2m^2_Z}h^{\gamma}_2(s_1)]
~ , \nonumber \\
\tilde{f}^{\gamma^*\gamma^*Z^*}_2(s_1,0,m^2_Z)
& = & \tilde{f}^{\gamma^*\gamma^*Z^*}_3(s_1,0,m^2_Z)=
-~{s_1\over4m^2_Z} h^{\gamma}_1(s_1)
~ , \nonumber \\
\tilde{f}^{\gamma^*\gamma^*Z^*}_4(s_1,0,m^2_Z) &=&
-~{s_1\over8m^4_Z} h^{\gamma}_2(s_1) ~,
\label{gammagammaZ-CPv-f-on}
\eqa
\noindent
and no "scalar" term.\par

Therefore, the general off-shell case for this vertex has
two more transverse
terms and two more scalar ones.\par

\section{The effective Lagrangian description}

The effective Lagrangian is an adequate formalism
 to  describe  the NP effects generated at a
 scale $\Lambda$, which is much higher than the actual energy
(or external mass) in the process considered ($\sqrt{s_i}$ or $M_Z$).
In this case,
it is natural  to restrict the set of operators to those with
the lowest possible dimensions; (the higher dimension contributions being
depressed by powers of $s_i/\Lambda^2 $),
and thus reducing somewhat the number
of free parameters. Of course, the dimension of the operators
needed to generate each specific form of interactions vertex, may
strongly depend on it.

Below, for each NAGC type of vertex ($Z^*Z^*Z^*$, $Z^*Z^*\gamma^*$,
$\gamma^*\gamma^* Z^*$),
we first establish a set of operators,
with the lowest possible dimension, which can generate the vertex forms
established in Section 2. Each such  lowest dimensional operator
generating a given vertex form ($I_i, ... $ or $ J_i ...$) produces a
coupling function ($f_i(s_1,s_2,s_3), ...$ or $ g_i(s_1,s_2,s_3), ...$)
characterized
by the lowest power of $s_i$ consistent with the
corresponding Bose constraints presented in Appendices A, B.
Thus, a constant $f_j$ appears in the case of a fully symmetric
function, a factor $(s_i-s_j)$ for a function antisymmetric in
the exchange of $s_i$ and $s_j$, ...etc.\par

 The lowest dimensional operators
contributing to NAGC have\footnote{For the CP-violating
$Z^*Z^*Z^*$ case, there exist a single operator of $dim=4$
which is of course also included, see below.} mainly $dim=6$.
We therefore start by enumerating all of them.
It turns out though, that this list operators is not
 sufficient to generate all  vertex forms. We therefore proceed
to  include also a minimal set of higher
dimensional operators which   generate the missing vertices. This
 constitutes what we call the basic effective Lagrangian
 expressed as
\bq
\L=e~(~\sum_i l_i\O_i+\sum_i \tilde{l}_i\tilde{\O}_i~) ~ ,
\label{NP-Lagrangian}
\eq
\noindent
where the operators $\O_i$ and $\tilde{\O}_i$ are CP-conserving
and CP-violating  respectively, while  $l_i$ and
$\tilde{l}_i$ are their corresponding (dimensional) coupling
constants.\par

\subsection{The $Z^*Z^*Z^*$ CP-conserving operators ($i=1,6$)}

Using the notation
\bq
\wtil{Z}_{\mu \nu}=\frac{1}{2} \epsilon_{\mu \nu \rho \sigma}Z^{\rho
\sigma} ~~ ,~~ Z_{\mu\nu}=\partial_\mu Z_\nu -\partial_\nu Z_\mu
~ ,
\eq
 and similarly for the photon tensor $F_{\mu\nu}$, the  set of
the $Z^*Z^*Z^*$ CP-conserving operators defined as said above, is
\bqa
\O^{Z^*Z^*Z^*}_1=\wtil{Z}_{\mu \nu}(\partial_{\sigma}{Z}^{\sigma\mu})
Z^{\nu} & , &
\O^{Z^*Z^*Z^*}_2= \square\wtil{Z}_{\mu \nu}Z^{\mu}\square Z^{\nu}
 ~~, \nonumber\\
\O^{Z^*Z^*Z^*}_3=(\square^2\wtil{Z}_{\mu \nu})(\square\partial^\sigma
Z^{\mu\nu})Z_{\sigma} & , &
\O^{Z^*Z^*Z^*}_4=\wtil{Z}_{\mu \nu}(\partial^\mu
Z^{\nu})(\partial^{\sigma}
Z_{\sigma}) ~ ~, \nonumber\\
\O^{Z^*Z^*Z^*}_5=\wtil{Z}_{\mu \nu}(\partial^\mu Z^{\nu})
(\square\partial^{\sigma} Z_{\sigma}) & , &
\O^{Z^*Z^*Z^*}_6=\wtil{Z}_{\mu \nu}(\partial^\mu\square
Z^{\nu})(\partial^{\sigma} Z_{\sigma}) ~ .
\label{ZZZ-CPc-op}
\eqa

The transverse terms are given by $\O^{Z^*Z^*Z^*}_1$ ($dim=6$),
$O^{Z^*Z^*Z^*}_2$ ($dim=8$), and $\O^{Z^*Z^*Z^*}_3$
($dim=12$). We note in particular that the  operator
$\O^{Z^*Z^*Z^*}_3$ is
required for generating the fully antisymmetric
structure of $f^{Z^*Z^*Z^*}_3(s_1,s_2,s_3)$, (see below).
The scalar terms are $\O^{Z^*Z^*Z^*}_4$ ($dim=6$) and
$O^{Z^*Z^*Z^*}_{5,6}$ ($dim=8$).\par

The corresponding coupling functions (see (\ref{ZZZ-CPc})) are
\bqa
f^{Z^*Z^*Z^*}_1(s_1,s_2,s_3)& = &-~{1\over2}(s_1+s_2-2s_3)l^{Z^*Z^*Z^*}_1
+{1\over2}(s_3(s_1+s_2)-2s_1s_2)l^{Z^*Z^*Z^*}_2\nonumber\\
&+& {1\over2}[s_1s_2(s_1-s_2)^2-s^2_3\{s_1(s_3-s_1)+s_2(s_3-s_2)\}]
l^{Z^*Z^*Z^*}_3 ~ ,
\nonumber\\
f^{Z^*Z^*Z^*}_2(s_1,s_2,s_3)&=& -~{3\over2}(s_1-s_2)l^{Z^*Z^*Z^*}_1
-{3\over2}s_3(s_1-s_2)l^{Z^*Z^*Z^*}_2\nonumber\\
&+& {s_2-s_1\over2}\Big[
s_3(s_1s_2-s^2_1-s^2_2+s_3(s_1+s_2))-2s_1s_2(s_1+s_2)\Big ]
l^{Z^*Z^*Z^*}_3 ~ , \nonumber\\
f^{Z^*Z^*Z^*}_3(s_1,s_2,s_3) &=& [s^2_1(s_2-s_3)+s^2_3(s_1-s_2)
+s^2_2(s_3-s_1)]l^{Z^*Z^*Z^*}_3 ~ , \nonumber\\
g^{Z^*Z^*Z^*}_1(s_1,s_2,s_3)& = & 2l^{Z^*Z^*Z^*}_1+2l^{Z^*Z^*Z^*}_4
-2s_1l^{Z^*Z^*Z^*}_5-(s_2+s_3)l^{Z^*Z^*Z^*}_6\nonumber\\
&+ & 2(s^2_3s_2+s^2_2s_1-s^2_1s_2)l^{Z^*Z^*Z^*}_3 ~ , \nonumber\\
g^{Z^*Z^*Z^*}_2(s_1,s_2,s_3)& = & 2l^{Z^*Z^*Z^*}_1+2l^{Z^*Z^*Z^*}_4
-2s_2l^{Z^*Z^*Z^*}_5-(s_1+s_3)l^{Z^*Z^*Z^*}_6\nonumber\\
&+ & 2(s^2_3s_1+s^2_1 s_2-s^2_2s_1)l^{Z^*Z^*Z^*}_3 ~ , \nonumber\\
g^{Z^*Z^*Z^*}_3(s_1,s_2,s_3) &= & 2l^{Z^*Z^*Z^*}_1+2l^{Z^*Z^*Z^*}_4
-2s_3l^{Z^*Z^*Z^*}_5-(s_2+s_1)l^{Z^*Z^*Z^*}_6\nonumber\\
& - & [s_1(s^2_3-s^2_2)-s_3(s^2_1+s^2_2)+s_2(s^2_3-s^2_1)]
l^{Z^*Z^*Z^*}_3 ~ . \label{ZZZ-CPc-Lan}
\eqa

We also remark that the on-shell coupling $f^Z_5$
defined in  \cite{Hagn, neut, modzg} for the
CP conserving $Z^*ZZ$ vertex, is related to the relevant
 three transverse couplings defined here for the off-shell case by
\bq
f^Z_5=m^2_Z[l^{Z^*Z^*Z^*}_1+m^2_Z(l^{Z^*Z^*Z^*}_2+s^2_3
l^{Z^*Z^*Z^*}_3)]~ . \label{ZZZ-CPc-l-on}
\eq
Thus, going from the on-shell treatment
of the CP conserving $Z^*ZZ$  NAGC case, to the
present effective Lagrangian
 off-shell one, we have to increase the number of parameters from
one to three.

\vspace{0.5cm}
\subsection{ The $Z^*Z^*Z^*$ CP-violating operators ($i=1,14$)}
The relevant set of operators  is
\bqa
\tilde{\O}^{Z^*Z^*Z^*}_1=
-Z_{\sigma}(\partial^{\sigma}Z_{\nu})(\partial_{\mu}
Z^{\mu\nu}) & ,&
\tilde{\O}^{Z^*Z^*Z^*}_2=(\square
Z_{\alpha})(\partial^{\alpha}Z_{\mu})(\square Z^{\mu}) ~ , \nonumber\\
\tilde{\O}^{Z^*Z^*Z^*}_3=
Z_{\alpha}(\partial^{\alpha}Z_{\mu})(\square^2 Z^{\mu}) & ,&
\tilde{\O}^{Z^*Z^*Z^*}_4=(\square^2\partial^{\alpha}Z_{\beta})
(\partial^{\mu}\square Z_{\alpha})
(\partial^{\beta}Z_{\mu}) ~ , \nonumber\\
\tilde{\O}^{Z^*Z^*Z^*}_5=Z^{\mu}Z_{\mu}(\partial^{\sigma}
Z_{\sigma}) & , & \nonumber\\
\tilde{\O}^{Z^*Z^*Z^*}_6=(\square Z^{\mu})Z_{\mu}(\partial^{\sigma}
Z_{\sigma}) & , &
\tilde{\O}^{Z^*Z^*Z^*}_7=Z^{\mu}Z_{\mu}\square(\partial^{\sigma}
Z_{\sigma}) ~ , \nonumber\\
\tilde{\O}^{Z^*Z^*Z^*}_8=
(\partial^{\sigma} Z_{\sigma})(\partial^{\nu} Z_{\mu})
(\partial^{\mu} Z_{\nu}) & , &
\tilde{\O}^{Z^*Z^*Z^*}_9=
(\partial^{\sigma} Z_{\sigma})(\square\partial^{\alpha} Z_{\beta})
(\partial^{\beta}Z_{\alpha}) ~ , \nonumber\\
\tilde{\O}^{Z^*Z^*Z^*}_{10}=(\square\partial^{\sigma} Z_{\sigma})
(\partial^{\alpha} Z_{\beta})(\partial^{\beta} Z_{\alpha})
& , &
\tilde{\O}^{Z^*Z^*Z^*}_{11}=
\square\partial^{\alpha}(\partial^{\sigma} Z_{\sigma})
(\partial^{\beta} Z_{\beta})Z_{\alpha} ~ , \nonumber\\
\tilde{\O}^{Z^*Z^*Z^*}_{12}=
\square\partial^{\alpha}(\partial^{\sigma} Z_{\sigma})
(\partial^{\beta} Z_{\beta}) (\square Z_{\alpha}) & , &
\tilde{\O}^{Z^*Z^*Z^*}_{13}=
\square^2\partial^{\alpha}(\partial^{\sigma} Z_{\sigma})
(\partial^{\beta} Z_{\beta})Z_{\alpha} ~ , \nonumber\\
\tilde{\O}^{Z^*Z^*Z^*}_{14}=
(\partial^{\sigma} Z_{\sigma})(\partial^{\mu} Z_{\mu})
(\partial^{\nu} Z_{\nu}) & . & \label{ZZZ-CPv-op}
\eqa
The transverse terms are given by
$\tilde{\O}^{Z^*Z^*Z^*}_{1}$ ($dim=6$),
$\tilde{\O}^{Z^*Z^*Z^*}_{2,3}$ ($dim=8$)
and $\tilde{\O}^{Z^*Z^*Z^*}_4$ ($dim=12$);
while the scalar ones are generated by
 $\tilde{\O}^{Z^*Z^*Z^*}_{5}$ ($dim=4$),
$\tilde{\O}^{Z^*Z^*Z^*}_{6-8,14}$ ($dim=6$),
$\tilde{\O}^{Z^*Z^*Z^*}_{9-11}$ ($dim=8$) and
$\tilde{\O}^{Z^*Z^*Z^*}_{12,13}$ ($dim=10$). Note the presence of
a $dim=4$ operator, $\tilde{\O}^{Z^*Z^*Z^*}_{5}$, multiplied by
a dimensionless coupling, which would
induce CP-violation when one $Z$ has a scalar component coupled
to a heavy quark pair.\par

The corresponding coupling functions are:
\bqa
\tilde{f}^{Z^*Z^*Z^*}_1(s_1,s_2,s_3) & = &{1\over2}(s_2-s_1)
(\tilde{l}^{Z^*Z^*Z^*}_1
+s_3\tilde{l}^{Z^*Z^*Z^*}_2)-{1\over2}(s^2_1-s^2_2)\tilde{l}^{Z^*Z^*Z^*}_3
 ~ , \nonumber\\
\tilde{f}^{Z^*Z^*Z^*}_2(s_1,s_2,s_3) & = &
{1\over2}(s_2-s_3)(\tilde{l}^{Z^*Z^*Z^*}_1+s_1\tilde{l}^{Z^*Z^*Z^*}_2)
-{1\over2}(s^2_3-s^2_2)\tilde{l}^{Z^*Z^*Z^*}_3 ~ , \nonumber\\
\tilde{f}^{Z^*Z^*Z^*}_3(s_1,s_2,s_3) & = &
{1\over2}(s_3-s_1)(\tilde{l}^{Z^*Z^*Z^*}_1+s_2\tilde{l}^{Z^*Z^*Z^*}_2)
-{1\over2}(s^2_1-s^2_3)\tilde{l}^{Z^*Z^*Z^*}_3 ~, \nonumber\\
\tilde{f}^{Z^*Z^*Z^*}_4(s_1,s_2,s_3) & = &
{1\over8}(a_1-a_2)\tilde{l}^{Z^*Z^*Z^*}_4 ~ , \nonumber\\
\tilde{g}^{Z^*Z^*Z^*}_1(s_1,s_2,s_3) & = &
-{1\over2}(s_1+s_2)(\tilde{l}^{Z^*Z^*Z^*}_1+s_3\tilde{l}^{Z^*Z^*Z^*}_2)
-{1\over2}(s^2_1+s^2_2)\tilde{l}^{Z^*Z^*Z^*}_3\nonumber\\
&+ & 2\tilde{l}^{Z^*Z^*Z^*}_5-(s_1+s_2)\tilde{l}^{Z^*Z^*Z^*}_6
-2s_3\tilde{l}^{Z^*Z^*Z^*}_7 ~ , \nonumber\\
\tilde{g}^{Z^*Z^*Z^*}_2(s_1,s_2,s_3) & = &
-{1\over2}(s_3+s_2)(\tilde{l}^{Z^*Z^*Z^*}_1+s_1\tilde{l}^{Z^*Z^*Z^*}_2)
-{1\over2}(s^2_3+s^2_2)\tilde{l}^{Z^*Z^*Z^*}_3\nonumber\\
&+ & 2\tilde{l}^{Z^*Z^*Z^*}_5
-(s_3+s_2)\tilde{l}^{Z^*Z^*Z^*}_6-2s_1\tilde{l}^{Z^*Z^*Z^*}_7 ~ ,
\nonumber\\
\tilde{g}^{Z^*Z^*Z^*}_3(s_1,s_2,s_3) & = &
-{1\over2}(s_1+s_3)(\tilde{l}^{Z^*Z^*Z^*}_1+s_2\tilde{l}^{Z^*Z^*Z^*}_2)
-{1\over2}(s^2_1+s^2_3)\tilde{l}^{Z^*Z^*Z^*}_3\nonumber\\
&+ & 2\tilde{l}^{Z^*Z^*Z^*}_5
-(s_1+s_3)\tilde{l}^{Z^*Z^*Z^*}_6-2s_2\tilde{l}^{Z^*Z^*Z^*}_7 ~ ,\nonumber\\
\tilde{g}^{Z^*Z^*Z^*}_4(s_1,s_2,s_3) &= &{1\over2}\tilde{l}^{Z^*Z^*Z^*}_1
+{1\over8}(a_1+a_2)\tilde{l}^{Z^*Z^*Z^*}_4
-{1\over2}\tilde{l}^{Z^*Z^*Z^*}_8 \nonumber\\
&+ & {1\over4}(s_2+s_3)\tilde{l}^{Z^*Z^*Z^*}_9
+{1\over2}s_1\tilde{l}^{Z^*Z^*Z^*}_{10} ~ , \nonumber\\
\tilde{g}^{Z^*Z^*Z^*}_5(s_1,s_2,s_3) &= &{1\over2}\tilde{l}^{Z^*Z^*Z^*}_1
+{1\over8}(a_1+a_2)\tilde{l}^{Z^*Z^*Z^*}_4
-{1\over2}\tilde{l}^{Z^*Z^*Z^*}_8
+{1\over4}(s_1+s_3)\tilde{l}^{Z^*Z^*Z^*}_9\nonumber\\
&+ &{1\over2}s_2\tilde{l}^{Z^*Z^*Z^*}_{10} ~ , \nonumber\\
\tilde{g}^{Z^*Z^*Z^*}_6(s_1,s_2,s_3) &= &{1\over2}\tilde{l}^{Z^*Z^*Z^*}_1
+{1\over8}(a_1+a_2)\tilde{l}^{Z^*Z^*Z^*}_4
-{1\over2}\tilde{l}^{Z^*Z^*Z^*}_8
+{1\over4}(s_2+s_1)\tilde{l}^{Z^*Z^*Z^*}_9\nonumber\\
&+ &{1\over2}s_3\tilde{l}^{Z^*Z^*Z^*}_{10} ~ , \nonumber\\
\tilde{g}^{Z^*Z^*Z^*}_7(s_1,s_2,s_3) &= &
{1\over8}(a_2-a_1)\tilde{l}^{Z^*Z^*Z^*}_4
+{1\over4}(s_2-s_1)\tilde{l}^{Z^*Z^*Z^*}_9
+{1\over2}(s_1-s_2)\tilde{l}^{Z^*Z^*Z^*}_{10}\nonumber\\
&+ &{1\over2}(s_1-s_2)\tilde{l}_{11}
+{s_3\over2}(s_2-s_1)\tilde{l}^{Z^*Z^*Z^*}_{12}
+{1\over2}(s^2_2-s^2_1)\tilde{l}^{Z^*Z^*Z^*}_{13} ~ , \nonumber\\
\tilde{g}^{Z^*Z^*Z^*}_8(s_1,s_2,s_3) & = &
{1\over8}(a_1-a_2)\tilde{l}^{Z^*Z^*Z^*}_4
+{1\over4}(s_2-s_3)\tilde{l}^{Z^*Z^*Z^*}_9
+{1\over2}(s_3-s_2)\tilde{l}^{Z^*Z^*Z^*}_{10}\nonumber\\
&+ & {1\over2}(s_3-s_2)\tilde{l}^{Z^*Z^*Z^*}_{11}
+{s_1\over2}(s_2-s_3)\tilde{l}^{Z^*Z^*Z^*}_{12}
+{1\over2}(s^2_2-s^2_3)\tilde{l}^{Z^*Z^*Z^*}_{13} ~ , \nonumber\\
\tilde{g}^{Z^*Z^*Z^*}_9(s_1,s_2,s_3) &= &
{1\over8}(a_1-a_2)\tilde{l}^{Z^*Z^*Z^*}_4
+{1\over4}(s_3-s_1)\tilde{l}^{Z^*Z^*Z^*}_9
+{1\over2}(s_1-s_3)\tilde{l}^{Z^*Z^*Z^*}_{10}\nonumber\\
&+ &{1\over2}(s_1-s_3)\tilde{l}^{Z^*Z^*Z^*}_{11}
+{s_2\over2}(s_3-s_1)\tilde{l}^{Z^*Z^*Z^*}_{12}
+{1\over2}(s^2_3-s^2_1)\tilde{l}^{Z^*Z^*Z^*}_{13} ~ , \nonumber\\
\tilde{g}^{Z^*Z^*Z^*}_{10}(s_1,s_2,s_3) &= &
-~{3\over2}\tilde{l}^{Z^*Z^*Z^*}_1
-{1\over8}(a_1+a_2)\tilde{l}^{Z^*Z^*Z^*}_4
-{3\over2}\tilde{l}^{Z^*Z^*Z^*}_8
+{1\over2}(s_1+s_2+s_3)\tilde{l}^{Z^*Z^*Z^*}_9\nonumber\\
&+ &{1\over2}(s_1+s_2+s_3)\tilde{l}^{Z^*Z^*Z^*}_{10}
-(s_1+s_2+s_3)\tilde{l}^{Z^*Z^*Z^*}_{11} \nonumber \\
&+ &(s_1s_3+s_2s_3+s_1s_2)\tilde{l}^{Z^*Z^*Z^*}_{12}
+ (s^2_1+s^2_2+s^2_3)\tilde{l}^{Z^*Z^*Z^*}_{13}
-\tilde{l}^{Z^*Z^*Z^*}_{14} ~, \label{ZZZ-CPv-Lan}
\eqa
\noindent
with
\bqa
a_1-a_2 & = & s^2_1(s_2-s_3)+s^2_2(s_3-s_1)+s^2_3(s_1-s_2) ~ ,
\nonumber \\
a_1+a_2 &= & s^2_1(s_2+s_3)+s^2_2(s_3+s_1)+s^2_3(s_1+s_2)
~ .
\eqa

We also  remark that
the on-shell single parameter $f^Z_4$,  defined
in \cite{Hagn, neut, modzg},  is related to the present
ones by
\bq
f^Z_4= m^2_Z[\tilde{l}^{Z^*Z^*Z^*}_1+m^2_Z\tilde{l}^{Z^*Z^*Z^*}_2
+(s_3+m^2_Z)\tilde{l}^{Z^*Z^*Z^*}_3] ~ . \label{ZZZ-CPv-l-on}
\eq
Thus,  going from the on-shell treatment
of the CP-violating $Z^*ZZ$  NAGC case, to the
present effective Lagrangian
 off-shell one, we have again to increase the number of parameters from
one to three.

\vspace{0.5cm}
\subsection{ The $Z^*Z^*\gamma^*$ CP-conserving operators ($i=1,5$)}

The operator set is
\bqa
\O^{Z^*Z^*\gamma^*}_1=
-\wtil{F}_{\mu \nu}Z^{\nu}(\partial_{\sigma}Z^{\sigma\mu}) & ,&
\O^{Z^*Z^*\gamma^*}_2=\wtil{Z}^{\mu \nu}Z_{\nu}
(\partial^{\sigma}F_{\sigma\mu}) ~ , \nonumber\\
\O^{Z^*Z^*\gamma^*}_3=(\square\partial^{\sigma}Z^{\rho\alpha})
Z_{\sigma}\wtil{F}_{\rho \alpha} & , &
\O^{Z^*Z^*\gamma^*}_4=\wtil{F}_{\mu \nu}Z^{\mu\nu}(\partial^{\sigma}
Z_{\sigma}) ~ , \nonumber\\
\O^{Z^*Z^*\gamma^*}_5  =
\wtil{F}_{\mu \nu}Z^{\mu\nu}\square(\partial^{\sigma}
Z_{\sigma}) & .& \label{ZZgamma-CPc-op}
\eqa
Here the transverse terms are given by $\O^{Z^*Z^*\gamma^*}_{1,2}$
($dim=6$) and
$\O^{Z^*Z^*\gamma^*}_{3}$ ($dim=8$); while the scalar ones
are induced by $\O^{Z^*Z^*\gamma^*}_4$ ($dim=6$) and
$\O^{Z^*Z^*\gamma^*}_{5}$ ($dim=8$).\par

The corresponding coupling functions are:
\bqa
f^{Z^*Z^*\gamma^*}_1(s_1,s_2,s_3) &= & s_3l^{Z^*Z^*\gamma^*}_2
-{1\over2}s_3(s_1+s_2)l^{Z^*Z^*\gamma^*}_3 ~ ,
\nonumber\\
f^{Z^*Z^*\gamma^*}_2(s_1,s_2,s_3) & = & (s_1-s_2)l^{Z^*Z^*\gamma^*}_1
+{1\over2}(s_2-s_1)(s_1+s_2)
l^{Z^*Z^*\gamma^*}_3 ~ , \nonumber\\
f^{Z^*Z^*\gamma^*}_3(s_1,s_2,s_3) & = &(s_1-s_2)l^{Z^*Z^*\gamma^*}_3
~ , \nonumber\\
g^{Z^*Z^*\gamma^*}_1(s_1,s_2,s_3) & =& -l^{Z^*Z^*\gamma^*}_1
+2s_2l^{Z^*Z^*\gamma^*}_3+2l^{Z^*Z^*\gamma^*}_4
-2s_1l^{Z^*Z^*\gamma^*}_5 ~ , \nonumber\\
g^{Z^*Z^*\gamma^*}_2(s_1,s_2,s_3)& = & -l^{Z^*Z^*\gamma^*}_1
+2s_1l^{Z^*Z^*\gamma^*}_3+2l^{Z^*Z^*\gamma^*}_4
-2s_2l^{Z^*Z^*\gamma^*}_5 ~. \label{ZZgamma-CPc-Lan}
\eqa

Comparing now to the  parameters defined in
\cite{Hagn, neut, modzg}, we remark that when
two $Z$'s are on-shell one obtains
\bq
f_5^{\gamma}= m^2_Z(l^{Z^*Z^*\gamma^*}_2-m^2_Zl^{Z^*Z^*\gamma^*}_3)
~ , \label{ZZgamma-CPc-l1-on}
\eq
while when one $\gamma$ and one $Z$ are on shell one
obtains\footnote{It is
important to note that, contrary to the case of the form
\[
I^{Z^*Z^*\gamma^*,3}_{\alpha\beta\mu}=
q^{\beta}_3~[q_1~q_2~\mu~\alpha]
+q^{\alpha}_3~[q_1~q_2~\mu~\beta] ~,
\]
the form
$q^{\alpha}_3~[q_1~q_2~\mu~\beta]$ associated to the $h_4^{Z,\gamma}$
couplings, defined in \cite{Hagn, neut}, has not a
well-defined Bose symmetry property. In fact under Bose
symmetry, $h_3^{Z,\gamma}$ and $h_4^{Z,\gamma}$ get mixed. The same
remark applies to the CP-violating coupling $h_2^{Z,\gamma}$.}
\bq
h_3^Z= m^2_Z(l^{Z^*Z^*\gamma^*}_1 -m^2_Z
l^{Z^*Z^*\gamma^*}_3)~~,~~h_4^Z= 2m^4_Zl^{Z^*Z^*\gamma^*}_3 ~ .
\label{ZZgamma-CPc-l23-on}
\eq

So when considering these two on-shell processes we have the
same number of transverse parameters as in the general off-shell case.

\vspace{0.5cm}
\subsection{ The $Z^*Z^*\gamma^*$ CP-violating operators ($i=1,9$)}
These operators are
\bqa
\tilde{\O}^{Z^*Z^*\gamma^*}_1=
-F^{\mu\beta} Z_{\beta} (\partial^{\sigma}Z_{\sigma\mu}) & , &
\tilde{\O}^{Z^*Z^*\gamma^*}_2=
-(\partial_{\alpha}\partial_{\beta}\square Z_{\mu})
Z^{\alpha}F^{\mu\beta} ~ , \nonumber\\
 \tilde{\O}^{Z^*Z^*\gamma^*}_3=
-(\partial_{\mu}F^{\mu\beta})Z_{\alpha}(\partial^{\alpha}Z_{\beta})
& , &  \tilde{\O}^{Z^*Z^*\gamma^*}_4=
\partial^{\mu}F_{\mu\nu}(\square\partial^{\nu}Z_{\alpha})Z^{\alpha}
 ~ , \nonumber\\
\tilde{\O}^{Z^*Z^*\gamma^*}_5=
(\partial^{\sigma}Z_{\sigma})F_{\mu\nu}(\partial^{\mu}Z^{\nu}) & ,&
\tilde{\O}^{Z^*Z^*\gamma^*}_6=
(\partial^{\sigma}Z_{\sigma})(\partial^{\mu}F_{\mu\nu})Z^{\nu}
~ , \nonumber\\
\tilde{\O}^{Z^*Z^*\gamma^*}_7=
\square(\partial^{\sigma}Z_{\sigma})F_{\mu\nu}(\partial^{\mu}Z^{\nu})
& , & \tilde{\O}^{Z^*Z^*\gamma^*}_8=
(\partial^{\sigma}Z_{\sigma})F_{\mu\nu}(\square\partial^{\mu}Z^{\nu})
~ , \nonumber\\
\tilde{\O}^{Z^*Z^*\gamma^*}_9=
\square\partial^{\nu}(\partial^{\sigma}Z_{\sigma})(\partial^{\beta}
Z_{\beta})\partial^{\mu}F_{\mu\nu} & . & \label{ZZgamma-CPv-op}
\eqa

The transverse terms are given by $\tilde{\O}^{Z^*Z^*\gamma^*}_{1,3}$
($dim=6$),
$\tilde{\O}^{Z^*Z^*\gamma^*}_{2,4}$ ($dim=8$); while the
scalar ones are generated by $\tilde{\O}^{Z^*Z^*\gamma^*}_{5,6}$ ($dim=6$),
$\tilde{\O}^{Z^*Z^*\gamma^*}_{7,8}$ ($dim=8$) and
$\tilde{\O}^{Z^*Z^*\gamma^*}_{9}$ ($dim=10$).\par

The corresponding coupling functions are:
\bqa
\tilde{f}^{Z^*Z^*\gamma^*}_1(s_1,s_2,s_3) & = &
{s_3\over2}(s_1-s_2)\tilde{l}^{Z^*Z^*\gamma^*}_4 ~ ,\nonumber\\
\tilde{f}^{Z^*Z^*\gamma^*}_2(s_1,s_2,s_3) &= &
-{1\over2}s_3\tilde{l}^{Z^*Z^*\gamma^*}_3
-{1\over8}[s_1(s_2-s_1-s_3)
+ s_2(s_1-s_2-s_3)]\tilde{l}^{Z^*Z^*\gamma^*}_2 ~ , \nonumber\\
\tilde{f}^{Z^*Z^*\gamma^*}_3(s_1,s_2,s_3) &= &{1\over2}(s_1-s_2)
\tilde{l}^{Z^*Z^*\gamma^*}_1
+{1\over8}[s_2(s_2+s_3)-s_1(s_1+s_3)]\tilde{l}^{Z^*Z^*\gamma^*}_2
~ , \nonumber\\
\tilde{f}^{Z^*Z^*\gamma^*}_4(s_1,s_2,s_3)& = &{1\over8}(s_1-s_2)
\tilde{l}^{Z^*Z^*\gamma^*}_2
~ , \nonumber\\
\tilde{g}^{Z^*Z^*\gamma^*}_1(s_1,s_2,s_3) & = &
{1\over2}s_3\tilde{l}^{Z^*Z^*\gamma^*}_1
-{1\over2}s_3\tilde{l}^{Z^*Z^*\gamma^*}_3
-{1\over8}[s_1(s_2-s_1-s_3)
+ s_2(s_1-s_2-s_3)]\tilde{l}^{Z^*Z^*\gamma^*}_2\nonumber\\
&+ &{1\over2}s_3\tilde{l}^{Z^*Z^*\gamma^*}_5-s_3
\tilde{l}^{Z^*Z^*\gamma^*}_6
+{1\over4}[(s_2-s_1)^2-s_3(s_1+s_2)]\tilde{l}^{Z^*Z^*\gamma^*}_7
\nonumber\\
&-&{1\over4}[(s_2-s_1)^2+s_3(s_1+s_2)]
\tilde{l}^{Z^*Z^*\gamma^*}_8 ~ , \nonumber\\
\tilde{g}^{Z^*Z^*\gamma^*}_2(s_1,s_2,s_3) &= &
{1\over8}[s_2(s_2+s_3)-s_1(s_1+s_3)]\tilde{l}^{Z^*Z^*\gamma^*}_2
\nonumber\\
&+ & {1\over2}(s_2-s_1)\tilde{l}^{Z^*Z^*\gamma^*}_5
+{1\over4}[(s^2_1-s^2_2)-s_3(s_1-s_2)]\tilde{l}^{Z^*Z^*\gamma^*}_7
\nonumber\\
&+&{1\over4}[(s^2_1-s^2_2)+s_3(s_1-s_2)]
\tilde{l}^{Z^*Z^*\gamma^*}_8 ~ , \nonumber\\
\tilde{g}^{Z^*Z^*\gamma^*}_3(s_1,s_2,s_3)& = &
-~{1\over4}\tilde{l}^{Z^*Z^*\gamma^*}_1+{s_1+s_2\over8}
\tilde{l}^{Z^*Z^*\gamma^*}_2
+{1\over4}\tilde{l}^{Z^*Z^*\gamma^*}_5
-{1\over8}(s_1+s_2)\tilde{l}^{Z^*Z^*\gamma^*}_7  \nonumber\\
&- &{1\over8}(s_1+s_2)\tilde{l}^{Z^*Z^*\gamma^*}_8 ~ , \nonumber\\
\tilde{g}^{Z^*Z^*\gamma^*}_4(s_1,s_2,s_3) &= &
-{1\over8}(s_1-s_2)\tilde{l}^{Z^*Z^*\gamma^*}_7
+{1\over8}(s_1-s_2)\tilde{l}_8 ~ , \nonumber\\
\tilde{g}^{Z^*Z^*\gamma^*}_5(s_1,s_2,s_3) &= &{s_2-s_1\over8}
\tilde{l}^{Z^*Z^*\gamma^*}_2
+{1\over4}(s_2-s_1)\tilde{l}^{Z^*Z^*\gamma^*}_7
+{1\over4}(s_1-s_2)\tilde{l}^{Z^*Z^*\gamma^*}_8\nonumber\\
&- &{1\over2}s_3(s_1-s_2)\tilde{l}^{Z^*Z^*\gamma^*}_9 ~ .
\label{ZZgamma-CPv-Lan}
\eqa

Comparing to the parameters defined in  \cite{Hagn, neut, modzg},
   when two $Z$'s are on shell, one obtains
\bq
f_4^{\gamma}= m^2_Z(\tilde{l}^{Z^*Z^*\gamma^*}_3-{m^2_Z\over2}
\tilde{l}^{Z^*Z^*\gamma^*}_2) ~, \label{ZZgamma-CPv-l-on}
\eq
while when one $\gamma$ and one $Z$ are on shell, we get
\bq
h_1^Z= m^2_Z(\tilde{l}^{Z^*Z^*\gamma^*}_1
- {m^2_Z\over2}\tilde{l}^{Z^*Z^*\gamma^*}_2)~~, ~~
h_2^Z= m^4_Z\tilde{l}^{Z^*Z^*\gamma^*}_2 ~.
\label{ZZgamma-CPv-l23-on}
\eq
\noindent
So the off-shell case has one more transverse parameter
($\tilde{l}^{Z^*Z^*\gamma^*}_4$) than the on-shell one.

\vspace{0.5cm}
\subsection{ The $\gamma^*\gamma^*Z^*$ CP-conserving
operators ($i=1,4$)}

The four operators of  this case are
\bqa
\O^{\gamma^*\gamma^*Z^*}_1=
-\wtil{F}_{\rho \alpha}(\partial_{\sigma}F^{\sigma\rho})Z^{\alpha} & ,&
\O^{\gamma^*\gamma^*Z^*}_2=\square\wtil{F}^{\mu \nu}
(\partial^{\sigma}F_{\sigma\mu})Z_{\nu} ~ , \nonumber\\
\O^{\gamma^*\gamma^*Z^*}_3=
(\square\partial^{\sigma}F^{\rho\alpha})Z_{\sigma}\wtil{F}_{\rho \alpha}
& ,& \O^{\gamma^*\gamma^*Z^*}_4=\wtil{F}_{\mu \nu}F^{\mu\nu}
(\partial^{\sigma} Z_{\sigma}) ~. \label{gammagammaZ-CPc-op}
\eqa
The transverse terms are given by
$\O^{\gamma^*\gamma^*Z^*}_1$ ($dim=6$) and
$\O^{\gamma^*\gamma^*Z^*}_{2,3}$ ($dim=8$); while the
 scalar term by  $\O^{\gamma^*\gamma^*Z^*}_4$ ($dim=6$).\par

The corresponding coupling functions are:
\bqa
f^{\gamma^*\gamma^*Z^*}_1(s_1,s_2,s_3) & = &
{1\over2}(s_1+s_2)l^{\gamma^*\gamma^*Z^*}_1
+s_1s_2l^{\gamma^*\gamma^*Z^*}_2
-{1\over2}(s_1-s_2)^2l^{\gamma^*\gamma^*Z^*}_3 ~ , \nonumber\\
f^{\gamma^*\gamma^*Z^*}_2(s_1,s_2,s_3) & = &
{1\over2}(s_1-s_2)l^{\gamma^*\gamma^*Z^*}_1
+{1\over2}(s_1-s_2)(s_3-2s_1-2s_2)l^{\gamma^*\gamma^*Z^*}_3 ~ ,
\nonumber\\
f^{\gamma^*\gamma^*Z^*}_3(s_1,s_2,s_3) &= &
(s_2-s_1)l^{\gamma^*\gamma^*Z^*}_3 ~ , \nonumber\\
g^{\gamma^*\gamma^*Z^*}_1(s_1,s_2,s_3)
&= &(s_1+s_2)l^{\gamma^*\gamma^*Z^*}_3+4l^{\gamma^*\gamma^*Z^*}_4
~ . \label{gammagammaZ-CPc-Lan}
\eqa

When one $\gamma$ and one $Z$ are on shell, one gets
\bq
h^{\gamma}_3= m^2_Z l^{\gamma^*\gamma^*Z^*}_1~~ , ~~
h^{\gamma}_4= 2m^4_Z l^{\gamma^*\gamma^*Z^*}_3 ~ ,
\label{gammagammaZ-CPc-l-on}
\eq
when comparing to the parameters of \cite{Hagn, neut, modzg}, and one
observes that there is one less transverse parameter
($l^{\gamma^*\gamma^*Z^*}_2$) than in the off-shell case.

\vspace{0.5cm}
\subsection{ The $\gamma^*\gamma^*Z^*$ CP-violating
operators ($i=1,6$).}
We now have
\bqa
\tilde{\O}^{\gamma^*\gamma^*Z^*}_1=
-(\partial^{\sigma}F_{\sigma\mu})Z_{\beta}F^{\mu\beta} & ,&
\tilde{\O}^{\gamma^*\gamma^*Z^*}_2=
(\square F^{\mu\nu})F_{\nu\alpha}(\partial_{\mu}Z^{\alpha})
~ , \nonumber\\
\tilde{\O}^{\gamma^*\gamma^*Z^*}_3=
-(\partial_{\alpha}\partial_{\beta}\partial^{\rho}F_{\rho\mu})Z^{\alpha}
F^{\mu\beta} & , & \tilde{\O}^{\gamma^*\gamma^*Z^*}_4=
(\square \partial_{\mu}F^{\mu\nu})
(\partial^{\sigma}F_{\sigma\alpha})(\partial_{\nu}Z^{\alpha})
~ , \nonumber\\
\tilde{\O}^{\gamma^*\gamma^*Z^*}_5=
(\partial_{\sigma}Z^{\sigma})F^{\mu\nu}F_{\mu\nu} & , &
\tilde{\O}^{\gamma^*\gamma^*Z^*}_6=
\partial_{\mu}(\partial_{\sigma}Z^{\sigma})F^{\mu\nu}
(\partial^{\beta}F_{\beta\nu}) ~ . \label{gammagammaZ-CPv-op}
\eqa
The transverse terms are given by
$\tilde{\O}^{\gamma^*\gamma^*Z^*}_1$ ($dim=6$),
$\tilde{\O}^{\gamma^*\gamma^*Z^*}_{2,3}$ ($dim=8$)
and $\tilde{\O}^{\gamma^*\gamma^*Z^*}_4$ ($dim=10$).
The scalar terms are $\tilde{\O}^{\gamma^*\gamma^*Z^*}_5$ ($dim=6$)
and $\tilde{\O}^{\gamma^*\gamma^*Z^*}_6$ ($dim=8$).\par

The corresponding coupling functions are
\bqa
\tilde{f}^{\gamma^*\gamma^*Z^*}_1(s_1,s_2,s_3) & = &
{1\over2}(s_1-s_2)\tilde{l}^{\gamma^*\gamma^*Z^*}_1
+{1\over4}[s_2(s_2+s_3)-s_1(s_1+s_3)]\tilde{l}^{\gamma^*\gamma^*Z^*}_2
\nonumber\\
&+ & {1\over4}(s_1-s_2)(s_3-s_1-s_2)
\tilde{l}^{\gamma^*\gamma^*Z^*}_3 ~ , \nonumber\\
\tilde{f}^{\gamma^*\gamma^*Z^*}_2(s_1,s_2,s_3) &= &
-~{1\over4}(s_1+s_2)\tilde{l}^{\gamma^*\gamma^*Z^*}_1
+{s^2_1+s^2_2\over4}\tilde{l}^{\gamma^*\gamma^*Z^*}_2
-{s_1s_2\over2}(s_1+s_2)\tilde{l}^{\gamma^*\gamma^*Z^*}_4 ~ , \nonumber\\
\tilde{f}^{\gamma^*\gamma^*Z^*}_3(s_1,s_2,s_3) & = &
{1\over4}(s_2-s_1)\tilde{l}^{\gamma^*\gamma^*Z^*}_1
+{s^2_1-s^2_2\over4}\tilde{l}^{\gamma^*\gamma^*Z^*}_2
+{s_1s_2\over2}(s_2-s_1)\tilde{l}^{\gamma^*\gamma^*Z^*}_4~ ,\nonumber\\
\tilde{f}^{\gamma^*\gamma^*Z^*}_4(s_1,s_2,s_3) & = &
{1\over8}(s_1-s_2)\tilde{l}^{\gamma^*\gamma^*Z^*}_2
-~{1\over8}(s_1-s_2)\tilde{l}^{\gamma^*\gamma^*Z^*}_3 ~ , \nonumber\\
\tilde{g}^{\gamma^*\gamma^*Z^*}_1(s_1,s_2,s_3) &= &
{1\over2}(s_1+s_2)\tilde{l}^{\gamma^*\gamma^*Z^*}_1
+{1\over4}[s_1(s_2-s_1-s_3)+s_2(s_1-s_2-s_3)]
\tilde{l}^{\gamma^*\gamma^*Z^*}_2 \nonumber\\
&- & {1\over4}(s_1+s_2)(s_3-s_1-s_2)\tilde{l}^{\gamma^*\gamma^*Z^*}_3
-2(s_3-s_1-s_2)\tilde{l}^{\gamma^*\gamma^*Z^*}_5
\nonumber\\
&+ & {1\over2}[s_1(s_1-s_2-s_3)+s_2(s_2-s_1-s_3)]
\tilde{l}^{\gamma^*\gamma^*Z^*}_6 ~ , \nonumber\\
\tilde{g}^{\gamma^*\gamma^*Z^*}_2(s_1,s_2,s_3) & = &
{1\over8}(s_1+s_2)\tilde{l}^{\gamma^*\gamma^*Z^*}_2
+{1\over8}(s_1+s_2)\tilde{l}^{\gamma^*\gamma^*Z^*}_3
\nonumber\\
&+ &\tilde{l}^{\gamma^*\gamma^*Z^*}_5
+{1\over4}(s_1+s_2)\tilde{l}^{\gamma^*\gamma^*Z^*}_6 ~ .
\label{gammagammaZ-CPv-Lan}
\eqa

When one $\gamma$ and one $Z$ are on shell, one obtains
\bq
h^{\gamma}_1= m^2_Z(\tilde{l}^{\gamma^*\gamma^*Z^*}_1-s_1
\tilde{l}^{\gamma^*\gamma^*Z^*}_2)~~~ , ~~~
h^{\gamma}_2= -m^4_Z(\tilde{l}^{\gamma^*\gamma^*Z^*}_2-
\tilde{l}^{\gamma^*\gamma^*Z^*}_3) ~ ,
\label{gammagammaZ-CPv-l-on}
\eq
which express the on-shell parameters of \cite{Hagn, neut, modzg},
in terms of the present ones. We observe that
$\tilde{l}^{\gamma^*\gamma^*Z^*}_4$ is not involved and that only
two transverse parameters appear instead of four in the off-shell case.

\vspace{0.5cm}
\subsection{ Comments about the lowest dimensional parametrization.}

As already said the effective Lagrangian of
(\ref{NP-Lagrangian}) is suitable for describing NP effects
generated at a very high scale. If this occurs, then it may
 turn out to be
adequate to restrict to operators of $dim = 6$.
Keeping  only  transverse terms, (which is absolutely
legitimate, provided that no events involving $Z\to t\bar t$
decays are considered), then we end up with just four CP conserving and
 four CP-violating couplings; namely
\bq
l^{Z^*Z^*Z^*}_1 ~~, ~~
\tilde{l}^{Z^*Z^*Z^*}_1 ~~, ~~
l^{Z^*Z^*\gamma^*}_1 ~~,~~
l^{Z^*Z^*\gamma^*}_2 ~~, ~~
\tilde{l}^{Z^*Z^*\gamma^*}_1~~,~~
\tilde{l}^{Z^*Z^*\gamma^*}_3 ~~, ~~
l^{\gamma^*\gamma^*Z^*}_1 ~~, ~~
\tilde{l}^{\gamma^*\gamma^*Z^*}_1 ~~. \label{l-tran-dim6}
\eq
If in addition,  the higher dimensional operators above are also
included, we have to add to this  set of parameters the ones
\bqa
&&l^{Z^*Z^*Z^*}_2 ~~,~~l^{Z^*Z^*Z^*}_3 ~~,~~
\tilde l^{Z^*Z^*Z^*}_2 ~~,~~
\tilde l^{Z^*Z^*Z^*}_3 ~~,~~
\tilde l^{Z^*Z^*Z^*}_4 ~~,~~
\nonumber \\
&& l^{Z^*Z^*\gamma^*}_3 ~~, ~~
\tilde l^{Z^*Z^*\gamma^*}_2 ~~, ~~
\tilde l^{Z^*Z^*\gamma^*}_4 ~~, ~~
\nonumber \\
&& l^{Z^*Z^*\gamma^*}_2~~,~~
l^{Z^*Z^*\gamma^*}_3~~,~~
\tilde{l}^{Z^*Z^*\gamma^*}_2~~,~~
\tilde{l}^{Z^*Z^*\gamma^*}_3~~,~~
\tilde{l}^{Z^*Z^*\gamma^*}_4~~. \label{l-tran-highdim}
\eqa

Thus, within the context of the effective Lagrangian of this Section
3, we need 21 parameters to describe the off-shell effects for
all "transverse" NAGC. These parameters would be related to those
defined on-shell in \cite{Hagn, neut, modzg} by
Eqs. (\ref{ZZZ-CPc-l-on}, \ref{ZZZ-CPv-l-on},
\ref{ZZgamma-CPc-l1-on}, \ref{ZZgamma-CPc-l23-on},
\ref{ZZgamma-CPv-l-on}, \ref{ZZgamma-CPv-l23-on},
\ref{gammagammaZ-CPc-l-on}, \ref{gammagammaZ-CPv-l-on}).
Furthermore, if the  NP scale is very high,
then it is natural to expect that the $dim=8$ terms, (which are
proportional to $1/\Lambda^4$), should be  strongly suppressed.
The suppression should even be stronger for
the  higher $dim=10,12$ terms. In this case the set of eight parameters
in (\ref{l-tran-dim6}) should be the dominant ones.\par

Let us insist on the merit of the effective Lagrangian
(\ref{NP-Lagrangian}) which allows
through eq.(\ref{ZZZ-CPc-Lan}, \ref{ZZZ-CPv-Lan},
\ref{ZZgamma-CPc-Lan}, \ref{ZZgamma-CPv-Lan},
\ref{gammagammaZ-CPc-Lan}, \ref{gammagammaZ-CPv-Lan}), to get
the precise off-shell $s_i$-dependence of the amplitudes
consistent with  Bose symmetry and CVC.
Provided the NP scale is high, these
 should the suitable expressions
for a model independent data analysis.

On the other hand, if  the NP scale inducing NAGC  is near the energy
scale of the measurements, then the effective Lagrangian description
becomes inadequate. In such a case, dynamical  models like those
considered in the next Section can be much more useful in
providing hints for the description of the possible New
Physics.

Finally, if $Z\to t\bar t$ decays are also
included in the  NAGC analysis;
then  the  "scalar" couplings should also be
included. Altogether, there exist 23 such
couplings in the effective Lagrangian listed above.
Eleven of them correspond to $dim=6$ operators,
and constitute a set of the  three  CP conserving
\[
  l^{Z^*Z^*Z^*}_4 ~~,~~  l^{Z^*Z^*\gamma^*}_4 ~~ , ~~
 l^{\gamma^*\gamma^*Z^*}_4 ~~,
\]
and the  eight CP-violating
\[
\tilde{l}^{Z^*Z^*Z^*}_5 ~~, ~~
\tilde{l}^{Z^*Z^*Z^*}_6 ~~, ~~
\tilde{l}^{Z^*Z^*Z^*}_7 ~~, ~~
\tilde{l}^{Z^*Z^*Z^*}_8 ~~, ~~
\tilde{l}^{Z^*Z^*Z^*}_{14} ~~,
\tilde{l}^{Z^*Z^*\gamma^*}_5 ~~, ~~
\tilde{l}^{Z^*Z^*\gamma^*}_6 ~~,
\tilde{l}^{\gamma^*\gamma^*Z^*}_5 ~~
\]
couplings;
while the remaining 12  describe higher dimensional
scalar NAGC.\par

\vspace*{1cm}
Before concluding this sub-section we add a few comments
concerning $SU(2)\times U(1)$ gauge invariance.
Strictly speaking the NP vertices introduced to
 the effective Lagrangian by the NP operators in
(\ref{ZZZ-CPc-op}, \ref{ZZZ-CPv-op}
\ref{ZZgamma-CPc-op}, \ref{ZZgamma-CPv-op},
\ref{gammagammaZ-CPc-op}, \ref{gammagammaZ-CPv-op}),
should only be used in the unitary
gauge\footnote{We would like to thank E. Boos for discussions
on this point.}.
This restriction can be easily cured though,
by making  the substitutions
\bqa
Z_{\mu \nu} & \longrightarrow &
- \sw B_{\mu\nu} -~\frac{2\cw}{v^2}
 (\Phi^\dagger \vec \tau \Phi)
\cdot \vec W_{\mu\nu} ~~, \nonumber \\
F_{\mu \nu} & \longrightarrow &
 \cw B_{\mu\nu} -~\frac{2\sw}{v^2}
 (\Phi^\dagger \vec \tau \Phi)
\cdot \vec W_{\mu\nu} ~~, \nonumber \\
Z_\mu  & \longrightarrow &
i~ \frac{4\sw \cw}{e v^2}~  (\Phi^\dagger D_\mu  \Phi)
 ~~, \label{SU2U1-substitution}
\eqa
which transforms them to a gauge invariant form.
In (\ref{SU2U1-substitution}) $\Phi$ is the SM Higgs doublet,
$v$ its vacuum expectation value,
 and  $D_\mu$ is the usual $SU(2)\times U(1)$
covariant derivative.\par

The substitutions (\ref{SU2U1-substitution})
generally change the dimensionality of the various operators.
If after performing them, we make the further restriction that
only the lowest $dim=8$ operators are retained, then we just end
up with the two operators
\bqa
\O_{SU(2)\times U(1)} & = &
i \tilde{B}_{\mu\nu} (\partial_\sigma B^{\sigma\mu})
(\Phi^\dagger D^\nu  \Phi) ~~ ,  \nonumber \\
\tilde{\O}_{SU(2)\times U(1)} & = &
i B_{\mu\nu} (\partial_\sigma B^{\sigma\mu})
(\Phi^\dagger D^\nu  \Phi) ~~ . \label{SU2U1-op}
\eqa
These  are  the only $dim=8$
$SU(2)\times U(1)$  invariant operators
which in  the unitary gauge only involve
 either  purely neutral \underline{triple} gauge
couplings, or  couplings  affecting three
neutral gauge bosons and a  Higgs field.
They are closely related to the $\O_1^{V_1V_2V_3}$ and
$\tilde{\O}_1^{V_1V_2V_3}$ defined in the various sub-sections
above.

\section{A toy model: the fermionic triangle loop}

In  \cite{modzg},
we have discussed the possible dynamical origin of the
 triple neutral gauge boson interactions, when two of the gauge
bosons are on-shell. The first conclusion there was that,
at the  1-loop level of
any  fundamental renormalizable gauge theory,
non-vanishing contributions could only arise if  fermions run
along the loop; the bosonic loop always giving a vanishing result.
The second point was that no CP-violating NAGC couplings are generated in
such a context.\par

Here we explore the consequences of
this model when all three neutral gauge bosons are taken
off-shell.\par

\subsection{General structure of 1-loop couplings}

The triangle diagram is depicted in Fig.\ref{trif-fig}. The
fermion couplings are defined through the gauge Lagrangian
\cite{modzg}
\bqa
\L & = & -eQ_F A^\mu\bar F \gamma_\mu F
-{e\over2s_Wc_W}  Z^\mu \bar F \left  ( \gamma_\mu g_{vF}
- \gamma_\mu \gamma_5 g_{aF} \right ) F ~ .
\label{VFF}
\eqa

The complete expressions of the resulting off-shell
CP-conserving NAGC are
given in Appendix C, where for simplicity
we take a single fermion running along
the triangular loop.  These expressions
are directly applicable to any fermionic
contributions. Thus, \eg~ the SM prediction for the neutral
gauge boson self-interactions is obtained by summing
the contributions of the leptons and of the quarks.\par

To present these  results,
we first observe that the 1-loop fermionic diagrams
strongly reduce the six independent forms
that could exist in the general case;
(compare the most general type of such forms in Appendix A).
More explicitly, the
only non-vanishing coupling-functions contained in the 1-loop
diagrams  are  the two non-vanishing  transverse  ones
called    $f_{1,2}(s_1,s_2,s_3)$, and
a\footnote{Depending on the NAGC coupling considered, there
may by additional scalar functions like
$g_{2}(s_1,s_2,s_3)$ and/or $g_{3}(s_1,s_2,s_3)$;
but these functions are related to $g_{1}(s_1,s_2,s_3)$ by
equations like (\ref{G123p-loop}), since
 $f_3(s_1, s_2, s_3)\equiv 0$ for the diagram
in Fig.\ref{trif-fig}.}
single  scalar function  called  $g_{1}(s_1,s_2,s_3)$.
In particular, no $h_4$-type of coupling
(compare \cite{Hagn, neut}),  is allowed by such
diagrams. This has already been noticed in the
on-shell case \cite{modzg}; where it has been remarked that
higher order or non-perturbative effects are
required for generating  $h_4$-couplings.\par

To establish contact with the  Effective Lagrangian  of
Section 3, we   consider the heavy
fermion limit of the above functions.
In such a limit, (retaining only the dominant $1/M_F^2$
contributions),
the  heavy fermion loop  predictions are
 identical to those of a CP-conserving  effective
Lagrangian in which the only non-vanishing couplings are
\[
l^{Z^*Z^*Z^*}_1 ~,~ l^{Z^*Z^*Z^*}_4~ , ~
l^{Z^*Z^*\gamma^*}_1 ~,~ l^{Z^*Z^*\gamma^*}_2 ~,~
l^{Z^*Z^*\gamma^*}_4 ~,~
l^{\gamma^*\gamma^* Z^*}_1 ~,~ l^{\gamma^*\gamma^* Z^*}_4 ~.
\]

Of course, if the mass of  the fermion in the loop of
Fig.\ref{trif-fig} is comparable to (or lighter than) the energies
considered,  additional structures appear in the $f_{1,2}$ and
$g_1$ functions, that cannot be described by the above effective
Lagrangian. If NAGC are ever observed, then the experimental search for
such  structures, will provide a very important means for
identifying the responsible  NP degrees of freedom.\par

\subsubsection{The $Z^*Z^*Z^*$ couplings at 1-loop.}

Following the results in Appendix C,
the fermionic triangle contribution is written as
\bqa
&&f^{Z^*Z^*Z^*}_1(s_1,s_2,s_3)=-~{e^2 g_{aF}\over32\pi^2s^3_Wc^3_W}
\{(3g^2_{vF}+g^2_{aF})\G_1(s_1,s_2,s_3)
-(g^2_{aF}-g^2_{vF})\G_3(s_1,s_2,s_3)\} , \nonumber\\
&&f^{Z^*Z^*Z^*}_2(s_1,s_2,s_3)={e^2 g_{aF}\over32\pi^2s^3_Wc^3_W}
\{(3g^2_{vF}+g^2_{aF})\G_2(s_1,s_2,s_3)
-(g^2_{aF}-g^2_{vF})\G_4(s_1,s_2,s_3)\}, \nonumber\\
&&g^{Z^*Z^*Z^*}_1(s_1,s_2,s_3)={e^2\over8\pi^2s^3_Wc^3_W}g_{aF}
(3g^2_{vF}+g^2_{aF})\G'_1(s_1,s_2,s_3) ~ , \label{ZZZ-loop-fg}
\eqa
where the functions  $\G_i(s_1,s_2,s_3)$ and $\G'_1$ are
given in Appendix C in terms of Passarino-Veltman
$B_0$ and $C_0$ functions \cite{PVHag}.\par

As required by the anomaly cancellation (and explained in Appendix C),
all the $\G_j$ and $\G'_j$ functions vanish in the  large
$M_F$ limit. Moreover, at the $1/ M^2_F$ level, they satisfy
\bqa
\G_1   \simeq 3\G_3   \simeq   {s_1+s_2-2s_3\over40M^2_F}~~~,~
& \G_2 \simeq 3\G_4 \simeq  {3(s_2-s_1) \over 40M^2_F} ~~~,
& ~\G'_1 \simeq {1\over 24M^2_F} ~ ~ , \label{Gi-a-asym}
\eqa
from which the leading contributions to  $f_{1,2}$ and $g_1$
are calculated using (\ref{ZZZ-loop-fg}).
As expected, these large $M_F$ results coincide
with those of the effective Lagrangian description, with the only
non zero parameters being
\bqa
l^{Z^*Z^*Z^*}_1 &= &\left ({g_{aF} \over 30M^2_F}\right)
\left ({e^2\over32\pi^2s^3_Wc^3_W}\right )
 (5g^2_{vF}+g^2_{aF})  ~~ , \nonumber\\
l^{Z^*Z^*Z^*}_4 & = &\left ({g_{aF} \over 60 M^2_F}\right)
\left ({e^2\over32\pi^2s^3_Wc^3_W}\right )(5g^2_{vF}+3 g^2_{aF})
~ ~ . \label{ZZZ-loop-l-asym}
\eqa\par

Combining this with (\ref{ZZZ-CPc-Lan}) for
the on-shell case $Z^*\to ZZ$ ($s_1=s_2=m^2_Z$), for which
(\ref{Gi-a-asym}) implies   $\G_{2,4}=0$,
we obtain
\bqa
&&f^{Z^*Z^*Z^*}_1(m^2_Z,m^2_Z,s_3)=
(s_3-m^2_Z)l^{Z^*Z^*Z^*}_1={s_3-m^2_Z\over m^2_Z}f^Z_5(s_3)
 ~ ~, \nonumber\\
&& f^{Z^*Z^*Z^*}_2(m^2_Z,m^2_Z,s_3)=0  ~~ ,
\eqa
which agrees with the expression given in \cite{modzg}.\par

\subsubsection{ The $Z^*Z^*\gamma^*$ couplings at 1-loop.}

The formalism in  Appendix C leads to
\bqa
&&f^{Z^*Z^*\gamma^*}_1(s_1,s_2,s_3)=-~{e^2Q_Fg_{aF}g_{vF}\over
8\pi^2s^2_Wc^2_W}~[\G_1(s_1,s_2,s_3)+\G_5(s_1,s_2,s_3)] ~,
\nonumber \\
&&f^{Z^*Z^*\gamma^*}_2(s_1,s_2,s_3)={e^2Q_Fg_{aF}g_{vF}\over
8\pi^2s^2_Wc^2_W}~[\G_2(s_1,s_2,s_3)+{1\over3}\G_4(s_1,s_2,s_3)]
~ , \nonumber \\
&&g^{Z^*Z^*\gamma^*}_1(s_1,s_2,s_3)=g^{Z^*Z^*\gamma^*}_2(s_2,s_1,s_3)
={e^2Q_Fg_{aF}g_{vF}\over
2\pi^2s^2_Wc^2_W}~ \G'_1(s_1,s_2,s_3) ~ , \label{ZZgamma-loop-fg}
\eqa
where the needed $\G_j$-functions are again given there.

To derive the leading contribution to these couplings in the large
$ M^2_F$ limit, we need first the leading contributions to the
$\G_j$ defined in Appendix C. Keeping terms only up to the
$1/M^2_F$ order, (as in the derivation of (\ref{Gi-a-asym})),
this is given by
\bq
\G_1+\G_5 \simeq -{s_3\over12M^2_F}~~~, ~~~
\G_2+{1\over3}\G_4 \simeq {(s_2-s_1)\over12M^2_F}~~~,
~~~\G'_1  \simeq {1\over 24M^2_F} ~, \label{Gi-b-asym}
\eq
which, substituted to (\ref{ZZgamma-loop-fg}), result to values
of the couplings functions consistent with those obtained in
(\ref{ZZgamma-CPc-Lan}), provided
\bq
- l^{Z^*Z^*\gamma^*}_1=
 l^{Z^*Z^*\gamma^*}_2=
 2l^{Z^*Z^*\gamma^*}_4= \left({1\over12M^2_F}\right )
{e^2Q_Fg_{aF}g_{vF}\over 8\pi^2s^2_Wc^2_W} ~ ,
\label{ZZgamma-loop-l-asym}
\eq
while all other $l_j^{Z^*Z^*\gamma^*}$ should vanish.

Comparing to the on-shell cases:
\begin{itemize}
\item[a)]
 $\gamma^*\to ZZ$, $s_1=s_2=m^2_Z$,  $\G_{2}+\G_{4}/3=0$
leads to
\bqa
&&f^{Z^*Z^*\gamma^*}_1(m^2_Z,m^2_Z,s_3)=s_3l^{Z^*Z^*\gamma^*}_2=
{s_3\over m^2_Z}f^{\gamma}_5(s_3)  \nonumber\\
&&f^{Z^*Z^*\gamma^*}_2(m^2_Z,m^2_Z,s_3)=0~~.
\eqa

\item[b) ]
 $Z^*\to Z\gamma$, $s_3=0~,~s_1=m^2_Z$,
$\G_{1}+\G_{5}=0$ implies   $h^{Z}_4=0$, and
\bqa
&&f^{Z^*Z^*\gamma^*}_1(m^2_Z,s_2,0) =0~~~, \nonumber\\
&&f^{Z^*Z^*\gamma^*}_2(m^2_Z,s_2,0)=(m^2_z-s_2)l^{Z^*Z^*\gamma^*}_1=
{m^2_z-s_2\over m^2_Z}h^Z_3(s_1) ~~,
\eqa
\end{itemize}
\noindent
which agree with the expressions given in \cite{modzg}.

\subsubsection{ The $\gamma^*\gamma^*Z^*$ couplings at 1-loop.}

The results of Appendix C give
\bqa
&&f^{\gamma^*\gamma^*Z^*}_1(s_1,s_2,s_3)=-~{e^2Q^2_Fg_{aF}\over
8\pi^2s_Wc_W}~ [\G_6(s_1,s_2,s_3)+\G_7(s_1,s_2,s_3)]
 ~ , \nonumber\\
&&f^{\gamma^*\gamma^*Z^*}_2(s_1,s_2,s_3)=-~{e^2Q^2_Fg_{aF}\over
8\pi^2s_Wc_W} ~[ \G_6(s_1,s_2,s_3)-\G_7(s_1,s_2,s_3)]
~, \nonumber\\
&&g^{\gamma^*\gamma^*Z^*}_3(s_1,s_2,s_3)=~{e^2Q^2_Fg_{aF}\over
2\pi^2s_Wc_W}\G'_1(s_1,s_2,s_3) ~ . \label{gammagammaZ-loop-fg}
\eqa
At the $1/ M^2_F$ level, the leading  heavy fermion values
of  the $\G_j$-combinations appearing in
(\ref{gammagammaZ-loop-fg}), are
\bqa
&&\G_6(s_1,s_2,s_3)+\G_7(s_1,s_2,s_3) \simeq
{s_2+s_1\over12M^2_F} ~ , \nonumber\\
&&\G_6(s_1,s_2,s_3)-\G_7(s_1,s_2,s_3)\simeq {(s_1-s_2)\over12M^2_F}
~ , \nonumber\\
&&  \G'_1 \simeq {1\over24M^2_F} ~~ ,
\eqa
which as expected  coincide
with the effective Lagrangian results of
(\ref{gammagammaZ-CPc-Lan}) provided the only non-vanishing
couplings are
\bq
- l^{\gamma^*\gamma^*Z^*}_1= 4l^{\gamma^*\gamma^*Z^*}_4=
\left({1\over6M^2_F}\right ){e^2Q^2_Fg_{aF}\over
8\pi^2s_Wc_W} ~~ .
\eq

When only one photon and one $Z$ are on-shell,
(\ie~ $s_2=0$, $s_3=m^2_Z$, $\G_7=0$ ) we obtain
\bq
f^{\gamma^*\gamma^*Z^*}_1(s_1,0,m^2_Z)=
f^{\gamma^*\gamma^*Z^*}_2(s_1,0,m^2_Z)=
{s_1\over2}~ l^{\gamma^*\gamma^*Z^*}_1=
{s_1\over2m^2_Z} ~h^{\gamma}_3(s_1)~~, ~~
\eq
\noindent
which agree with the expressions given in \cite{modzg}.

\subsection{Quantitative discussion of the 1-loop off-shell effects.}

After having shown the structure of the NAGC generated at
1-loop, we now make a quantitative discussion
of the off-shell effects. These effects
are described below by three sets of  ratios which
quantify the following features:
\begin{itemize}
\item[a)]
The ratios $R^{5Z}_1$, $R^{5Z}_3$, $R^{5\gamma}$, $R^{3Z}$, $R^{3\gamma}$
 are sensitive to the $s_i$-dependences of the type of couplings
 existing already on-shell.

\item[b)]
The ratios $R'^{5Z}_1$, $R'^{5Z}_3$, $R'^{5\gamma}$,
$R'^{3Z}$, $R'^{3\gamma}$ study the relative size
(versus $s_i$) of  new types of couplings
as compared to those  already existing on-shell.

\item[c)]
The ratios $R^{ZZZ}_1$,
$R^{ZZZ}_2$, $R^{ZZ\gamma}$, $R^{Z\gamma Z}$, $R^{\gamma\gamma Z}$
aim to quantify the range of the mass  $M_F$ of the fermion
running along the loop, for which  the effective
Lagrangian structure (which already contains some
$s_i$-dependence) is adequate.
\end{itemize}

For each ratio, we indicate below their value in the large $M_F$
limit. As shown in the previous Section, these values agree with
the predictions of the effective Lagrangian.
We have compared these values
to a numerical computation done with the exact expressions
for finite $M_F$ values and for some choices of $s_i$ values
falling inside the range accessible at LEP2 ($0.2~TeV$) or at
LC ($0.5~TeV$). In Fig.\ref{fig3}-\ref{fig6}
we have selected some typical
examples of the $s_i$ and $M_F$ behaviours.\\

The above three points are discussed in turn for each NAGC vertex:

\subsubsection{ The off-shell 1-loop effects in $Z^*Z^*Z^*$
compared to $Z^*\to ZZ$.}
\begin{itemize}
\item[a)]
 The ratios $R^{5Z}_1$ and $R^{5Z}_3$ show the evolution of the
contributions to the $f_1^{Z^*Z^*Z^*}(s_1, s_2, s_2)$ -type
of coupling as defined in (\ref{ZZZ-loop-fg}),
from $s_1=s_2=m^2_Z$ up to some off-shell value:
\bqa
&&R^{5Z}_1={\G_1(s_1,s_2,s_3)\over \G_1(m^2_Z,m^2_Z,s_3)}
\to{2s_3-s_1-s_2\over2(s_3-m^2_Z)} ~~, \nonumber\\
&&R^{5Z}_3={\G_3(s_1,s_2,s_3)\over \G_3(m^2_Z,m^2_Z,s_3)}
\to{2s_3-s_1-s_2\over2(s_3-m^2_Z)} ~~. \label{Rj5Z}
\eqa
Note from (\ref{ZZZ-CPc-f-on}) the way that
$f_1^{Z^*Z^*Z^*}(s_1, s_2, s_2)$ is related to the on-shell
$f_5^Z$ coupling of \cite{Hagn, neut}.

\item[b)]
 The ratios $R'^{5Z}_1$ and $R'^{5Z}_3$ give the relative size of
the new $f^{Z^*Z^*Z^*}_2$ coupling as compared to
$f^{Z^*Z^*Z^*}_1$ already existing on-shell
\bqa
&&R'^{5Z}_1=-~{\G_2(s_1,s_2,s_3)\over \G_1(s_1,s_2,s_3)}
\to ~{3(s_1-s_2)\over s_1+s_2-2s_3} ~ , \nonumber\\
&&R'^{5Z}_3={\G_4(s_1,s_2,s_3)\over \G_3(s_1,s_2,s_3)}
\to-~{3 (s_1-s_2)\over s_1+s_2-2s_3} ~ . \label{Rpj5Z}
\eqa
\end{itemize}

The four  ratios in
(\ref{Rj5Z}, \ref{Rpj5Z}) are plotted versus $\sqrt{s_2}$
in Fig.\ref{fig3}a and Fig.\ref{fig3}b,    for
$\sqrt{s_3}=0.2$, and $0.5~TeV$
respectively. The  fixed values of $\sqrt{s_1}$ and  $M_F$
are indicated in the figures.
It can be seen there, that
the quadratic $s_2$-dependence predicted for the large
$M_F$ limit, starts to be valid already at a  rather low $M_F$;
 apart from   threshold violations  at
$s_2 \sim 4M^2_F$.

\begin{itemize}
\item[c)]
 The ratios $R^{ZZZ}_1$ and $R^{ZZZ}_2$,
\bqa
R^{ZZZ}_1 &= & {(2s_3-s_1-s_2)\G_2\over 3 \G_1(s_1-s_2)}\to 1
~~ , \nonumber \\
R^{ZZZ}_2 & = &{(2s_3-s_1-s_2)\G_4\over 3\G_3(s_1-s_2)}\to 1
~~ , \label{RjZZZ}
\eqa
\end{itemize}
which are equal to $1$ at large $M_f$,
show how much the exact 1-loop contribution
at finite values of $M_F$, differs from
the effective Lagrangian prediction.
They are presented in
 Fig.\ref{fig3}c versus $M_F$,  for $\sqrt{s_3}=0.2,~0.5~TeV$, and fixed
typical values of $\sqrt{s_{1,2}}$. For these ratios also,
we observe that they  are close to their large $M_F$ limits,
provided that $M_F$ is away from the threshold $\sqrt{s_3}/2$.\\

Similar ratios are next constructed for the other NAGC processes.

\subsubsection{The off-shell 1-loop effects in
$Z^*Z^*\gamma^*$  compared to  $\gamma^*\to ZZ$}

The corresponding ratios are
\bqa
R^{5\gamma} &= & {\G_1(s_1,s_2,s_3)+\G_5(s_1,s_2,s_3)
\over \G_1(m^2_Z,m^2_Z,s_3)+\G_5(m^2_Z,m^2_Z,s_3)}\to 1
~ ~ , \label{R5gamma} \\
R'^{5\gamma}& = & -~{3\G_2(s_1,s_2,s_3)+\G_4(s_1,s_2,s_3)
\over 3\G_1(s_1,s_2,s_3)+3\G_5(s_1,s_2,s_3)}\to {s_2-s_1\over s_3}
~~ , \label{Rp5gamma}
\eqa
illustrated versus $\sqrt{s_2}$ in Fig.\ref{fig4}a
for $\sqrt{s_3}=0.2,~0.5~TeV$ and fixed $\sqrt{s_1}$,
$M_F$; and the ratio
\bq
R^{ZZ\gamma}={s_3(3\G_2+\G_4)\over3(s_1-s_2)(\G_1+\G_5)}\to 1
~~, \label{RZZgamma}
\eq
presented versus $M_F$ in Fig.\ref{fig4}b,
for $\sqrt{s_3}=0.2,~0.5~TeV$, and
typical values of $\sqrt{s_{1,2}}$.

\subsubsection{The off-shell 1-loop effects
in $Z^*Z^*\gamma^*$ compared   to $Z^*\to Z\gamma$}

The relevant ratios (together with their large $M_F$ limits)
are
\bqa
R^{3Z}&=&{3\G_2(s_1,s_2,s_3)+\G_4(s_1,s_2,s_3)-3\G_1(s_1,s_2,s_3)
-3\G_5(s_1,s_2,s_3)
\over 3\G_2(m^2_Z,s_2,0)+\G_4(m^2_Z,s_2,0)-3\G_1(m^2_Z,s_2,0)
-3\G_5(m^2_Z,s_2,0)} \nonumber \\
& \to & {s_2+s_3-s_1 \over s_2-m^2_Z} ~~ , \label{R3Z} \\
R'^{3Z}&=&-~{3\G_1(s_1,s_2,s_3)+3\G_5(s_1,s_2,s_3)
\over 3\G_2(s_1,s_2,s_3)+\G_4(s_1,s_2,s_3)-3\G_1(s_1,s_2,s_3)
-3\G_5(s_1,s_2,s_3)}\nonumber\\
&\to & {s_3\over s_2-s_1+s_3} ~~ , \label{Rp3Z}
\eqa
presented versus $\sqrt{s_3}$ in  Fig.\ref{fig5}a,b,
 for $\sqrt{s_2}=0.2$, $0.5~TeV$, and
fixed values of $\sqrt{s_1}$, $M_F$.\par

On the other hand, the ratio
\bq
R^{Z\gamma Z} =
{s_3 [3\G_2(s_1, s_3, s_2)+\G_4(s_1, s_3, s_2)]
\over3(s_1-s_2)[\G_1(s_1, s_3, s_2)+\G_5(s_1, s_3, s_2)]}~ \to ~1
~~ , \label{RZgammaZ}
\eq
is shown versus $M_F$  in  Fig.\ref{fig5}c
for $\sqrt{s_2}=0.2,~0.5~TeV$,
and  fixed values of $\sqrt{s_{1,3}}$.

\subsubsection{The off-shell 1-loop
effects in $\gamma^*\gamma^*Z^*$ compared
to $\gamma^*\to Z\gamma$.}

We now have
\bqa
R^{3\gamma}& = & {\G_6(s_1,s_2,s_3) \over \G_6(s_1,0,m^2_Z)}\to 1
~~ , \label{R3gamma} \\
R'^{3\gamma} & = & {\G_7(s_1,s_2,s_3) \over
\G_6(s_1,s_2,s_3)}\to {s_2\over s_1} ~~ , \label{Rp3gamma}
\eqa
presented versus $\sqrt{s_2}$ in Fig.\ref{fig6}a
for $\sqrt{s_1}=0.2,~0.5~TeV$,
and  $\sqrt{s_3}$, $M_F$; while the ratio
\bq
R^{\gamma\gamma Z}={s_1\G_7(s_1, s_2, s_3)
\over s_2\G_6(s_1, s_2, s_3)}\to 1 ~~ ,
\eq
versus $M_F$ in Fig.\ref{fig6}b for
$\sqrt{s_1}=0.2,~0.5~TeV$ and  fixed
typical values of $\sqrt{s_{2,3}}$.

\subsubsection{\bf General comments:}

We have made many other runs with different $s_i$ and $M_F$ values.
The following  are the general conclusions we  draw from these:
\begin{itemize}
\item
The first is that the off-shell effects cannot be ignored in
detail experiments like those performed at
LEP2, where data with a fermion-pair invariant mass
 ranging from very low values
up  to about $m_Z$, have been collected. That will be even more true
at a Linear Collider in the future.
\item
Our 1-loop calculations indicate
that the large $M_F$ predictions are quite adequate, even
at low $M_F$ values, so long as $M_F$ is not too
close to a threshold. This is the same situation as in the previous
on-shell analysis,  \cite{modzg}.
It is furthermore a welcome situation, since it encourages us to
analyze the data, by using
the effective Lagrangian formalism, in which only  operators
of $dim \leq 6$ are retained. Ignoring $Z \to t\bar t$ events,
this means that the 8 parameters in (\ref{l-tran-dim6})
may be adequate, provided of course that
 we are not too close to an NP threshold.
\item
If on the other hand we are close to an NP threshold,
then we might even have direct production of new particles.
In such a case, the study of NAGC will provide
 useful complementary information on their
nature. Particularly because the set of new particle
 parameters entering their loop NAGC contribution,
is certainly different from
the one determining \eg~ their decay. This is obviously true \eg~
for NP of the  SUSY type.
\end{itemize}

\section{General off-shell NAGC contribution to \\
$e^-e^+\to f \bar f f'\bar f'$ }

The NAGC contribution to the $e^+e^-\to (f\bar f)+(f'\bar f')$
process is depicted
in Fig.\ref{eeffff-fig}. The complete Feynman
amplitude has the general form:
\bq
\A=-~{e\over m^2_Z}\sum_{ijk} {\V^{\sigma}_i(f\bar f)\over D_i}~
 {\V^{\tau}_j(f'\bar f')\over D_j}~
\Gamma^{ijk}_{\sigma\tau\rho}~ {\V^{\rho}_k(e^+e^-)\over D_k}
\eq
\noindent
where the summation over $ijk$ covers all possible off-shell
combinations of $\gamma^*$ and $Z^*$, namely $Z^*Z^*Z^*$,
$Z^*Z^*\gamma^*$, $Z^*\gamma^*Z^*$,
$Z^*\gamma^*\gamma^*$,
$\gamma^*Z^*Z^*$, $\gamma^*Z^*\gamma^*$,and
$\gamma^*\gamma^*Z^*$,
 with the propagators
\[
D_{i,j,k} = q^2_{i,j,k}~~ \mbox{for a}~ \gamma^*,~~~ \mbox{or}~~~
q^2_{i,j,k}-m^2_Z+im_Z\Gamma_Z ~~ \mbox{for a}~ Z^*
\]
\noindent
and the initial and
final fermionic vertices
\bqa
\V^{\sigma}_i(f\bar f)& = &\bar
u(f)\gamma^{\sigma}(g^i_{vf}-g^i_{af}\gamma^5) v(\bar f)
~ ~ , \nonumber\\
\V^{\tau}_j(f'\bar f')& = &\bar
u(f')\gamma^{\tau}(g^j_{vf'}-g^j_{af'}\gamma^5) v(\bar f')
~~, \nonumber\\
\V^{\rho}_k(e^+e^-)& = &
\bar v(e^+)\gamma^{\rho}(g^k_{ve}-g^k_{ae}\gamma^5) u(e^-)
\label{Vff-off}
\eqa
\noindent
 with $g^i_{vf},~g^i_{af}$ being the vector and axial, photon or $Z$,
couplings to the fermion $f$ (including the factor $-e$ or
$-e/2s_Wc_W$)\footnote{We mention for completeness
that conventions  are such that the effective Lagrangian
for a  gauge boson fermion interaction is
\[
\L= V_{i\mu}\V^{\mu}_i(f\bar f).
\]}
. In (\ref{Vff-off}),
$\Gamma^{ijk}_{\sigma\tau\rho}$ are the general vertices given in
Appendices A,B and discussed throughout the paper.\par

One should be careful in reordering the indices and momenta in the
various $(i,j,k)$ combinations in order to use the formulae written
for $Z^*Z^*\gamma^*$ and
$\gamma^*\gamma^* Z^*$ in Appendix A,B; so for
clarity we list them explicitly:
\bqa
\Gamma^{Z^*\gamma^*\gamma^*}_{\sigma\tau\rho}(q_1,q_2,q_3=-P)
& = &
\Gamma^{\gamma^*\gamma^* Z^*}_{\rho\tau\sigma}(q_3=-P,q_2,q_1)
~~ , \\
\Gamma^{\gamma^*Z^*\gamma^*}_{\sigma\tau\rho}(q_1,q_2,q_3=-P)
& = &
\Gamma^{\gamma^*\gamma^* Z^*}_{\sigma\rho\tau}(q_1,q_3=-P,q_2)
~~ , \\
\Gamma^{Z^*\gamma^*Z^*}_{\sigma\tau\rho}(q_1,q_2,q_3=-P)
& = &
\Gamma^{Z^* Z^*\gamma^*}_{\sigma\rho\tau}(q_1,q_3=-P,q_2)
~~ , \\
\Gamma^{\gamma^*Z^*Z^*}_{\sigma\tau\rho}(q_1,q_2,q_3=-P)
& = &
\Gamma^{Z^* Z^*\gamma^*}_{\rho\tau\sigma}(q_3=-P,q_2,q_1)
~~ .
\eqa
The basic SM (or MSSM) contributions are assumed to
 be included in
the $\Gamma$ vertices expressed in terms of $f_i$ and
$\tilde{f}_i$ defined in Section 2, using the
analytic expressions given in Appendix C.\par

For an experimental determination of  possible unknown additional
contributions, a simple parametrization of  the
$f_i(s_1,s_2,s_3)$ and $g_i(s_1,s_2,s_3)$ is needed.
If the NP effects arise at a high scale,
then the the effective Lagrangian of Section 3,
in which only the lowest dimensional operators are retained,
may  be adequate.

\section{Conclusions}

We have established the general Lorentz  and $U(1)_{em}$
invariant form of the off-shell three
neutral gauge boson self-couplings $V_1^*V_2^*V_3^*$, with applications
to $Z^*Z^*Z^*$, $Z^*Z^*\gamma^*$, and
$\gamma^*\gamma^*Z^*$.
In it, we have kept all types of transverse and scalar
off-shell vector boson components; and
considered both CP-conserving and CP-violating couplings.
They are given in Appendix A and B, respectively.
We have pointed out the new coupling forms  which do not
exist when two particles are
on-shell, thus making contact with the previous description
valid only when two gauge bosons are on-shell,
\cite{Hagn, neut, modzg}.

In the  $Z^*Z^*Z^*$ case, we have found (3 transverse + 3 scalar)
CP-conserving
and (4 transverse + 10 scalar) CP-violating coupling forms;
which reduce in the previously considered $Z^*\to ZZ$ on-shell case
 to (1+1)+(1+3).

In the $Z^*Z^*\gamma^*$ case we have found (3+2)+(4+5) coupling forms.
They reduce to
(1+0)+(1+0) in $\gamma^* \to ZZ$, and to (2+1)+(2+2)
in $Z^* \to Z\gamma$.

Finally in the $\gamma^*\gamma^*Z^*$ case we found (3+1)+(4+2)
coupling forms,
which reduce to (2+1)+(2+0) in $\gamma^* \to Z\gamma$.\par

These vertex forms apply to any kind of standard or non standard
dynamics (SM, MSSM,....). In general the functions which multiply these
coupling forms depend on the three off-shell masses ($s_1,s_2,s_3$).
If the  NP  scale inducing NAGC is very high
 ($\Lambda \gg m_Z$), then we have found that  an
effective Lagrangian involving a minimal set of operators
should be adequate for generating
 all  possible vertex forms consistent with Bose symmetry
and CVC. Some of these vertex forms
can be generated by dim=6 operators, while other ones require
higher $(dim=8,~ 10,~ 12)$ operators.
So a hierarchy is obtained among the various
possible off-shell effects. In each of the
$Z^*Z^*Z^*$, $Z^*Z^*\gamma^*$ and
$\gamma^*\gamma^*Z^*$ cases, this allows   us a simple
description in terms of a limited set of
constant parameters.
This should constitute a useful tool
for data analysis. For that
purpose we
have explicitly written the vertices with
both the complete set as well as with the set
restricted to the $dim=6$ operators. They are given in
eq.(\ref{ZZZ-CPc-Lan}, \ref{ZZZ-CPv-Lan},
\ref{ZZgamma-CPc-Lan}, \ref{ZZgamma-CPv-Lan},
\ref{gammagammaZ-CPc-Lan}, \ref{gammagammaZ-CPv-Lan}).\par

As an illustration of the SM and NP contributions,
we have considered the
neutral anomalous gauge couplings
generated by a fermionic triangle loop.
In Appendix C we have given the complete analytic expression of the
coupling functions arising at 1-loop, using general
gauge couplings to any fermion.
The use of this is twofold.
First, it allows to make an exact computation
of the SM  contribution.
And second, it provides an illustration of what type of
off-shell effects can appear for any kind of
NP fermion generating NAGC.

To this aim we have quantitatively discussed
through  Fig.\ref{fig3}-\ref{fig6},
the dependence of the neutral anomalous couplings
on the off-shell masses; as well as
the relative size of the new NAGC as
compared to those already existing in the on-shell case.
The $1/M^2_F$ limit of the heavy
fermion contribution appears to coincide with the
effective Lagrangian description restricted to the $dim=6$
operators. Thus, we have found that the effective Lagrangian
description is also valid, so long
the  fermion mass $M_F$ is not too close to $M_Z$ or
the energy threshold  $\sqrt{s}/2$ of the process considered.\par

We   emphasize though, that the 1-loop results
should   also be very useful in analyzing possible
 NAGC data   \underline{close to the threshold} for actually
producing the new
particles responsible for these NAGC.
In such a case  the effective Lagrangian formalism is not
applicable, and the  NAGC analysis must be done taking into
account the above 1-loop predictions; thus,
providing important complementary
information on the nature of the responsible NP particles.

Finally we have written the complete structure of the off-shell
$V_1^*V_2^*V_3^*$ contribution
to the $e^+e^- \to (f\bar f)+(f'\bar f')$ amplitude, which should be
used in the analysis of the events observable at present and
future $e^+e^-$ colliders.

As an overall conclusion we should stress
 that the off-shell effects in the neutral gauge boson
self-interactions cannot be ignored in
detail experiments like those performed at
LEP2,  and will be performed in the
future at  a Linear $e^-e^+$ Collider.
This is certainly related to the
fact that these couplings have to vanish whenever
all three gauge bosons participating in the vertex are
 on-shell.\\

{\bf Acknowledgments}:
It is a pleasure to thank Robert Sekulin
for discussions and suggestions.\\

\newpage

\renewcommand{\theequation}{A.\arabic{equation}}
\renewcommand{\thesection}{A.\arabic{section}}
\setcounter{equation}{0}
\setcounter{section}{0}

{\Large \bf Appendix A: The  CP-conserving
$V^*_1V^*_2V^*_3$ vertex.}\\

\begin{figure}[th]
\[
\epsfig{file=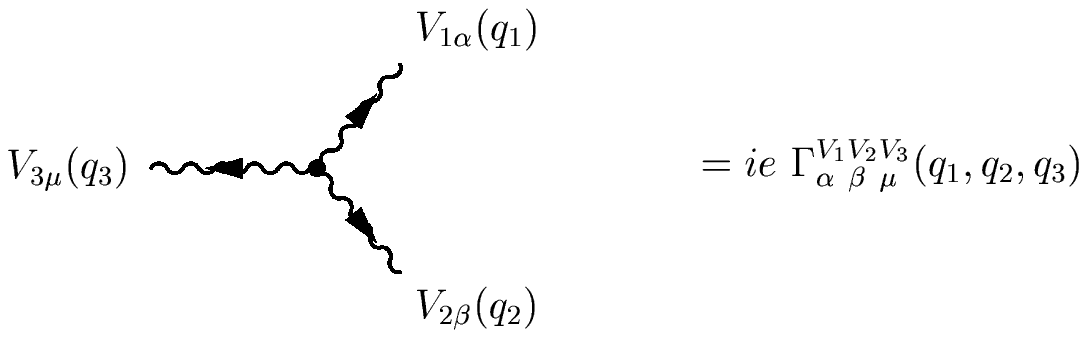,height=3cm,width=12cm}
\]
\vspace*{0.5cm}
\caption[1]{The general neutral gauge boson vertex
 $V_1V_2V_3$.}
\label{vvv-fig}
\end{figure}

The  general interaction among three,
possibly off-shell neutral gauge bosons (NAGC), is defined following
 the notation of Fig.\ref{vvv-fig} and  $s_i\equiv q^2_i$.
Note that all $q_i$ momenta are outgoing, so that
$q_1+q_2+q_3=0$. Since a vertex involving three
neutral gauge bosons is necessarily C-violating,
the construction of CP-conserving couplings
requires the use of   P-violating forms involving the
$\epsilon^{\mu\nu\rho\sigma}$ tensor, conveniently
denoted as
\bq
\epsilon^{\mu\nu\rho\sigma}A_{\mu}B_{\nu}C_{\rho}D_{\sigma}
=[ABCD] ~~ .
\eq
The most general Lorentz-invariant CP-conserving $V^*_1V^*_2V^*_3$
vertex involves at most six independent forms;
two of which are linear in the $q_i$ momenta,
while the rest are cubic. For an easy comparison with the forms
written in the on-shell case (\cite{Hagn, neut, modzg}) we
choose the  basis:
\bqa
&&[q_1-q_2~\mu~\alpha~\beta],~~~ [q_3~\mu~\alpha~\beta]~~ ,\nonumber\\
&&q^{\beta}_3~[q_1~q_2~\mu~\alpha]+q^{\alpha}_3~[q_1~q_2~\mu~\beta],
\nonumber\\
&&q^{\alpha}_1~[\beta~q_3~\mu~q_2] ~~,~~
q^{\beta}_2~[\alpha~q_3~\mu~q_1] ~~, ~~
q^{\mu}_3~[\beta~q_1~\alpha~q_2] ~~ ~~ .
\label{basisv1v2v3}
\eqa
The last three   forms in (\ref{basisv1v2v3}) imply  at least one
scalar $q_\mu V^\mu$ term and they are called  "scalar",
in contrast  to the other forms called "transverse".\\

\noindent
{\bf  The $Z^*Z^*Z^*$ case.}\\
\noindent
Here the additional constraint of full Bose symmetry among the
quantum numbers $(q_1,\alpha)$, $(q_2,\beta)$, $(q_3,\mu)$,
describing the three off-shell $Z^*$ should be imposed.
Writing thus
\bqa
\Gamma^{Z^*Z^*Z^*}_{\alpha\beta\mu}(q_1,q_2,q_3) &= &
 i  \sum_{j=1}^3  I^{Z^*Z^*Z^*,j}_{\alpha\beta\mu}
f^{Z^*Z^*Z^*}_j(s_1,s_2,s_3)~ \nonumber \\
& + &
i \sum_{j=1}^3  J^{Z^*Z^*Z^*,j}_{\alpha\beta \mu}
g^{Z^*Z^*Z^*}_j(s_1,s_2,s_3)~,
\label{ZZZ-CPc-ap}
\eqa
with
\bqa
&&I^{Z^*Z^*Z^*,1}_{\alpha\beta\mu}=[q_1-q_2~\mu~\alpha~\beta]~~, ~~
I^{Z^*Z^*Z^*,2}_{\alpha\beta\mu}=[q_3~\mu~\alpha~\beta]\nonumber\\
&&I^{Z^*Z^*Z^*,3}_{\alpha\beta\mu}= q^{\beta}_3~[q_1~q_2~\mu~\alpha]
+q^{\alpha}_3~[q_1~q_2~\mu~\beta] ~~ , \nonumber\\
&&J^{Z^*Z^*Z^*,1}_{\alpha\beta\mu}=q^{\alpha}_1~[\beta~q_3~\mu~q_2]~~,~~
J^{Z^*Z^*Z^*,2}_{\alpha\beta \mu}=q^{\beta}_2~[\alpha~q_3~\mu~q_1]~~,~~
\nonumber \\
&& J^{Z^*Z^*Z^*,3}_{\alpha\beta \mu}=q^{\mu}_3~[\beta ~q_1~\alpha~q_2]
~~ , \label{ZZZ-CPc-forms}
\eqa
we obtain that $f^{Z^*Z^*Z^*}_3(s_1,s_2,s_3)$ is a fully
\underline{antisymmetric} function of $(s_1,~s_2, ~s_3)$, while the other
transverse and scalar functions satisfy the Bose  relations
\bqa
f^{Z^*Z^*Z^*}_1(s_1,s_2,s_3)&=&f^{Z^*Z^*Z^*}_1(s_2,s_1,s_3)~~, ~~
f^{Z^*Z^*Z^*}_2(s_1,s_2,s_3)=-f^{Z^*Z^*Z^*}_2(s_2,s_1,s_3)~ ,\nonumber\\
f^{Z^*Z^*Z^*}_1(s_1,s_3,s_2)&=&{1\over2}\Big [-f^{Z^*Z^*Z^*}_1(s_1,s_2,s_3)
+f^{Z^*Z^*Z^*}_2(s_1,s_2,s_3) \nonumber \\
 &- &{s_2+s_1-s_3\over2}
f^{Z^*Z^*Z^*}_3(s_1,s_2,s_3)\Big ] ~ , \nonumber\\
f^{Z^*Z^*Z^*}_2(s_1,s_3,s_2)&=&{3\over2}f^{Z^*Z^*Z^*}_1(s_1,s_2,s_3)
+{1\over2}f^{Z^*Z^*Z^*}_2(s_1,s_2,s_3) \nonumber \\
&+ &{s_2-s_3-3s_1\over4} f^{Z^*Z^*Z^*}_3(s_1,s_2,s_3)~, \nonumber \\
g^{Z^*Z^*Z^*}_1(s_1,s_2,s_3)& = & g^{Z^*Z^*Z^*}_2(s_2,s_1,s_3)~~ , ~~
g^{Z^*Z^*Z^*}_3(s_1,s_2,s_3)=g^{Z^*Z^*Z^*}_3(s_2,s_1,s_3) ~,\nonumber\\
g^{Z^*Z^*Z^*}_1(s_1,s_2,s_3) &=&g^{Z^*Z^*Z^*}_1(s_1,s_3,s_2) +
 2 f^{Z^*Z^*Z^*}_3(s_1,s_3,s_2) ~, \nonumber \\
g^{Z^*Z^*Z^*}_2(s_1,s_2,s_3) &=&g^{Z^*Z^*Z^*}_3(s_1,s_3,s_2) -
  f^{Z^*Z^*Z^*}_3(s_1,s_3,s_2) ~, \nonumber \\
g^{Z^*Z^*Z^*}_3(s_1,s_2,s_3) &=&g^{Z^*Z^*Z^*}_2(s_1,s_3,s_2) -
  f^{Z^*Z^*Z^*}_3(s_1,s_3,s_2) ~ .
\label{ZZZ-CPc-Bose}
\eqa\\
Note that (\ref{ZZZ-CPc-Bose}) together with the antisymmetry of
 $f^{Z^*Z^*Z^*}_3(s_1,s_2,s_3)$  imply
\bq
g^{Z^*Z^*Z^*}_3(s_1,s_2,s_3)={1\over2}
\left [g^{Z^*Z^*Z^*}_1(s_3,s_2,s_1)
+g^{Z^*Z^*Z^*}_2(s_1,s_3,s_2)\right ]~.
\eq

\noindent
{\bf 2) The $Z^*Z^*\gamma^*$ case.}\\
\noindent
Restarting from the general $V^*_1V^*_2V^*_3$ vertex in
(\ref{basisv1v2v3}), with $(q_3,\mu)$ corresponding to the photon
and imposing the CVC constraint
$q^{\mu}_3~\Gamma^{Z^*Z^*\gamma^*}_{\alpha \beta\mu}(s_1,s_2,s_3)=0$
and Bose symmetry for $Z^*Z^*$, we end up with general vertex
containing  the five independent forms, namely
\bqa
\Gamma^{Z^*Z^*\gamma^*}_{\alpha\beta\mu}(q_1, q_2, q_3) &= & i
\sum_{j=1}^3 I^{Z^*Z^*\gamma^*,j}_{\alpha\beta\mu}
f^{Z^*Z^*\gamma^*}_j(s_1,s_2,s_3) \nonumber \\
&+& i\sum_{j=1,2}J^{Z^*Z^*\gamma^*,j}_{\alpha\beta\mu}
g^{Z^*Z^*\gamma^*}_j(s_1,s_2,s_3) ~ , \label{ZZgamma-CPc-ap}
\eqa
where
\bqa
I^{Z^*Z^*\gamma^*,1}_{\alpha\beta\mu} &= &
[q_1-q_2~\mu~\alpha~\beta]~+~{2q^{\mu}_3\over s_3}
[q_1~q_2~\alpha~\beta]~,\nonumber\\
I^{Z^*Z^*\gamma^*,2}_{\alpha\beta\mu}&=&[q_3~\mu~\alpha~\beta]
~,\nonumber\\
I^{Z^*Z^*\gamma^*,3}_{\alpha\beta\mu}&=&
q^{\beta}_3~[q_1~q_2~\mu~\alpha]
+q^{\alpha}_3~[q_1~q_2~\mu~\beta]~,\nonumber\\
J^{Z^*Z^*\gamma^*,1}_{\alpha\beta\mu}&=&
q^{\alpha}_1~[\beta~q_3~\mu~q_2]
~, \nonumber\\
J^{Z^*Z^*\gamma^*,2}_{\alpha\beta\mu}&=&
q^{\beta}_2~[\alpha~q_3~\mu~q_1]~. \label{ZZgamma-CPc-forms}
\eqa
\noindent
Bose symmetry imposes the constraints
\bqa
f^{Z^*Z^*\gamma^*}_1(s_1,s_2,s_3)=f^{Z^*Z^*\gamma^*}_1(s_2,s_1,s_3)
& , &
f^{Z^*Z^*\gamma^*}_2(s_1,s_2,s_3)=-f^{Z^*Z^*\gamma^*}_2(s_2,s_1,s_3)
~~~~~~~~~~ \nonumber\\
f^{Z^*Z^*\gamma^*}_3(s_1,s_2,s_3)=-f^{Z^*Z^*\gamma^*}_3(s_2,s_1,s_3)
&,&
g^{Z^*Z^*\gamma^*}_1(s_1,s_2,s_3)=g^{Z^*Z^*\gamma^*}_2(s_2,s_1,s_3)
~.~  \label{ZZgamma-CPc-Bose}
\eqa\\

\noindent
{\bf 3) The $\gamma^*\gamma^*Z^*$ case.}\\
\noindent
In the general $V^*_1V^*_2V^*_3$ vertex of
(\ref{basisv1v2v3}), $(q_3,\mu)$ corresponds now to  $Z^*$.
Imposing then the  two CVC constraints
$q^{\alpha}_1~
\Gamma^{\gamma^*\gamma^*Z^*}_{\alpha\beta\mu}(s_1,s_2,s_3)=
q^{\beta}_2~
\Gamma^{\gamma^*\gamma^*Z^*}_{\alpha\beta\mu}(s_1,s_2,s_3)=0$
and Bose symmetry for $\gamma^*~\gamma^*$, we end up with
the four independent vertex  forms
\bqa
\Gamma^{\gamma^*\gamma^*Z^*}_{\alpha\beta\mu}(q_1,q_2,q_3) &= &
i \sum_{j=1}^3 I^{\gamma^*\gamma^*Z^*,j}_{\alpha\beta\mu}
f^{\gamma^*\gamma^*Z^*}_j(s_1,s_2,s_3) \nonumber \\
&+& i  J^{\gamma^*\gamma^*Z^*,1}_{\alpha\beta\mu}
g^{\gamma^*\gamma^*Z^*}_1(s_1,s_2,s_3) ~~,
\label{gammagammaZ-CPc-ap}
\eqa
with
\bqa
I^{\gamma^*\gamma^*Z^*,1}_{\alpha\beta\mu}& = &
[q_1-q_2~\mu~\alpha~\beta]~-~{q^{\alpha}_1\over s_1}
[\beta~q_3~\mu~q_2]
~-~{q^{\beta}_2\over s_2}([\alpha~q_3~\mu~q_1])\nonumber\\
I^{\gamma^*\gamma^*Z^*,2}_{\alpha\beta\mu}&=&
([q_3~\mu~\alpha~\beta]
~-~{q^{\alpha}_1\over s_1}
[\beta~q_3~\mu~q_2]
~+~{q^{\beta}_2\over s_2}([\alpha~q_3~\mu~q_1])
\nonumber\\
I^{\gamma^*\gamma^*Z^*,3}_{\alpha\beta\mu} &= &
q^{\beta}_3~[q_1~q_2~\mu~\alpha]
+q^{\alpha}_3~[q_1~q_2~\mu~\beta]
+{s_2-s_1-s_3\over2s_1}q^{\alpha}_1~[\beta~q_3~\mu~q_2]
\nonumber \\
&- &{s_1-s_2-s_3\over2s_2}q^{\beta}_2~[\alpha~q_3~\mu~q_1]
\nonumber\\
J^{\gamma^*\gamma^*Z^*,1}_{\alpha\beta\mu}&=& q^{\mu}_3~
[\beta~q_1~\alpha~q_2] ~, \label{gammagammaZ-CPc-forms}
\eqa
\noindent
and the Bose symmetry constraints
\bqa
f^{\gamma^*\gamma^*Z^*}_1(s_1,s_2,s_3)
=f^{\gamma^*\gamma^*Z^*}_1(s_2,s_1,s_3)
& , &
f^{\gamma^*\gamma^*Z^*}_2(s_1,s_2,s_3)
=-f^{\gamma^*\gamma^*Z^*}_2(s_2,s_1,s_3)
\nonumber\\
f^{\gamma^*\gamma^*Z^*}_3(s_1,s_2,s_3)
=-f^{\gamma^*\gamma^*Z^*}_3(s_2,s_1,s_3)
& ,&
g^{\gamma^*\gamma^*Z^*}_1(s_1,s_2,s_3)
=g^{\gamma^*\gamma^*Z^*}_1(s_2,s_1,s_3)~.\nonumber\\
&&\label{gammagammaZ-CPc-Bose}
\eqa

\newpage
\renewcommand{\theequation}{B.\arabic{equation}}
\renewcommand{\thesection}{B.\arabic{section}}
\setcounter{equation}{0}
\setcounter{section}{0}

{\large \bf Appendix B:  The  CP-violating forms for
the $V^*_1V^*_2V^*_3$ vertex. }\\

These  vertices are P-conserving and  C-violating,
and can most generally be expressed in terms
of the following 14 independent Lorentz invariant forms:
(Indices $i,~,j,~k$ run from 1 to 3.)
\begin{itemize}
\item
3 terms like  $(V_i.V_j)(V_k.(q_i-q_j))$,
\item
3 terms like  $(V_i.V_j)(V_k.q_k)$,
\item
8 terms like $[V_k.(q_i-q_j)~ \mbox{or} ~V_k.q_k] \cdot
[V_j.(q_k-q_i)~ \mbox{or} ~V_j.q_j] \cdot
[V_i.(q_j-q_k) ~\mbox{or} ~V_i.q_i]$.
\end{itemize}
Four of these terms are "transverse", while the other 10  contain
at least one "scalar" $q.V$ coefficient.

\noindent
{\bf 1) The $Z^*Z^*Z^*$ case.}\\
  Applying  full Bose symmetry among the three $Z^*$, we obtain the
structure
\bqa
\Gamma^{Z^*Z^*Z^*}_{\alpha\beta\mu}(q_1, q_2, q_3) &= & i\sum_{j=1}^4
\tilde{I}^{Z^*Z^*Z^*,j}_{\alpha\beta\mu}
\tilde{f}^{Z^*Z^*Z^*}_j(s_1,s_2,s_3) \nonumber \\
&+& i \sum_{j=1}^{10} \tilde{J}^{Z^*Z^*Z^*,j}_{\alpha\beta\mu}
\tilde{g}^{Z^*Z^*Z^*}_j(s_1,s_2,s_3) ~ , \label{ZZZ-CPv-ap}
\eqa
where the transverse forms are
\bqa
\tilde{I}^{Z^*Z^*Z^*,1}_{\alpha\beta\mu}=
g^{\alpha\beta}(q_1-q_2)^{\mu} & , &
\tilde{I}^{Z^*Z^*Z^*,2}_{\alpha\beta\mu}=
g^{\beta\mu}(q_3-q_2)^{\alpha} ~ , \nonumber \\
\tilde{I}^{Z^*Z^*Z^*,3}_{\alpha\beta\mu} =
g^{\alpha\mu}(q_1-q_3)^{\beta}& ,
& \tilde{I}^{Z^*Z^*Z^*,4}_{\alpha\beta\mu}=
(q_2-q_3)^{\alpha}(q_1-q_3)^{\beta}(q_1-q_2)^{\mu} ~,
\label{ZZZ-CPv-forms-t}
\eqa
\noindent
while the scalar ones are
\bqa
&&\tilde{J}^{Z^*Z^*Z^*,1}_{\alpha\beta\mu}=g^{\alpha\beta}q_3^{\mu}~~,~~
\tilde{J}^{Z^*Z^*Z^*,2}_{\alpha\beta\mu}=g^{\beta\mu}q_1^{\alpha}~~,~~
\tilde{J}^{Z^*Z^*Z^*,3}_{\alpha\beta\mu}=
g^{\alpha\mu}q_2^{\beta}\nonumber\\
&&\tilde{J}^{Z^*Z^*Z^*,4}_{\alpha\beta\mu}=
q_1^{\alpha}(q_1-q_3)^{\beta}(q_1-q_2)^{\mu}~~,~~
\tilde{J}^{Z^*Z^*Z^*,5}_{\alpha\beta\mu}=
q_2^{\beta}(q_2-q_3)^{\alpha}(q_2-q_1)^{\mu}\nonumber\\
&&
\tilde{J}^{Z^*Z^*Z^*,6}_{\alpha\beta\mu}=
q_3^{\mu}(q_3-q_1)^{\beta}(q_3-q_2)^{\alpha} ~~ , ~~
\tilde{J}^{Z^*Z^*Z^*,7}_{\alpha\beta\mu}=
q_1^{\alpha}q_2^{\beta}(q_1-q_2)^{\mu}
\nonumber\\
&& \tilde{J}^{Z^*Z^*Z^*,8}_{\alpha\beta\mu}=
q_3^{\mu}q_2^{\beta}(q_3-q_2)^{\alpha}
~~, ~~\tilde{J}^{Z^*Z^*Z^*,9}_{\alpha\beta\mu}=
q_1^{\alpha}q_3^{\mu}(q_1-q_3)^{\beta}
\nonumber\\
&&\tilde{J}^{Z^*Z^*Z^*,10}_{\alpha\beta\mu}=
q_1^{\alpha}q_2^{\beta}q_3^{\mu} ~~. \label{ZZZ-CPv-forms-s}
\eqa
\noindent
The  Bose relations obtained from them for the transverse forms
are
\bqa
&&\tilde{f}^{Z^*Z^*Z^*}_1(s_1,s_2,s_3)
 =-\tilde{f}^{Z^*Z^*Z^*}_1(s_2,s_1,s_3)
 =\tilde{f}^{Z^*Z^*Z^*}_2(s_3,s_2,s_1) ~~~~
\nonumber \\
&&=-\tilde{f}^{Z^*Z^*Z^*}_2(s_3,s_1,s_2)
 =\tilde{f}^{Z^*Z^*Z^*}_3(s_1,s_3,s_2)
=-\tilde{f}^{Z^*Z^*Z^*}_3(s_2,s_3,s_1) ~~~~,
\nonumber \\
&&\tilde{f}^{Z^*Z^*Z^*}_4(s_1,s_2,s_3)
=-\tilde{f}^{Z^*Z^*Z^*}_4(s_2,s_1,s_3) \nonumber \\
&& =-\tilde{f}^{Z^*Z^*Z^*}_4(s_1,s_3,s_2)
=\tilde{f}^{Z^*Z^*Z^*}_4(s_3,s_2,s_1) ~,
\label{ZZZ-CPv-Bose-t}
\eqa
while for the scalar ones we get
\bqa
\tilde{g}^{Z^*Z^*Z^*}_1(s_1,s_2,s_3)
&=\tilde{g}^{Z^*Z^*Z^*}_1(s_2,s_1,s_3)
&=\tilde{g}^{Z^*Z^*Z^*}_2(s_3,s_2,s_1) \nonumber \\
 =\tilde{g}^{Z^*Z^*Z^*}_2(s_3,s_1,s_2)
& =\tilde{g}^{Z^*Z^*Z^*}_3(s_1,s_3,s_2)
& =\tilde{g}^{Z^*Z^*Z^*}_3(s_2,s_3,s_1) ~ , \nonumber \\
\tilde{g}^{Z^*Z^*Z^*}_{6}(s_1,s_2,s_3)
&=\tilde{g}^{Z^*Z^*Z^*}_{6}(s_2,s_1,s_3)
& =\tilde{g}^{Z^*Z^*Z^*}_4(s_3,s_2,s_1) \nonumber \\
=\tilde{g}^{Z^*Z^*Z^*}_4(s_3,s_1,s_2)
 &=\tilde{g}^{Z^*Z^*Z^*}_5(s_1,s_3,s_2)
& =\tilde{g}^{Z^*Z^*Z^*}_5(s_2,s_3,s_1) ~ , \nonumber \\
\tilde{g}^{Z^*Z^*Z^*}_{7}(s_1,s_2,s_3)
& =-\tilde{g}^{Z^*Z^*Z^*}_{7}(s_2,s_1,s_3)
& =\tilde{g}^{Z^*Z^*Z^*}_{8}(s_3,s_2,s_1) \nonumber \\
=-\tilde{g}^{Z^*Z^*Z^*}_{8}(s_3,s_1,s_2)
 &=\tilde{g}^{Z^*Z^*Z^*}_{9}(s_1,s_3,s_2)
& =-\tilde{g}^{Z^*Z^*Z^*}_{9}(s_2,s_3,s_1) ~ , \nonumber
\eqa
\bq
\tilde{g}^{Z^*Z^*Z^*}_{10}(s_1,s_2,s_3)
=\tilde{g}^{Z^*Z^*Z^*}_{10}(s_2,s_1,s_3)
 =\tilde{g}^{Z^*Z^*Z^*}_{10}(s_3,s_2,s_1)
 =\tilde{g}^{Z^*Z^*Z^*}_{10}(s_1,s_3,s_2) ~.
\label{ZZZ-CPv-Bose-s}
\eq \\

\noindent
{\bf 2) The $Z^*Z^*\gamma^*$ case}\\
Restarting from the initial list of CP-violating  $V^*_1V^*_2V^*_3$
forms, with $(q_3,\mu)$ corresponding to the photon,
and imposing the CVC constraint
$q^{\mu}_3~\Gamma^{Z^*Z^*\gamma^*}_{\alpha\beta\mu}(s_1,s_2,s_3)=0$
and Bose symmetry for $Z^*Z^*$, we get
\bqa
\Gamma^{Z^*Z^*\gamma^*}_{\alpha\beta\mu}(q_1, q_2, q_3)
& =&i \sum_{j=1}^4
\tilde{I}^{Z^*Z^*\gamma^*,j}_{\alpha\beta\mu}
\tilde{f}^{Z^*Z^*\gamma^*}_j(s_1,s_2,s_3) \nonumber \\
& + &i \sum_{j=1}^5 \tilde{J}^{Z^*Z^*\gamma^*,j}_{\alpha\beta\mu}
\tilde{g}^{Z^*Z^*\gamma^*}_j(s_1,s_2,s_3) ~ ~,
\label{ZZgamma-CPv-ap}
\eqa
involving four transverse and five scalar forms. These are
\bqa
\tilde{I}^{Z^*Z^*\gamma^*,1}_{\alpha\beta\mu}&=&
g^{\alpha\beta}\left ((q_1-q_2)^{\mu}-{(s_2-s_1)
\over s_3}q_3^\mu \right ) ~ , \nonumber\\
\tilde{I}^{Z^*Z^*\gamma^*,2}_{\alpha\beta\mu} &= &
g^{\mu\beta}(q_3-q_2)^{\alpha}+g^{\mu\alpha}(q_3-q_1)^{\beta}
-{q_3^{\mu}(q_3-q_1)^{\beta}(q_3-q_2)^{\alpha}\over s_3}\nonumber\\
&+ & {q_3^{\mu}\over 2 s_3}  [ q_2^{\beta}(q_3-q_2)^{\alpha}
+q_1^{\alpha}(q_3-q_1)^{\beta}  ] ~ , \nonumber\\
\tilde{I}^{Z^*Z^*\gamma^*,3}_{\alpha\beta\mu} &= &
g^{\mu\beta}(q_3-q_2)^{\alpha}-g^{\mu\alpha}(q_3-q_1)^{\beta}
+{q_3^{\mu}\over2 s_3} [q_2^{\beta}(q_3-q_2)^{\alpha}
-q_1^{\alpha}(q_3-q_1)^{\beta}] ~ , \nonumber\\
\tilde{I}^{Z^*Z^*\gamma^*,4}_{\alpha\beta\mu} &= &
(q_2-q_3)^{\alpha}(q_1-q_3)^{\beta}(q_1-q_2)^{\mu}
-{s_2-s_1\over s_3}q_3^{\mu}(q_3-q_1)^{\beta}(q_3-q_2)^{\alpha} ~,
\nonumber \\
\tilde{J}^{Z^*Z^*\gamma^*,1}_{\alpha\beta\mu} &= &
g^{\mu\beta}q_1^{\alpha}+g^{\mu\alpha}q_2^{\beta}
-{q_3^{\mu}\over2 s_3}[q_1^{\alpha}(q_3-q_1)^{\beta}
+q_2^{\beta}(q_3-q_2)^{\beta}]
+{q_1^{\alpha}q_2^{\beta}q_3^{\mu}\over s_3} ~, \nonumber\\
\tilde{J}^{Z^*Z^*\gamma^*,2}_{\alpha\beta\mu} &= &
g^{\mu\beta}q_1^{\alpha}-g^{\mu\alpha}q_2^{\beta}
-{q_3^{\mu}\over2 s_3}[q_1^{\alpha}(q_3-q_1)^{\beta}
-q_2^{\beta}(q_3-q_2)^{\alpha}] ~, \nonumber\\
\tilde{J}^{Z^*Z^*\gamma^*,3}_{\alpha\beta\mu}& = &
q_1^{\alpha}(q_1-q_3)^{\beta}(q_1-q_2)^{\mu}
+q_2^{\beta}(q_2-q_3)^{\alpha}(q_2-q_1)^{\mu}\nonumber\\
&+& {s_2-s_1\over s_3}q_3^{\mu} [q_1^{\alpha}(q_3-q_1)^{\beta}
-q_2^{\beta}(q_3-q_2)^{\alpha}] ~, \nonumber\\
\tilde{J}^{Z^*Z^*\gamma^*,4}_{\alpha\beta\mu} &= &
q_1^{\alpha}(q_1-q_3)^{\beta}(q_1-q_2)^{\mu}
-q_2^{\beta}(q_2-q_3)^{\alpha}(q_2-q_1)^{\mu}\nonumber\\
& + &{s_2-s_1\over s_3}q_3^{\mu}[q_1^{\alpha}(q_3-q_1)^{\beta}
+q_2^{\beta}(q_3-q_2)^{\alpha}] ~, \nonumber\\
\tilde{J}^{Z^*Z^*\gamma^*,5}_{\alpha\beta\mu} &= &
q_1^{\alpha}q_2^{\beta}(q_1-q_2)^{\mu}
-{s_2-s_1\over s_3}q_1^{\alpha}q_2^{\beta}q_3^{\mu} ~ ,
\label{ZZgamma-CPv-forms}
\eqa
implying the  Bose relations
\bqa
\tilde{f}^{Z^*Z^*\gamma^*}_1(s_1,s_2,s_3)
=-\tilde{f}^{Z^*Z^*\gamma^*}_1(s_2,s_1,s_3) &, &
\tilde{f}^{Z^*Z^*\gamma^*}_2(s_1,s_2,s_3)
=\tilde{f}^{Z^*Z^*\gamma^*}_2(s_2,s_1,s_3) ~ , \nonumber \\
\tilde{f}^{Z^*Z^*\gamma^*}_3(s_1,s_2,s_3)
=-\tilde{f}^{Z^*Z^*\gamma^*}_3(s_2,s_1,s_3) &,&
\tilde{f}^{Z^*Z^*\gamma^*}_4(s_1,s_2,s_3)
=-\tilde{f}^{Z^*Z^*\gamma^*}_4(s_2,s_1,s_3) ~ , \nonumber \\
\tilde{g}^{Z^*Z^*\gamma^*}_1(s_1,s_2,s_3)
=\tilde{g}^{Z^*Z^*\gamma^*}_1(s_2,s_1,s_3) & , &
\tilde{g}^{Z^*Z^*\gamma^*}_2(s_1,s_2,s_3)
=-\tilde{g}^{Z^*Z^*\gamma^*}_2(s_2,s_1,s_3) ~, \nonumber \\
\tilde{g}^{Z^*Z^*\gamma^*}_3(s_1,s_2,s_3)
=\tilde{g}^{Z^*Z^*\gamma^*}_3(s_2,s_1,s_3) & ,&
\tilde{g}^{Z^*Z^*\gamma^*}_4(s_1,s_2,s_3)
=-\tilde{g}^{Z^*Z^*\gamma^*}_4(s_2,s_1,s_3) ~, \nonumber \\
\tilde{g}^{Z^*Z^*\gamma^*}_5(s_1,s_2,s_3)
=-\tilde{g}^{Z^*Z^*\gamma^*}_5(s_2,s_1,s_3) & . &
\label{ZZgamma-CPv-Bose}
\eqa \\

\noindent
{\bf 3) The $\gamma^*\gamma^*Z^*$ case}\\
Imposing on the general $V^*_1V^*_2V^*_3$ vertex
the two CVC constraints and Bose symmetry
for the two photons
leaves 6 invariant forms,
\bqa
\Gamma^{\gamma^*\gamma^*Z^*}_{\alpha\beta\mu}(q_1, q_2, q_3)
& = &i \sum_{i=1}^ 4
\tilde{I}^{\gamma^*\gamma^*Z^*,i}_{\alpha\beta\mu}
\tilde{f}^{\gamma^*\gamma^*Z^*}_i(s_1,s_2,s_3) \nonumber \\
&+ & i \sum_{i=1,2}\tilde{J}^{\gamma^*\gamma^*Z^*,i}_{\alpha\beta\mu}
\tilde{g}^{\gamma^*\gamma^*Z^*}_i(s_1,s_2,s_3)~,
\label{gammagammaZ-CPv-ap}
\eqa
\noindent
where $(q_3,\mu)$ correspond to  $Z^*$ and
\bqa
&& \tilde{I}^{\gamma^*\gamma^*Z^*,1}_{\alpha\beta\mu} =
g^{\alpha\beta}(q_1-q_2)^{\mu}
-{q_1^{\alpha}\over2s_1}(q_1-q_3)^{\beta}(q_1-q_2)^{\mu}\nonumber\\
&& +{q_2^{\beta}\over2s_2}(q_2-q_3)^{\alpha}(q_2-q_1)^{\mu}
 + {s_3\over2s_1s_2}q_1^{\alpha}q_2^{\beta}(q_1-q_2)^{\mu} ~, \nonumber\\
&& \tilde{I}^{\gamma^*\gamma^*Z^*,2}_{\alpha\beta\mu} =
g^{\mu\beta}(q_3-q_2)^{\alpha}+g^{\mu\alpha}(q_3-q_1)^{\beta}
-{s_2-s_3\over s_1}q_1^{\alpha}g^{\mu\beta}
-{s_1-s_3\over s_2}q_2^{\beta}g^{\mu\alpha}\nonumber\\
&& + {q_2^{\beta}\over2 s_2}(q_2-q_3)^{\alpha}(q_2-q_1)^{\mu}
+{q_1^{\alpha}\over2 s_1}(q_1-q_3)^{\beta}(q_1-q_2)^{\mu}
+{s_1-s_2\over 2s_1s_2}q_1^{\alpha}q_2^{\beta}(q_1-q_2)^{\mu}
\nonumber\\ &&
+{q_3^{\mu}\over2 s_2}q_2^{\beta}(q_3-q_2)^{\alpha}
+{q_3^{\mu}\over2 s_1}q_1^{\alpha}(q_3-q_1)^{\beta}
+{2s_3-s_1-s_2\over2s_1s_2}q_1^{\alpha}q_2^{\beta}q_3^{\mu}
~ , \nonumber\\
&&\tilde{I}^{\gamma^*\gamma^*Z^*,3}_{\alpha\beta\mu}=
g^{\mu\beta}(q_3-q_2)^{\alpha}-g^{\mu\alpha}(q_3-q_1)^{\beta}
-{s_2-s_3\over s_1}q_1^{\alpha}g^{\mu\beta}
+{s_1-s_3\over s_2}q_2^{\beta}g^{\mu\alpha}\nonumber\\
&&
+{q_2^{\beta}\over2 s_2}(q_2-q_3)^{\alpha}(q_2-q_1)^{\mu}
-{q_1^{\alpha}\over2 s_1}(q_1-q_3)^{\beta}(q_1-q_2)^{\mu}
+{s_1-s_2\over 2s_1s_2}q_1^{\alpha}q_2^{\beta}q_3^{\mu}\nonumber\\
&&
+{q_3^{\mu}\over2 s_2}q_2^{\beta}(q_3-q_2)^{\alpha}
-{q_3^{\mu}\over2 s_1}q_1^{\alpha}(q_3-q_1)^{\beta}
+{2s_3-s_1-s_2\over2s_1s_2}q_1^{\alpha}q_2^{\beta}(q_1-q_2)^{\mu}
~ , \nonumber\\
&&\tilde{I}^{\gamma^*\gamma^*Z^*,4}_{\alpha\beta\mu}=
(q_2-q_3)^{\alpha}(q_1-q_3)^{\beta}(q_1-q_2)^{\mu}
-{s_3-s_2\over s_1}q_1^{\alpha}(q_1-q_3)^{\beta}(q_1-q_2)^{\mu}
\nonumber\\
&&+{s_3-s_1\over s_2}q_2^{\beta}(q_2-q_3)^{\alpha}(q_2-q_1)^{\mu}
+{(s_3-s_2)(s_3-s_1)\over s_1s_2}q_1^{\alpha}q_2^{\beta}(q_1-q_2)^{\mu}
~ , \nonumber\\
&&\tilde{J}^{\gamma^*\gamma^*Z^*,1}_{\alpha\beta\mu}=
g^{\alpha\beta}q_3^{\mu}
+{q_3^{\mu}\over2 s_2}q_2^{\beta}(q_3-q_2)^{\alpha}
+{q_3^{\mu}\over2 s_1}q_1^{\alpha}(q_3-q_1)^{\beta}
+{s_3\over2s_1s_2}q_1^{\alpha}q_2^{\beta}q_3^{\mu}
~ , \nonumber\\
&&\tilde{J}^{\gamma^*\gamma^*Z^*,2}_{\alpha\beta\mu}=
q_3^{\mu}(q_3-q_1)^{\beta}(q_3-q_2)^{\alpha}
-{s_1-s_3\over s_2}q_3^{\mu}q_2^{\beta}(q_3-q_2)^{\alpha}\nonumber\\
&&
-~ {s_2-s_3\over s_1}q_3^{\mu}q_1^{\alpha}(q_3-q_1)^{\beta}
+{(s_2-s_3)(s_1-s_3)\over s_1s_2}q_1^{\alpha}q_2^{\beta}q_3^{\mu}
 ~ , \label{gammagammaZ-CPv-forms}
\eqa
with the Bose relations
\bqa
\tilde{f}^{\gamma^*\gamma^*Z^*}_1(s_1,s_2,s_3)
=-\tilde{f}^{\gamma^*\gamma^*Z^*}_1(s_2,s_1,s_3) & ,&
\tilde{f}^{\gamma^*\gamma^*Z^*}_2(s_1,s_2,s_3)
=\tilde{f}^{\gamma^*\gamma^*Z^*}_2(s_2,s_1,s_3)
~,~~~~~~~~~~ \nonumber \\
\tilde{f}^{\gamma^*\gamma^*Z^*}_3(s_1,s_2,s_3)
=-\tilde{f}^{\gamma^*\gamma^*Z^*}_3(s_2,s_1,s_3) & ,&
\tilde{f}^{\gamma^*\gamma^*Z^*}_4(s_1,s_2,s_3)
=-\tilde{f}^{\gamma^*\gamma^*Z^*}_4(s_2,s_1,s_3)
~,~~~~~~~~~~ \nonumber \\
\tilde{g}^{\gamma^*\gamma^*Z^*}_1(s_1,s_2,s_3)
=\tilde{g}^{\gamma^*\gamma^*Z^*}_1(s_2,s_1,s_3) & ,&
\tilde{g}^{\gamma^*\gamma^*Z^*}_2(s_1,s_2,s_3)
=\tilde{g}^{\gamma^*\gamma^*Z^*}_2(s_2,s_1,s_3)
~. \label{gammagammaZ-CPv-Bose}
\eqa

\newpage
\renewcommand{\theequation}{C.\arabic{equation}}
\renewcommand{\thesection}{C.\arabic{section}}
\setcounter{equation}{0}
\setcounter{section}{0}

{\large \bf Appendix C:  Fermionic triangle 1-loop contributions
to the off-shell $V^*_1V^*_2V^*_3$ couplings}\\

\begin{figure}[hb]
\vspace*{2cm}
\[
\epsfig{file=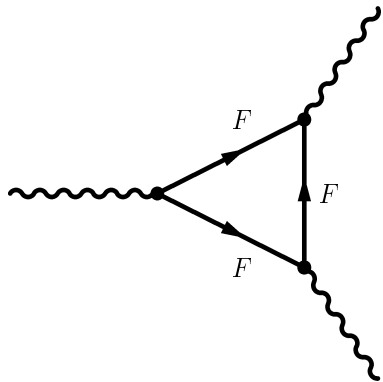,height=5cm,width=5cm}
\]
\vspace*{0cm}
\caption[2]{The fermionic triangle.}
\label{trif-fig}
\end{figure}

\begin{figure}[p]
\vspace*{-3cm}
\[
\epsfig{file=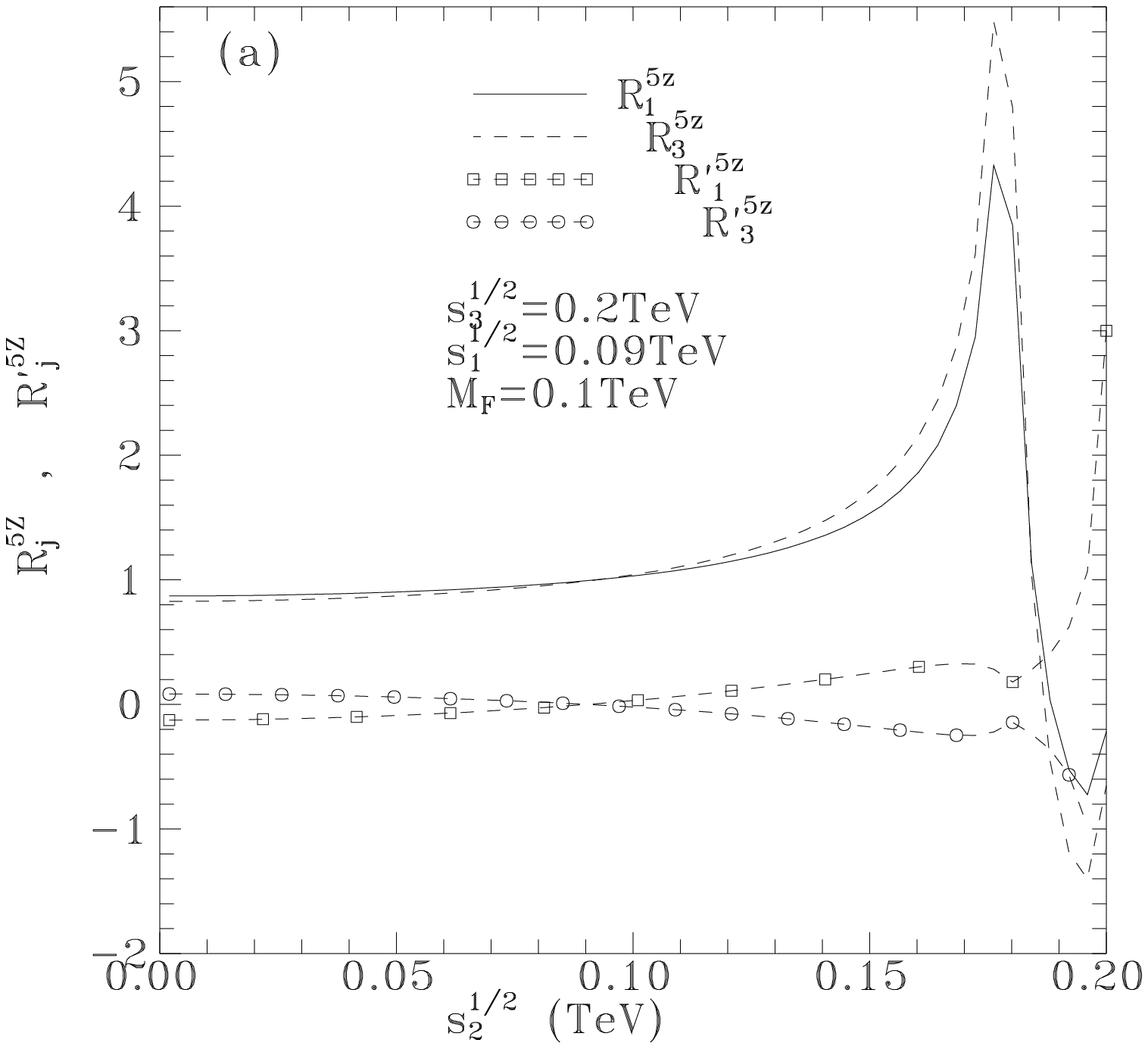,height=7.5cm}\hspace{0.5cm}
\epsfig{file=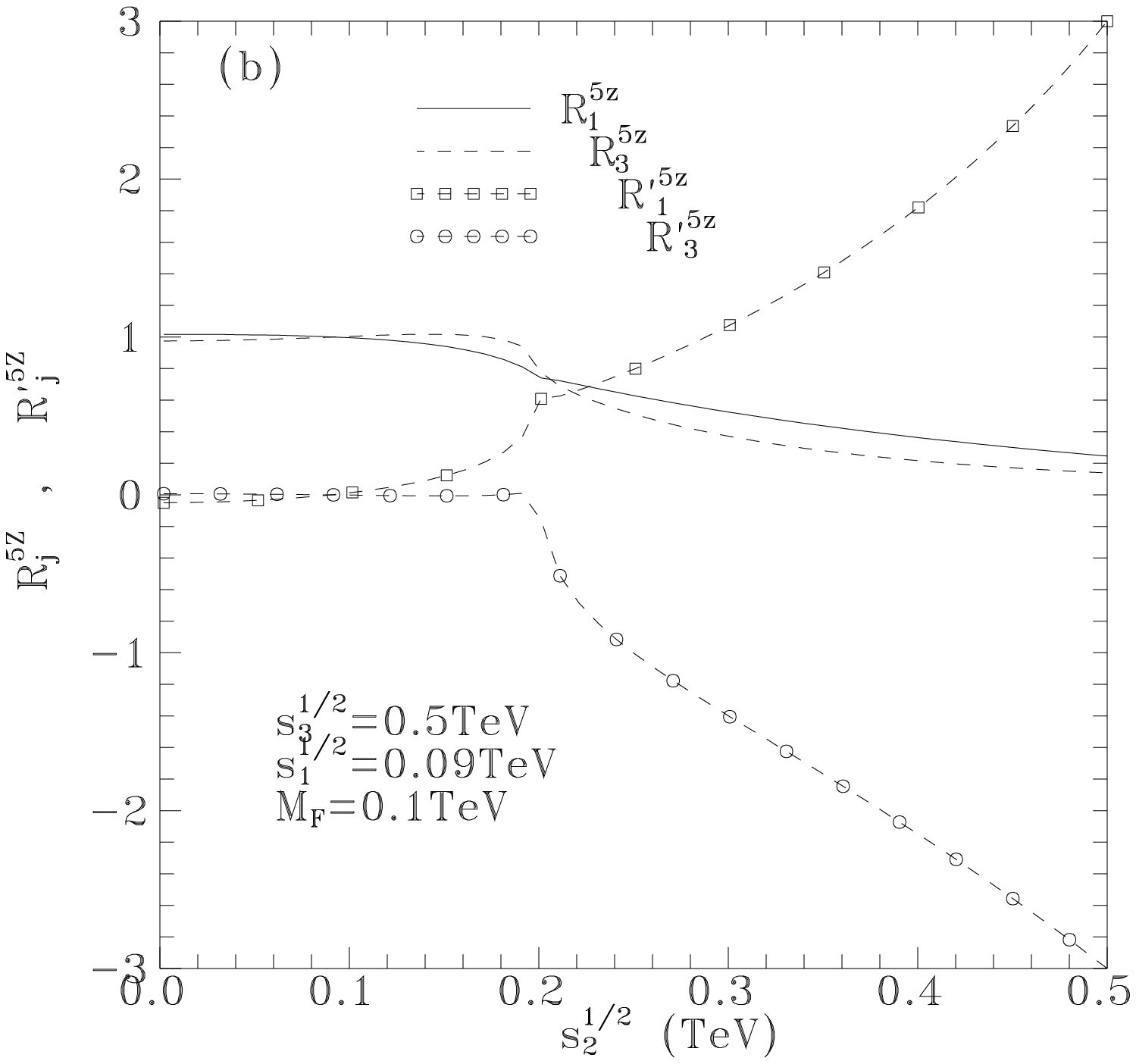,height=7.5cm}
\]
\vspace*{0.5cm}
\[
\epsfig{file=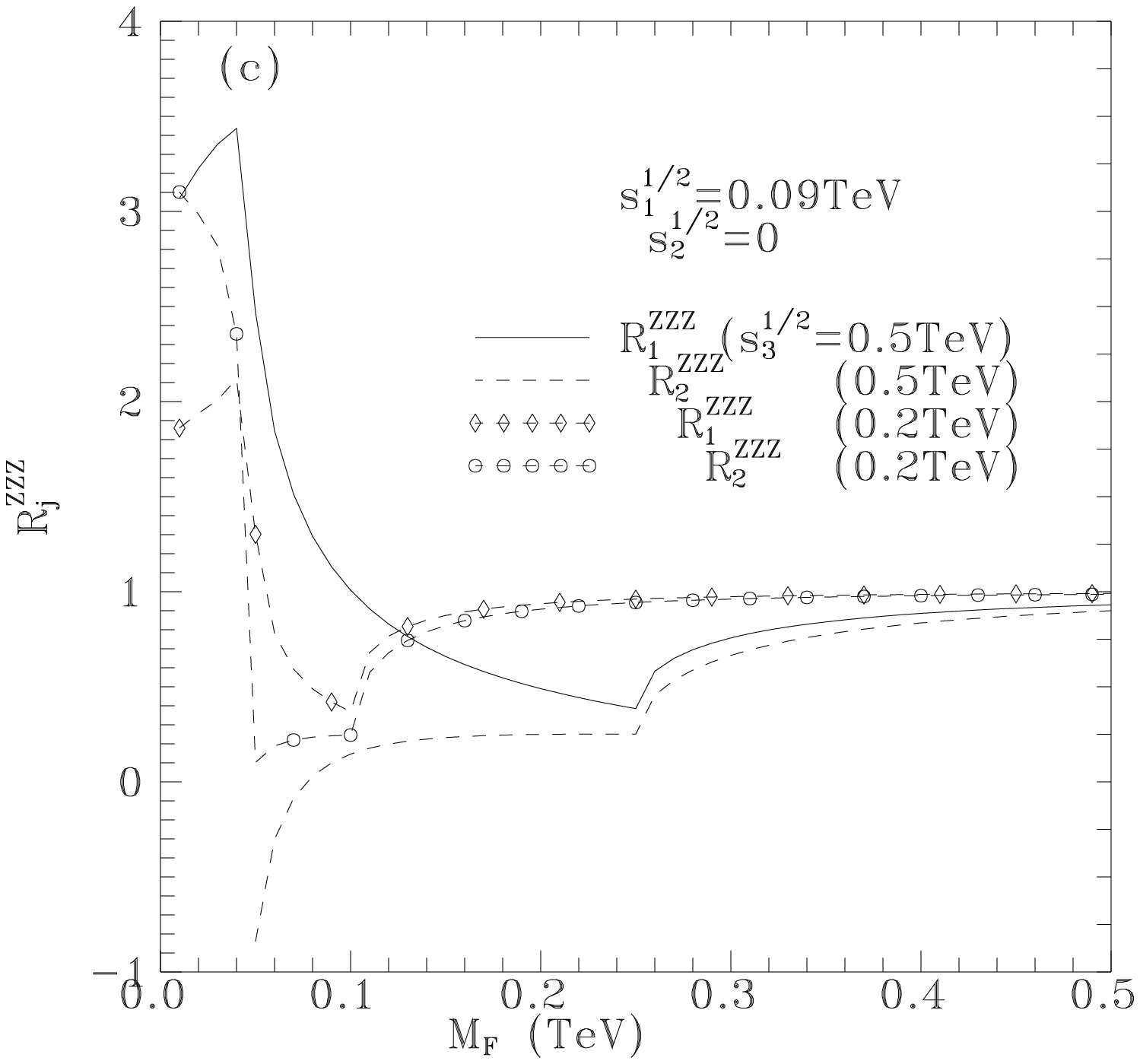,height=7.5cm,width=8.5cm}
\]
\vspace*{0.5cm}
\caption[1]{$Z^*Z^*Z^*$ off-shell effects compared to
$Z^*\to ZZ$:
Ratios $R^{5Z}_1$ and $R^{5Z}_3$ show
the $\sqrt{s_2}$-dependence of the contributions to the
$f^Z_5$-type of coupling; ratios
$R'^{5Z}_1$ and $R'^{5Z}_3$ give the relative size, versus $\sqrt{s_2}$,
of the new contributions as compared to
the ones already existing on-shell; (a) at
$\sqrt{s_3}=0.2~TeV$, (b) at $\sqrt{s_3}=0.5~TeV$.
Ratios $R^{ZZZ}_1$ and $R^{ZZZ}_2$ show the departure versus $M_F$
of the exact 1-loop contribution, as compared to
the effective Lagrangian prediction at $\sqrt{s_3}=0.2~TeV$ and
$0.5~TeV$, (c). The definitions of $s_1, s_2, s_3$ are given in
the text.}
\label{fig3}
\end{figure}

\begin{figure}[p]
\vspace*{-4cm}
\[
\epsfig{file=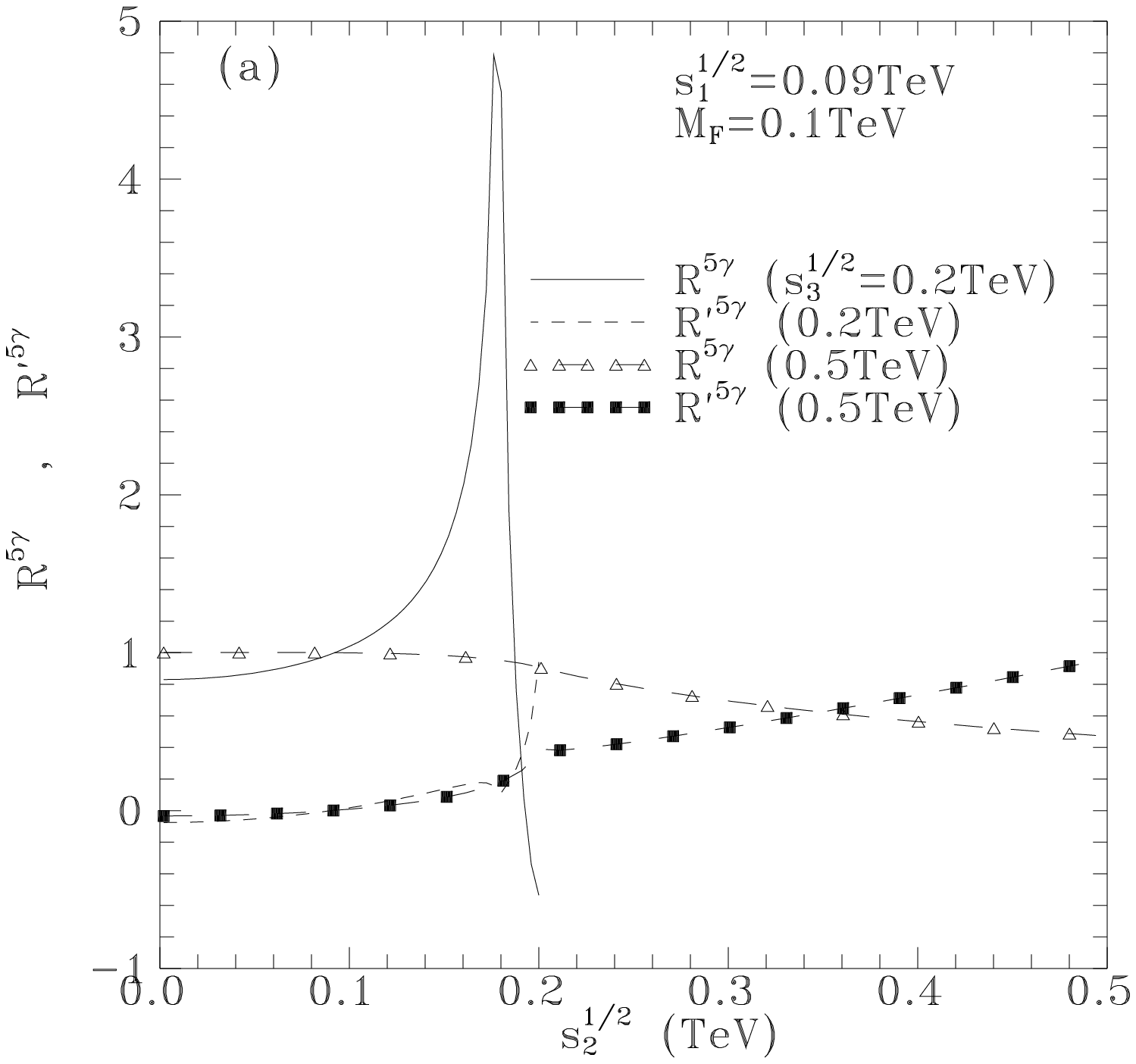,height=7.5cm}\hspace{0.5cm}
\epsfig{file=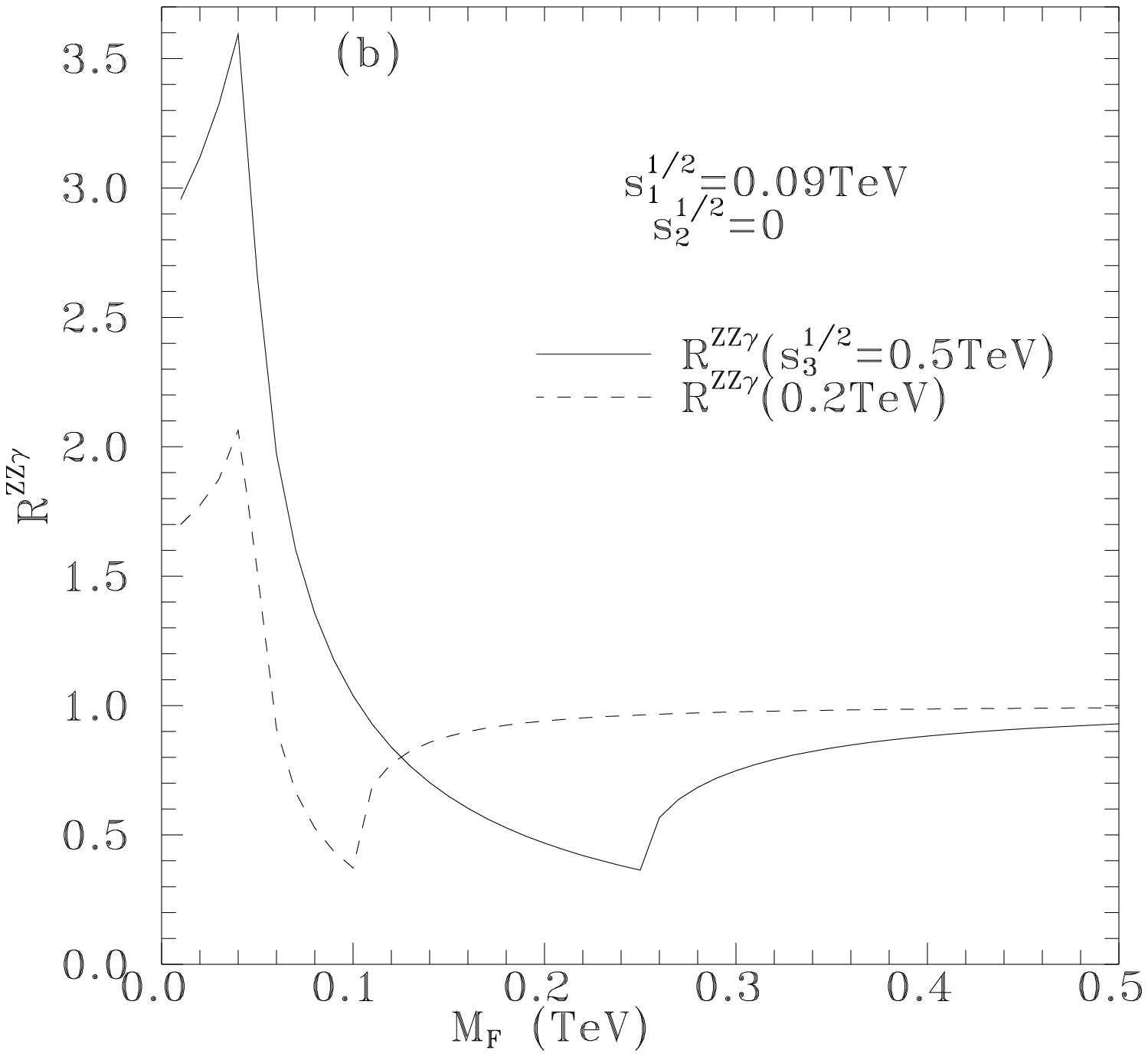,height=7.5cm}
\]
\vspace*{0.5cm}
\caption[1]{$Z^*Z^*\gamma^*$ off-shell effects compared to
$\gamma^*\to ZZ$:
Ratio $R^{5\gamma}$  show
the $\sqrt{s_2}$-dependence of the contributions to the
$f^{\gamma}_5$-type of coupling,
while  $R'^{5\gamma}$ gives the relative size, versus $\sqrt{s_2}$, of
the new contributions as compared to
the ones already existing on-shell; at
$\sqrt{s_3}=0.2~TeV$ and at $\sqrt{s_3}=0.5~TeV$, (a).
Ratios $R^{ZZ\gamma}$ show the departure versus $M_F$
of the exact 1-loop contribution, as compared to
the effective Lagrangian prediction at $\sqrt{s_3}=0.2~TeV$ and
$0.5~TeV$, (b). The definitions of $s_1, s_2, s_3$ are given in
the text.}
\label{fig4}
\end{figure}

\begin{figure}[p]
\vspace*{-3cm}
\[
\epsfig{file=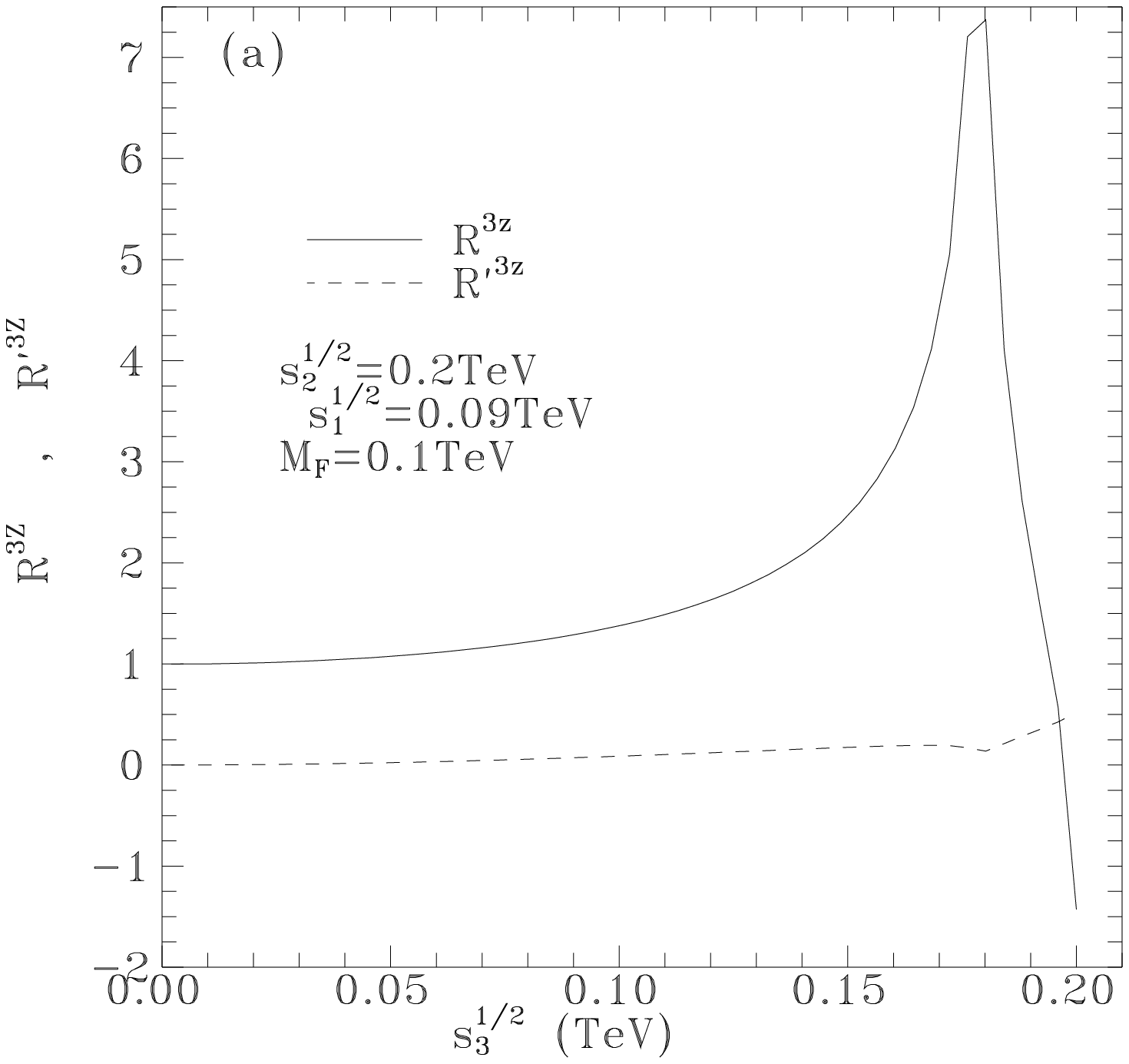,height=7.5cm}\hspace{0.5cm}
\epsfig{file=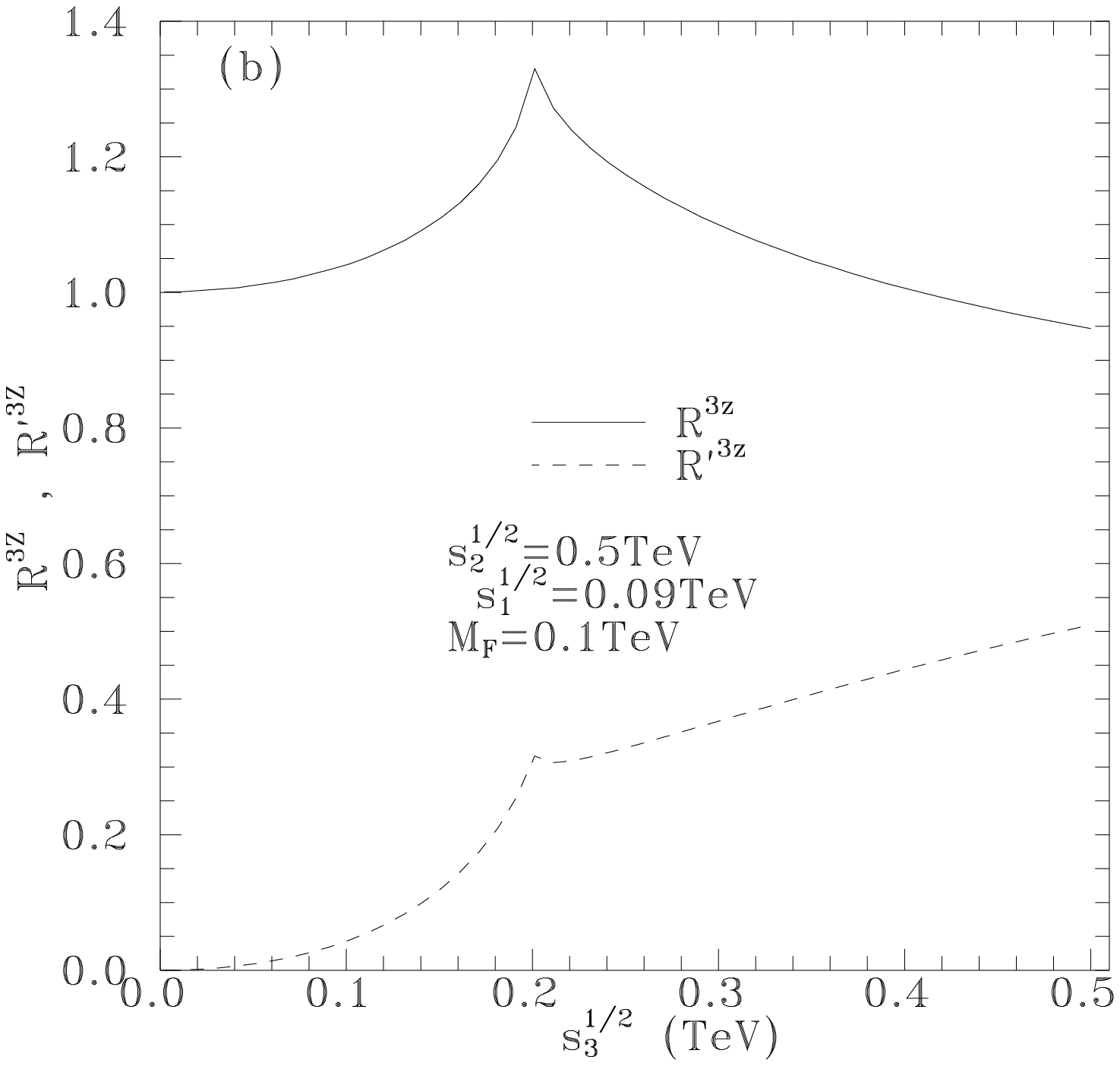,height=7.5cm}
\]
\vspace*{0.5cm}
\[
\epsfig{file=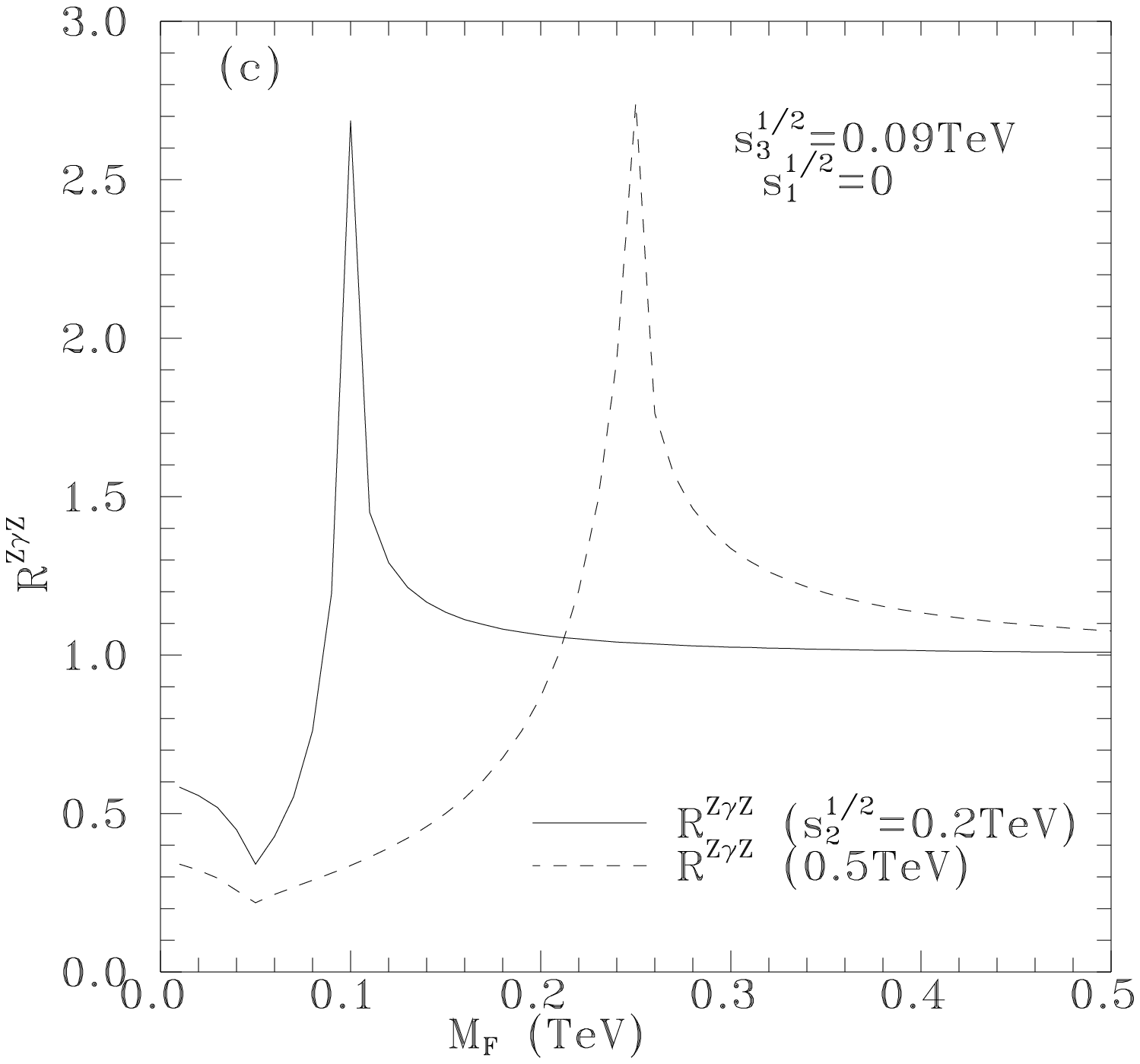,height=7.5cm,width=8.5cm}
\]
\vspace*{0.5cm}
\caption[1]{$Z^*Z^*\gamma^*$ off-shell effects as compared to
$Z^*\to Z\gamma$:
Ratio $R^{3Z}$ shows
the $\sqrt{s_3}$-dependence of the contributions to the
$h^Z_3$-type of coupling,
and ratio $R'^{3Z}$ gives the relative size, versus $\sqrt{s_3}$, of
the new contributions as compared to
the ones already existing on-shell; (a) at
$\sqrt{s_2}=0.2~TeV$, (b) at $\sqrt{s_2}=0.5~TeV$.
Ratio $R^{Z\gamma Z}$ shows the departure versus $M_F$
of the exact 1-loop contribution, as compared to
the effective Lagrangian prediction at $\sqrt{s_2}=0.2~TeV$ and
$0.5~TeV$, (c). The definitions of $s_1, s_2, s_3$ are given in
the text.}
\label{fig5}
\end{figure}

\begin{figure}[p]
\vspace*{-3cm}
\[
\epsfig{file=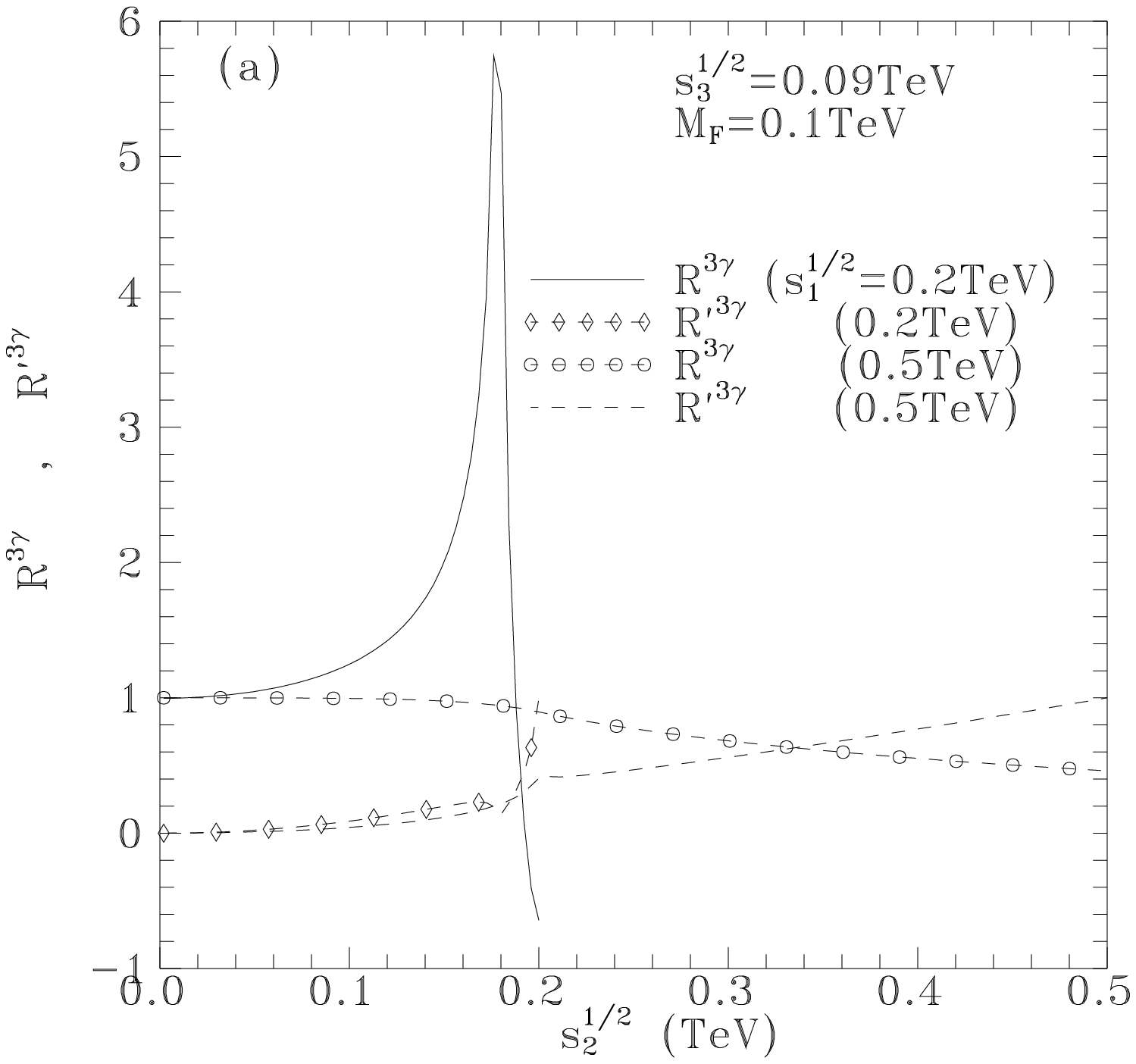,height=7.5cm}\hspace{0.5cm}
\epsfig{file=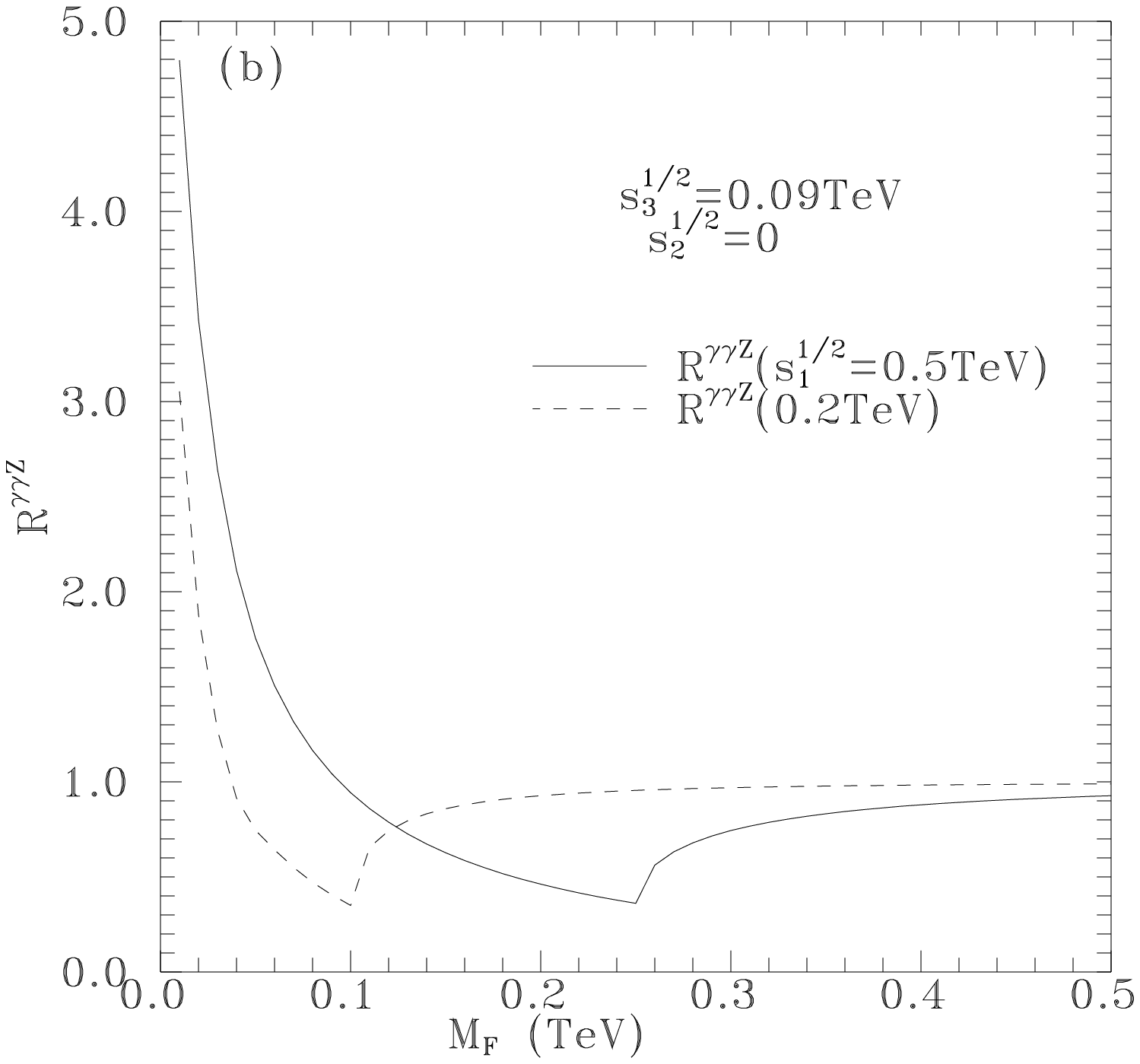,height=7.5cm}
\]
\vspace*{0.5cm}
\caption[1]{$\gamma^*\gamma^*Z^*$ off-shell effects as compared to
$\gamma^*\to Z\gamma$:
Ratio $R^{3\gamma}$  shows
the $\sqrt{s_2}$-dependence of the contributions to the
$h^{\gamma}_3$-type of coupling,
and ratio $R'^{3\gamma}$ gives the relative size, versus $\sqrt{s_3}$, of
the new contributions as compared to
the ones already existing on-shell; (a) at
$\sqrt{s_1}=0.2~TeV$ and at $\sqrt{s_1}=0.5~TeV$.
Ratio $R^{\gamma\gamma Z}$ shows the departure versus $M_F$
of the exact 1-loop contribution, as compared to
the effective Lagrangian prediction at $\sqrt{s_1}=0.2~TeV$ and
$0.5~TeV$, (b). The definitions of $s_1, s_2, s_3$ are given in
the text.}
\label{fig6}
\end{figure}

\begin{figure}[htb]
\vspace*{-1cm}
\[
\epsfig{file=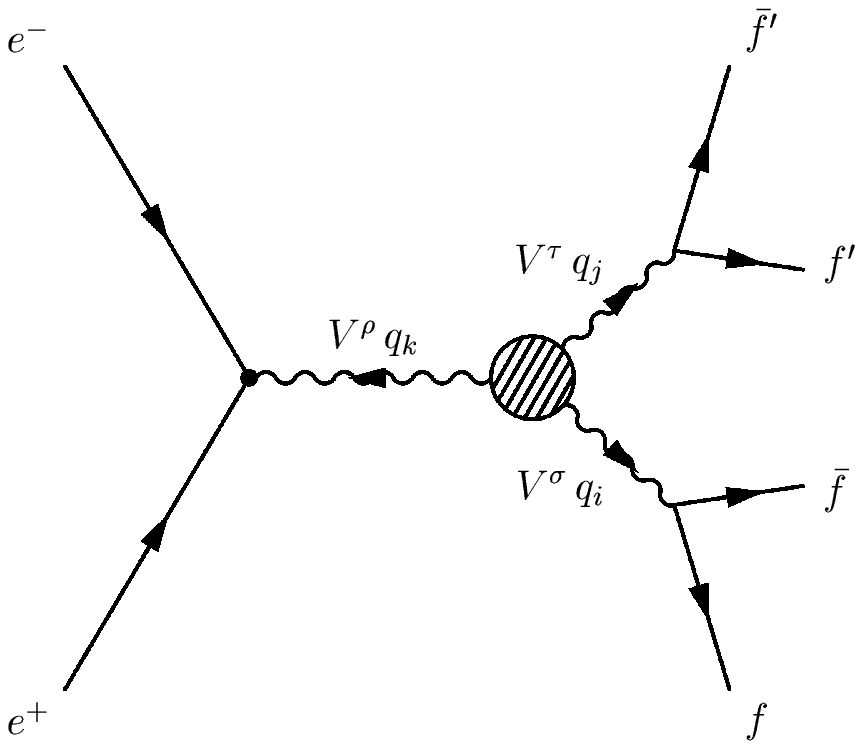,height=5cm,width=5cm}
\]
\vspace*{0cm}
\caption[2]{The $VVV$ contribution to the $e^+e^-\to(f\bar
f)(f'\bar f')$ process.}
\label{eeffff-fig}
\end{figure}

The basic triangle diagram is depicted in Fig.\ref{trif-fig}.
For simplicity we only consider the case that a single fermion is
running along the loop with the
couplings defined in\footnote{ In models like SUSY
we could also have fermion loops, where two different charginos
mix through their $Z$ couplings, while running  along the loop.
Such contributions were calculated in
\cite{modzg} in the case that
only one of the neutral gauge bosons were
off shell, and they were found to be rather small.
Here they are neglected.}  (\ref{VFF}). Only a restricted set of
CP-conserving NAGC are generated by this triangle loop (no
CP-violating coupling appear). They are explicitly given below
in terms of the Passarino-Veltman 1-loop
functions\footnote{We follow  the same notation as in the last paper
in \cite{PVHag}, but we omit the common fermion mass $M_F$ from
the arguments of the 1-loop $B_0$ and $C_0$ functions.
We also note that in this case $C_0(s_1, s_2, s_3)$ is a fully
symmetric function of $s_1, s_2, s_3$ .}  \cite{PVHag}.\\

\vspace{0.5cm}
\noindent
{\bf 1) Application to $Z^*Z^*Z^*$}\\
\noindent
\bqa
&&f^{Z^*Z^*Z^*}_1(s_1,s_2,s_3)=-~{e^2 g_{aF}\over32\pi^2s^3_Wc^3_W}
\{(3g^2_{vF}+g^2_{aF})\G_1(s_1,s_2,s_3)
-(g^2_{aF}-g^2_{vF})\G_3(s_1,s_2,s_3)\} , \nonumber\\
&&f^{Z^*Z^*Z^*}_2(s_1,s_2,s_3)={e^2 g_{aF}\over32\pi^2s^3_Wc^3_W}
\{(3g^2_{vF}+g^2_{aF})\G_2(s_1,s_2,s_3)
-(g^2_{aF}-g^2_{vF})\G_4(s_1,s_2,s_3)\}, \nonumber\\
&&f^{Z^*Z^*Z^*}_3(s_1,s_2,s_3)= 0 ~ , \nonumber \\
&&g^{Z^*Z^*Z^*}_j(s_1,s_2,s_3)={e^2\over8\pi^2s^3_Wc^3_W}g_{aF}
(3g^2_{vF}+g^2_{aF})\G'_j(s_1,s_2,s_3) ~ , \label{ZZZ-loop-fg-ap}
\eqa
where
\bqa
&&\G_1(s_1,s_2,s_3)={1\over\lambda^2}
\{C_0(s_1,s_2,s_3)[s_3(2s_3-s_1-s_2)
-(s_1-s_2)^2](\lambda M^2_F+2s_1s_2s_3)\nonumber\\
&&-{1\over2}[B_0(s_1)-B_0(s_2)](s_1-s_2)
[\lambda(2M^2_F+s_3)+12s_1s_2s_3]\nonumber\\
&&-{s_3\over2}[B_0(s_1)+B_0(s_2)-2B_0(s_3)][2\lambda
M^2_F+s^2_3(s_1+s_2)-2s_3(s^2_1+s^2_2-4s_1s_2)\nonumber\\
&&+(s_1+s_2)(s_1-s_2)^2]\}
+{2s^2_3-s_3(s_1+s_2)-(s_1-s_2)^2\over3\lambda}
\label{G1-loop}
\eqa
\bqa
&&\G_2(s_1,s_2,s_3)=-~{(s_1-s_2)(s_3-s_1-s_2)\over\lambda}
\nonumber \\
&& +
{1\over\lambda^2}\{-3(s_1-s_2)(s_3-s_1-s_2)(\lambda
M^2_F+2s_1s_2s_3)C_0(s_1,s_2,s_3)\nonumber\\
&&-{1\over2}[B_0(s_1)-B_0(s_2)] [2\lambda M^2_F(s_3-2s_1-2s_2)-s_3
(s_1+s_2)(s^2_3+s^2_1+s^2_2+14s_1s_2)\nonumber\\
&&+2s^2_3(s^2_1+s^2_2+6s_1s_2)-4s_1s_2(s_1-s_2)^2]\nonumber\\
&&+{1\over2}[B_0(s_1)+B_0(s_2)-2B_0(s_3)](s_1-s_2)(2\lambda
M^2_F+\lambda s_3+12s_1s_2s_3)\} ~ , \label{G2-loop}
\eqa
\bqa
&&\G_3(s_1,s_2,s_3)={M^2_F\over\lambda}
\{-[2s^2_3-s_3(s_1+s_2)-(s_1-s_2)^2]C_0(s_1,s_2,s_3)\nonumber\\
&&+3(s_1-s_2)[B_0(s_1)-B_0(s_2)]+3s_3[B_0(s_1)+B_0(s_2)-2B_0(s_3)]\}
\label{G3-loop}
\eqa
\bqa
&&\G_4(s_1,s_2,s_3)={3M^2_F\over\lambda}
\{(s_1-s_2)[(s_3-s_1-s_2)C_0(s_1,s_2,s_3)\nonumber\\
&& -B_0(s_1)-B_0(s_2)+2B_0(s_3)]+(s_3-2s_1-2s_2)[B_0(s_1)-B_0(s_2)]\} ~ ,
\label{G4-loop}
\eqa
while the scalar functions are determined through
\bqa
&&\G'_1(s_1,s_2,s_3)={1\over\lambda^2}\{-C_0(s_3,s_2,s_1)[\lambda
M^2_F(s_1-s_3-s_2)+s_3s_2(2s^2_1-s_1(s_3+s_2)-(s_3-s_2)^2)]\nonumber\\
&&+{1\over2}[B_0(s_3)-B_0(s_1)]s_3[s^2_1+2s_1(2s_2-s_3)+s^2_3
+4s_3s_2-5s^2_2]\nonumber\\
&&+{1\over2}[B_0(s_2)-B_0(s_1)]s_2[s^2_1+2s_1(2s_3-s_2)
+s^2_2+4s_3s_2-5s^2_3]\nonumber\\
&&-~{\lambda \over 2}(s_1-s_3-s_2)\} ~ \label{G1p-loop}
\eqa
and the  Bose result
\bqa
&&\G'_3(s_1,s_2,s_3)=\G'_3(s_2,s_1,s_3)=\G'_2(s_1,s_3,s_2)=
\G'_2(s_2,s_3,s_1) \nonumber \\
&& =\G'_1(s_3,s_2,s_1)=\G'_3(s_3,s_1,s_2) ~~ , \label{G123p-loop}
\eqa
derived  from (\ref{ZZZ-CPc-Bose}) and $f_3^{Z^*Z^*Z^*}=0$.
 In all cases we define
\bq
\lambda=s^2_3+s^2_1+s^2_2-2s_1s_2-2s_3(s_1+s_2) ~ .
\label{lampda-loop}
\eq \\

\vspace{0.5cm}
\noindent
{\bf 2) Application to $Z^*Z^*\gamma^*$}\\
\noindent
\bqa
&&f^{Z^*Z^*\gamma^*}_1(s_1,s_2,s_3)=-~{e^2Q_Fg_{aF}g_{vF}\over
8\pi^2s^2_Wc^2_W}~[\G_1(s_1,s_2,s_3)+\G_5(s_1,s_2,s_3)] ~,
\nonumber \\
&&f^{Z^*Z^*\gamma^*}_2(s_1,s_2,s_3)={e^2Q_Fg_{aF}g_{vF}\over
8\pi^2s^2_Wc^2_W}~[\G_2(s_1,s_2,s_3)+{1\over3}\G_4(s_1,s_2,s_3)]
~ , \nonumber \\
&&g^{Z^*Z^*\gamma^*}_1(s_1,s_2,s_3)=
g^{Z^*Z^*\gamma^*}_2(s_2,s_1,s_3)={e^2Q_Fg_{aF}g_{vF}\over
2\pi^2s^2_Wc^2_W}~ \G'_1(s_1,s_2,s_3) ~ , \label{ZZgamma-loop-fg-ap}
\eqa
where the only new function not already appearing in the
$Z^*Z^*Z^*$ case is
\bqa
&&\G_5(s_1,s_2,s_3)={M^2_F\over \lambda}
\Big \{-[s_3(s_1+s_2)-(s_1-s_2)^2]C_0(s_1,s_2,s_3)\nonumber\\
&&
+[B_0(s_1)-B_0(s_2)](s_1-s_2)
+s_3[B_0(s_1)+B_0(s_2)-2B_0(s_3)]\Big \}+{1\over3} ~.
\label{G5-loop}
\eqa \\

\vspace{0.5cm}
\noindent
{\bf 3) Application to $\gamma^*\gamma^*Z^*$}\\
\noindent
\bqa
&&f^{\gamma^*\gamma^*Z^*}_1(s_1,s_2,s_3)=-~{e^2Q^2_Fg_{aF}\over
8\pi^2s_Wc_W}~ [\G_6(s_1,s_2,s_3)+\G_7(s_1,s_2,s_3)]
 ~ , \nonumber\\
&&f^{\gamma^*\gamma^*Z^*}_2(s_1,s_2,s_3)=-~{e^2Q^2_Fg_{aF}\over
8\pi^2s_Wc_W} ~[ \G_6(s_1,s_2,s_3)-\G_7(s_1,s_2,s_3)]
~, \nonumber\\
&&g^{\gamma^*\gamma^*Z^*}_1(s_1,s_2,s_3)=~{e^2Q^2_Fg_{aF}\over
2\pi^2s_Wc_W}\G'_1(s_3,s_2,s_1) ~ , \label{gammagammaZ-loop-fg-ap}
\eqa
\bqa
&&\G_6(s_1,s_2,s_3)={1\over \lambda^2}
\Big \{-2s_1C_0(s_3,s_2,s_1)[\lambda
M^2_F(s_1-s_3-s_2)-s^3_3s_2+s^2_3s_2(2s_2-s_1)\nonumber\\
&&+s_3s_2(2s^2_1-s_1s_2-s^2_2)]-{s_1\over2}[B_0(s_1)-B_0(s_2)]
[s^3_3-2s^2_3(s_1+3s_2)+s_3(s^2_1+12s_1s_2+3s^2_2)\nonumber\\
&&+2s_2(s_1-s_2)^2]-{s_3s_1\over2}[B_0(s_1)+B_0(s_2)-2B_0(s_3)]
[s^2_3 \nonumber\\
&& +2s_3(2s_2-s_1)+s^2_1+4s_1s_2-5s^2_2] \Big \}
-{s_1(s_1-s_3-s_2)\over\lambda} ~ , \label{G6-loop}
\eqa
\bqa
&&\G_7(s_1,s_2,s_3)={1\over \lambda^2}\Big
\{-2s_2C_0(s_3,s_2,s_1)[\lambda
M^2_F(s_2-s_3-s_1)-s^3_3s_1+s^2_3s_1(2s_1-s_2)\nonumber\\
&&-s_3s_1(s^2_1+s_1s_2-2s^2_2)]+{s_2\over2}[B_0(s_1)-B_0(s_2)]
[s^3_3-2s^2_3(s_2+3s_1)+s_3(s^2_2+12s_1s_2+3s^2_1)\nonumber\\
&&+2s_1(s_1-s_2)^2]-{s_2s_3\over2}[B_0(s_1)+B_0(s_2)-2B_0(s_3)]
[s^2_3+2s_3(2s_1-s_2)\nonumber\\
&& +s^2_2+4s_1s_2-5s^2_1]\Big\}-~{s_2(s_2-s_3-s_1)\over\lambda} ~.
\label{G7-loop}
\eqa

In principle the triangular  graph in
Fig.\ref{trif-fig} (with a single fermion of mass $M_F$ running
along it), could also include ambiguous  axial anomaly
contributions. Such contributions do not have the structure of
a  self interaction among three neutral gauge bosons, and
they are presumably cancelled   by other (possibly extremely heavy)
 fermions. The cancellation of these
anomalous contributions is easily imposed by requiring that all
$\G_j$ and $\G'_j$ functions defined above vanish in the limit
$(M_F^2 \gg |s_1|,~ |s_2|,~ |s_3|)$. Thus, for cancelling the anomaly
in the actual calculation of the functions above, we occasionally needed
to subtract an appropriate $M_F$-independent term.

\clearpage
\newpage

\newpage

\clearpage

\clearpage
\newpage

\clearpage

\newpage


\begin{thebibliography}{99}

%
\bibitem{Hagn} K. Hagiwara, R.D. Peccei and D. Zeppenfeld,
\np{B282}{253}{1987}.
%
\bibitem{GG} K.J.F. Gaemers and G.J. Gounaris,
\zp{C1}{259}{1979}.
%
\bibitem{Baur} U. Baur, E.L. Berger, \pr{D47}{4889}{1993};
U. Baur, T Han anf J. Ohnemus, \pr{D57}{2823}{1998} and references
therein.
%
\bibitem{Wudka}   J.Wudka, Int. J. Mod. Phys. {\bf A9}, 2301 (1994);
J. Ellison and J. Wudka, \arnps{48}{33}{1988}.
%
\bibitem{work}  H. Aihara et. al., Summary of the Working Subgroup on
Anomalous Gauge Boson Interactions of the DPF Long-Range Planning Study,
to be published in ``Electroweak Symmetry Breaking and Beyond the
Standard Model'', Editors T. Barklow, S. Dawson,
H. Haber and J. Siegrist,
LBL-37155, hep-ph/9503425.
%
\bibitem{Biebel} J. Biebel, \pl{B448}{125}{1999}.
%
\bibitem{Alcaraz} J. Alcaraz, report presented at the ECFA meeting,
Oxford, March 1999.
%
\bibitem{neut}G.J. Gounaris, J. Layssac and F.M. Renard,
hep-ph/9910395, \pr{D61}{073013}{2000}.
%
\bibitem{modzg} G.J. Gounaris, J. Layssac and F.M. Renard,
hep-ph/0003143; to appear in Phys.Rev.D.
%
\bibitem{TEVZg} CDF Collaboration, F. Abe \etal \prl{74}{1936}{1995};
D0 Collaboration, S. Abachi \etal\pr{D56}{6742}{1997}; Report of the
Working Group on Photon and Weak Boson Production, U.Baur et al,
hep-ph/0005226.
%
\bibitem{expLEP2}
ALEPH coll,CERN-EP-99-141, \pl{B469}{287}{1999};
CERN-OPEN-99-275,
prepared for Int Eur, Conf. on HEP, Tampere,
Finland Jul 1999.\\
DELPHI coll (C. Matteuzzi et al) CERN-OPEN-2000-025;
(R.Jacobson) Lake Louise 1999, Electroweak Physics 405.\\
L3 Collaboration, M. Acciari \etal,\pl{B436}{187}{1998};
\pl{B450}{281}{1999};
(M.A. Falagan for L3 Coll.) Lake Louise 1999, Electroweak Physics 371.\\
OPAL Coll (G.Abbiendi et al) CERN-EP-2000-017, \pl{B476}{256}{2000}.
%
\bibitem{LC} Opportunities
and Requirements for Experimentation at a Very High Energy
$e^{+}e^{-}$ Collider, SLAC-329(1928); Proc. Workshops on Japan
Linear Collider, KEK Reports, 90-2, 91-10 and 92-16;
P.M. Zerwas, DESY 93-112, Aug. 1993; Proc. of the Workshop on
$e^{+}e^{-}$ Collisions at 500 GeV: The Physics Potential, DESY
92-123A,B,(1992), C(1993), D(1994), E(1997) ed. P. Zerwas;
E. Accomando \etal\@ \prep{C299}{1}{1998}.
%
\bibitem{CLIC} " The CLIC study of a multi-TeV $e^+e^-$ linear
collider", CERN-PS-99-005-LP (1999).
%
\bibitem{LEP2off} L. Conti, L. Pieri, R. Sekulin,
private communications.
%
\bibitem{AFS} J. Alcaraz, M.A. Falagan and E. Sanchez,
\pr{D61}{075006}{2000}.
%
\bibitem{FMR} F.M. Renard,\np{B196}{93}{1982}.
%
\bibitem{BBCD} A. Barroso, F. Boudjema,
J. Cole and N. Dombey,\zp{C28}{149}{1985}.
\bibitem{PVHag} G. Passarino and M. Veltman, \np{B160}{151}{1979};
 K. Hagiwara, S. Matsumoto, D. Haidt and C.S. Kim,
\zp{C64}{559}{1995}.
%




\end{thebibliography}
\end{document}


\pagenumbering{arabic}
\thispagestyle{empty}
\def\thefootnote{\fnsymbol{footnote}}
\setcounter{footnote}{1}

\begin{center}
{\Large\bf Addendum to\\ 
Off-shell structure of the anomalous $Z$ and
$\gamma$ self-couplings
\footnote{Partially
supported by EU contract  HPRN-CT-2000-00149.}}\\
\vspace{0.2cm}
Phys. Rev. {\bf D62},073012(2000)\\
\vspace{0.5cm}
{\large G.J. Gounaris$^a$, J. Layssac$^b$  and F.M. Renard$^b$}\\
\vspace{0.7cm}
$^a$Department of Theoretical Physics, Aristotle
University of Thessaloniki,\\
Gr-54006, Thessaloniki, Greece.\\
\vspace{0.2cm}
$^b$Physique
Math\'{e}matique et Th\'{e}orique,
UMR 5825\\
Universit\'{e} Montpellier II,
 F-34095 Montpellier Cedex 5.\\

\end{center}

\vspace{2cm}

The experimental search for direct couplings among three neutral gauge
bosons has recently received much attention \cite{DELPHI}. 
In a first step
the studies were done in the case one off-shell gauge boson ($\gamma$ or
$Z$) couples to two on-shell ones, using the description
established in \cite{Hagiwara}  with the modifications made in 
\cite{ntgc1}.
In a second step the studies were extended to the case in which 
the three gauge bosons are off-shell, allowing to treat a larger number
of available events corresponding to the decays of the final 
gauge bosons into various types of fermion pairs. 
The general description of the
couplings among three off-shell neutral gauge bosons has been
established in  \cite{ntgc2}. In the most general case a large set of
independent Lorentz invariant $ZZZ$, $ZZ\gamma$, $\gamma\gamma Z$
coupling forms appear, making the experimental analysis rather
involved. 

However, in \cite{ntgc2} it was already noticed that if one
demands that these couplings preserve the $SU(2)\times U(1)$ invariance,
and if we only retain the lowest dimension ($dim=8$) operators
 affecting exclusively the neutral gauge boson  and/or Higgs interactions, 
then  only one CP-conserving  and one
CP-violating operator are allowed, namely (see eq.(41) of \cite{ntgc2}):

\bqa
{\cal O}_{SU(2)\times U(1)} & = &
i \tilde{B}_{\mu\nu} (\partial_\sigma B^{\sigma\mu})
(\Phi^\dagger D^\nu  \Phi) ~~ ,  \nonumber \\
\tilde{{\cal O}}_{SU(2)\times U(1)} & = &
i B_{\mu\nu} (\partial_\sigma B^{\sigma\mu})
(\Phi^\dagger D^\nu  \Phi) ~~ . \label{SU2U1-op}
\eqa

The effective Lagrangian
\bq
{\cal L}=  f {\cal O}_{SU(2)\times U(1)}
+  \tilde{f}\tilde{{\cal O}}_{SU(2)\times U(1)} 
\label{NP-Lagrangian}
\eq 
\noindent
generates simultaneously the following $ZZZ$, $ZZ\gamma$ and 
$\gamma\gamma Z$ couplings, in the notations of \cite{ntgc2}:

\bq
l^{ZZ\gamma}_1=-l^{ZZ\gamma}_2={c_W\over s_W}l^{ZZZ}_1
=-{s_W\over c_W}l^{\gamma\gamma Z}_1={ v^2 \over4} f
\label{l1}\eq
\bq
\tilde{l}^{ZZ\gamma}_1=-\tilde{l}^{ZZ\gamma}_3={c_W\over s_W}
\tilde{l}^{ZZZ}_1
=-{s_W\over c_W}\tilde{l}^{\gamma\gamma Z}_1
={ v^2 \over4} \tilde{f}
\label{l1t}
\eq
The analysis of the experimental results for  off-shell 
 (as well as  on-shell)  events
can then be performed with only two arbitrary parameters.

If we further restrict to on-shell case, then the couplings 
defined in [2,3,4] may be used constrained as  
\bq
h^Z_3=-{s_W\over c_W} h^{\gamma}_3=-f^{\gamma}_5={c_W\over s_W}f^Z_5
=m^2_Z{ v^2 \over4} f
\label{hc}
\eq

\bq
h^Z_1=-{s_W\over c_W} h^{\gamma}_1=-f^{\gamma}_4={c_W\over s_W}f^Z_4
=m^2_Z{ v^2 \over4} \tilde{f}
\label{hv}\eq

Finally a few comments may be added concerning the motivation for
restricting to the two operators in (\ref{SU2U1-op}).
If we  take the point of view that there is no new physics contribution
to the anomalous couplings affecting  $W^\pm$,
then we can assume that it might only affect the neutral gauge boson
and/or the Higgs interactions.
Assuming further that   the responsible  effective scale 
$\Lambda_{NP}$ is much larger than the $Z$ mass and the  
energy range of the present colliders, we are then led  to
considering only the $SU(2)\times U(1)$ gauge  invariant forms,
which induce exclusively neutral gauge boson couplings and have 
 the lowest  possible dimension  $dim=8$.

The use of relations (\ref{l1},\ref{l1t}) or (\ref{hc},\ref{hv})
when analyzing experimental data should allow to get much more stringent
constraints on the couplings.
To our knowledge this had not been pointed out in the literature.\\

We thank Robert Sekulin for having drawn our attention on this
point.